\documentclass[prx,aps,amsfonts,amsmath,amssymb,twocolumn,superscriptaddress]{revtex4-2}
\usepackage{graphicx}
\usepackage{dcolumn}
\usepackage{appendix}
\usepackage{url}  
\usepackage{amsmath}
\usepackage{color}
\usepackage{bm}
\usepackage{multirow}
\usepackage{float}

\usepackage{physics}
\usepackage{amsfonts, amsmath, mathrsfs}
\usepackage[dvipsnames]{xcolor}
\usepackage{tikz}
\usepackage{braket}
\usepackage{svg}
\usepackage[version=4]{mhchem}
\usepackage[colorlinks=true, linkcolor=blue]{hyperref}
\usepackage{xspace}

\newcommand{\CGT}{\ce{Cr_2Ge_2Te_6}\xspace} 
\newcommand{\CrMoment}{\ce{Cr^{3+}}\xspace}

\newcommand{\eff}{\mathrm{eff}}

\newcommand{\EdelZeeman}{\mathrm{EZ}}

\begin{document}

\title{Manipulation of ferromagnetism with a light-driven nonlinear Edelstein–Zeeman field}

\author{Yinchuan Lv}
\thanks{These authors contributed equally to this work.}
\affiliation{Department of Physics, The Grainger College of Engineering, University of Illinois Urbana-Champaign, Urbana, Illinois 61801, USA}
\affiliation{Materials Research Laboratory, The Grainger College of Engineering, University of Illinois Urbana-Champaign, Urbana, Illinois 61801, USA}

\author{W. Joe Meese}
\thanks{These authors contributed equally to this work.}
\affiliation{Department of Physics, The Grainger College of Engineering, University of Illinois Urbana-Champaign, Urbana, Illinois 61801, USA}
\affiliation{Anthony J. Leggett Institute for Condensed Matter Theory, The Grainger College of Engineering, University of Illinois Urbana-Champaign, Urbana, Illinois 61801, USA}

\author{Azel Murzabekova}
\affiliation{Department of Physics, The Grainger College of Engineering, University of Illinois Urbana-Champaign, Urbana, Illinois 61801, USA}
\affiliation{Materials Research Laboratory, The Grainger College of Engineering, University of Illinois Urbana-Champaign, Urbana, Illinois 61801, USA}

\author{Jennifer Freedberg}
\affiliation{Department of Physics, The Grainger College of Engineering, University of Illinois Urbana-Champaign, Urbana, Illinois 61801, USA}
\affiliation{Materials Research Laboratory, The Grainger College of Engineering, University of Illinois Urbana-Champaign, Urbana, Illinois 61801, USA}

\author{Changjun Lee}
\affiliation{Department of Physics, The Grainger College of Engineering, University of Illinois Urbana-Champaign, Urbana, Illinois 61801, USA}
\affiliation{Materials Research Laboratory, The Grainger College of Engineering, University of Illinois Urbana-Champaign, Urbana, Illinois 61801, USA}

\author{Yiming Sun}
\affiliation{Department of Physics, The Grainger College of Engineering, University of Illinois Urbana-Champaign, Urbana, Illinois 61801, USA}
\affiliation{Materials Research Laboratory, The Grainger College of Engineering, University of Illinois Urbana-Champaign, Urbana, Illinois 61801, USA}

\author{Joshua Wakefield}
\affiliation{Department of Physics, Massachusetts Institute of Technology, Cambridge, MA 02139, USA}

\author{Takashi Kurumaji}
\affiliation{Division of Physics, Mathematics and Astronomy,
California Institute of Technology, Pasadena, California 91125, USA}

\author{Joseph Checkelsky}
\affiliation{Department of Physics, Massachusetts Institute of Technology, Cambridge, MA 02139, USA}

\author{Fahad Mahmood}
\affiliation{Department of Physics, The Grainger College of Engineering, University of Illinois Urbana-Champaign, Urbana, Illinois 61801, USA}
\affiliation{Materials Research Laboratory, The Grainger College of Engineering, University of Illinois Urbana-Champaign, Urbana, Illinois 61801, USA}

\date{March 5, 2026}

\maketitle

\noindent \textbf{Optical control of magnetization is often symmetry-forbidden because electric fields and magnetization transform differently under inversion and time-reversal. However, through even-order nonlinear response, optical excitation can generate a nonequilibrium magnetic density (the nonlinear Edelstein effect) that acts as an internal Edelstein–Zeeman field coupling to slower magnetic degrees of freedom. Here we demonstrate non-thermal, ultrafast optical control of ferromagnetism in the centrosymmetric van der Waals semiconductor Cr$_2$Ge$_2$Te$_6$ via a resonant nonlinear Edelstein effect. Using time-domain THz emission spectroscopy under near-infrared excitation, we directly observe magnetic dipole radiation arising from optically driven magnetization dynamics. The polarization, fluence, and temperature dependences of the THz emission are quantitatively captured by a mean-field description of a weakly anisotropic Heisenberg ferromagnet subject to an Edelstein–Zeeman field. Our results establish a general nonequilibrium route to optical control of magnetism in centrosymmetric materials.}

\bigskip

The Edelstein effect describes the generation of magnetization by an applied electric field through spin–orbit coupling. In its linear form, the Edelstein effect is captured phenomenologically as $\delta M_i = \gamma_{ij}^{(1)}J_j^{\phantom{1}}$, where $\delta M$ is the induced magnetization, $\bm J$ is the applied current, and $\gamma_{ij}^{(1)}$ is a material-dependent response tensor. Given that the current density $\bm J$ is directly proportional to the electric field $\bm{E}$, the linear Edelstein effect is a nonequilibrium version of a linear magneto-electric: $\delta M_i = \alpha_{ik}^{(1)} E_k^{\phantom{1}}$ \cite{edelsteinSpinPolarizationConduction1990, schmidMagnetoelectricClassificationMaterials1973, fiebigRevivalMagnetoelectricEffect2005}. Crucially, this effect can only exist in non-centrosymmetric materials since a current-driven nonequilibrium spin density arises only when electronic states at opposite momenta are inequivalent \cite{edelsteinSpinPolarizationConduction1990, xuLightinducedStaticMagnetization2021,oikeImpactElectronCorrelations2024}.

\begin{figure}
    \centering    \includegraphics[width=\columnwidth]{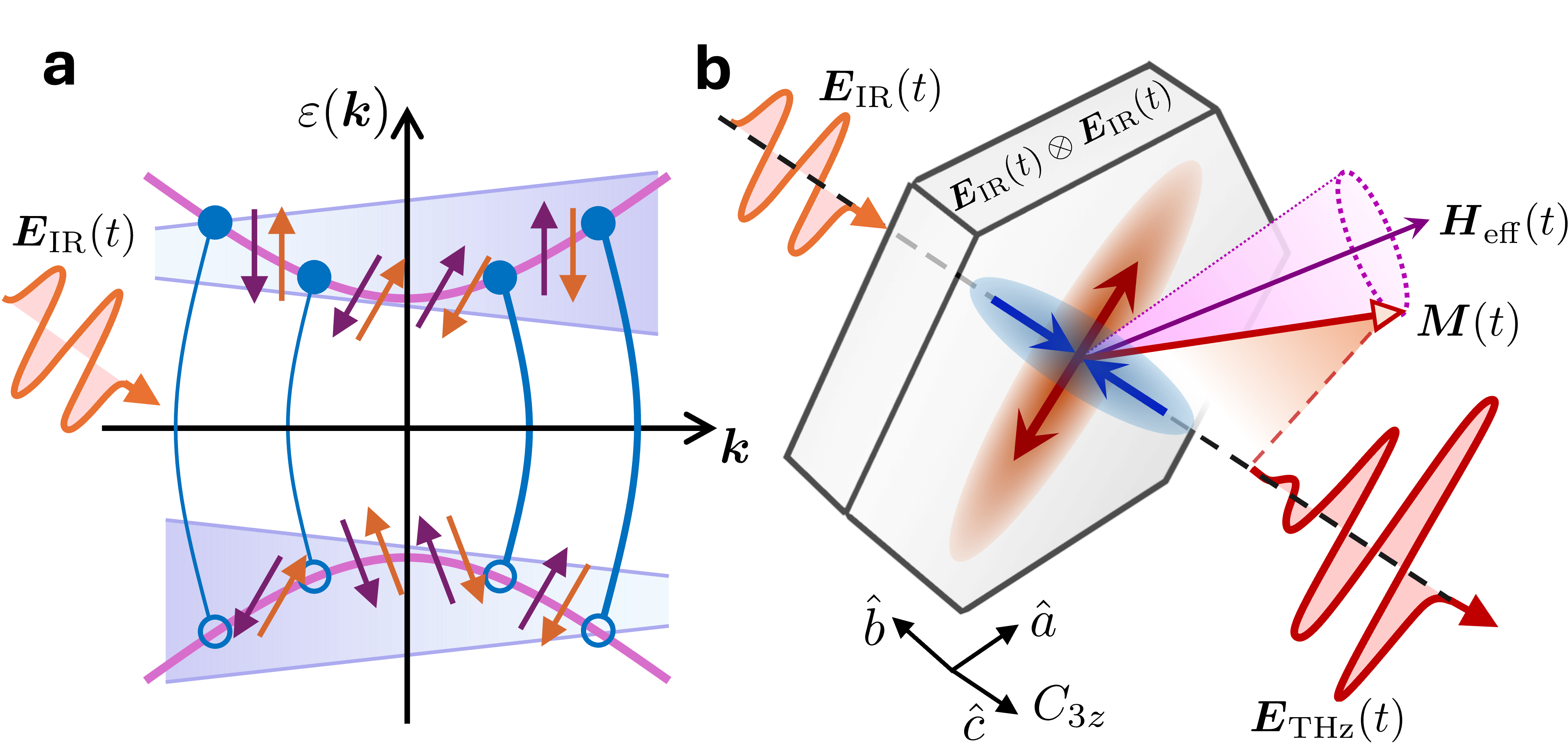}
    \caption{\textbf{Illustration of ferromagnetic control through the Edelstein-Zeeman field.} In the presence of intense above-gap infrared electric field, $\bm{E}_{\rm{IR}}(t)$, \textbf{a.} the nonlinear Edelstein effect is produced in a globally centrosymmetric semiconducting sample, leading to a nonequilibrium inter-band coherence that rectifies with $\bm{E}_{\rm{IR}}(t)$ at second-order. Spin-orbit coupling at locally non-centrosymmetric sites in the unit cell will produce a hidden spin texture in the bands indicated by the purple and orange arrows which form Kramers' pairs at each momentum. The nonequilibrium electric dipole currents and electric quadrupole transitions dynamically split the Kramers' degeneracy, contributing to the \textbf{b.} total external field $\bm{H}_\eff(t)$ capable of driving the slower dynamics of the ferromagnetic moment, $\bm{M}(t)$. Consequently, low-frequency THz radiation is emitted through the magnetic dipole channel. }
    \label{fig:general idea}
\end{figure}

In contrast, the nonlinear Edelstein effect can induce magnetization even in centrosymmetric crystals. Here the electric field appears at second order as $\delta M_i =  \alpha_{ijk}^{(2)} E_j^{\phantom{2}} E_k^{\phantom{2}}$, where $\alpha^{(2)}_{ijk}$ is a material-dependent nonlinear response tensor \cite{xuLightinducedStaticMagnetization2021, xiaoIntrinsicNonlinearElectric2022,xiaoTimeReversalEvenNonlinearCurrent2023, oikeImpactElectronCorrelations2024, xueNonlinearOpticsDrivenSpin2024, arnoldiRevealingHiddenSpin2024}. 
Unlike its linear counterpart, the nonlinear Edelstein effect arises from the interaction of the photoexcited carriers with the external electric field, rather than due to the equilibrium distribution of carriers. As illustrated in Fig.~\ref{fig:general idea}a, intense above-gap infrared excitation generates a transient inter-band coherence that rectifies with the driving electric field at second order. The nonequilibrium photo-excited state breaks time-reversal symmetry outright, and can dynamically generate either spin or orbital magnetic density when they flow through a crystal with local non-centrosymmetry or nontrivial quantum geometry \cite{xiaoBerryPhaseEffects2010, zhangHiddenSpinPolarization2014, ryooHiddenOrbitalPolarization2017, liElectronicStructureFerromagnetic2018,dongBerryPhaseEffects2020, huangHiddenSpinPolarization2020}. 

This itinerant magnetic density contributes to an effective internal field, the Edelstein–Zeeman field, which enters the total effective magnetic field acting on localized moments (Fig.~\ref{fig:general idea}b). The resulting time-dependent $\bm{H}_{\eff}(t)$ drives low-frequency dynamics of the ferromagnetic order parameter, leading to magnetic dipole radiation in the THz regime. Because the rectification process depends on both the intensity and polarization of the incident light, the Edelstein-Zeeman field, and thus the magnetic response can be tuned optically. As we show in this work, this mechanism enables non-thermal optical control of localized magnetic moments in insulating or semiconducting ferromagnets, even when the crystal structure is globally centrosymmetric.

\begin{figure*}
    \centering
    \includegraphics[width=17.5cm]{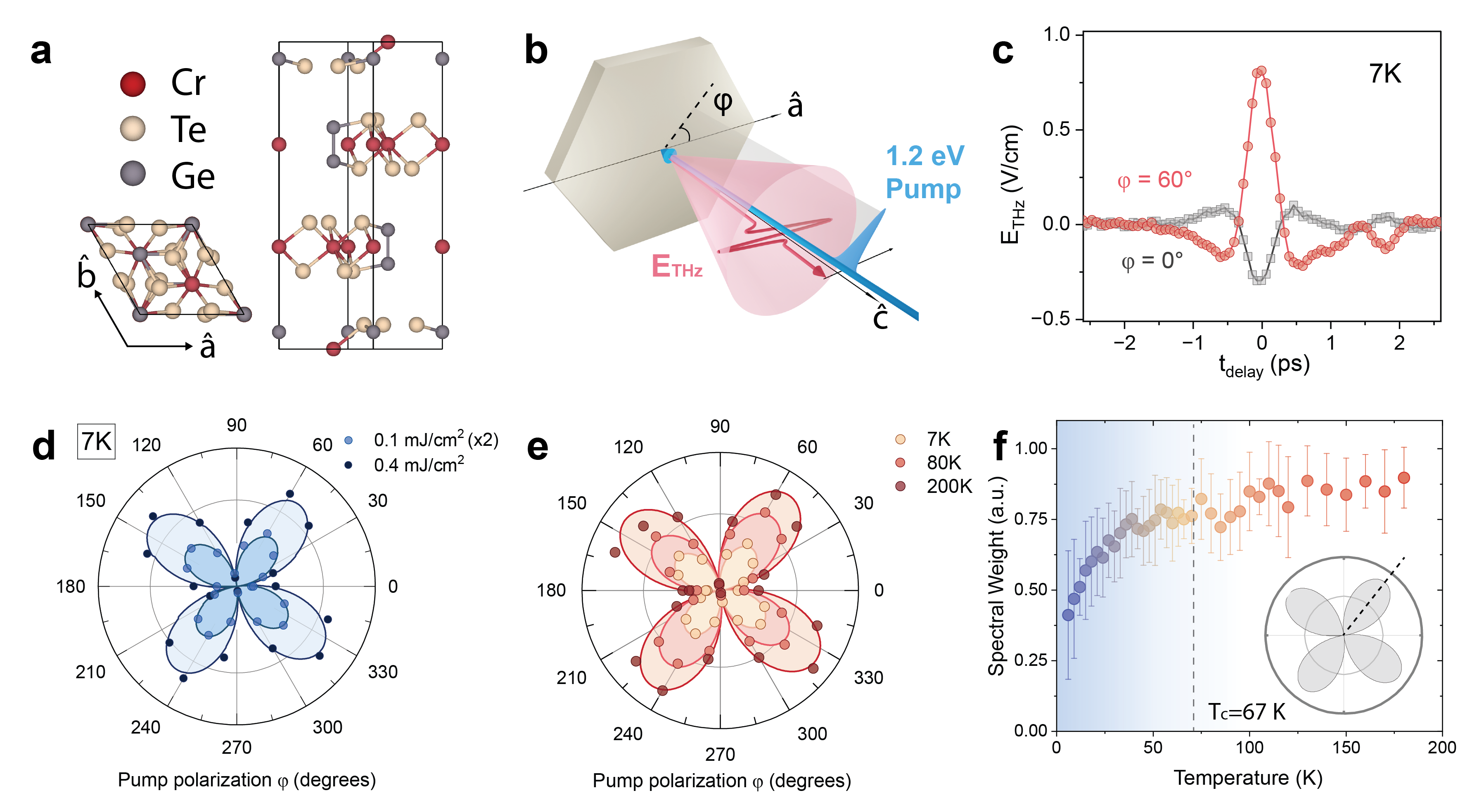}
    \caption{\textbf{THz emission from Cr$_2$Ge$_2$Te$_6$.} \textbf{a.} Crystal structure of bulk Cr$_2$Ge$_2$Te$_6$. \textbf{b.} Schematic of the THz emission experiment. Near-infrared pump pulses (1.2~eV) are incident normal to the crystallographic \textit{ab}-plane. The pump polarization angle $\varphi$, defined with respect to the crystal \textit{a}-axis, is controlled using a half-wave plate, and the emitted THz field component parallel to the \textit{a}-axis is detected. \textbf{c.} Time-domain THz emission waveform $E{\mathrm{THz}}$ as a function of pump-delay time $t{\mathrm{delay}}$. \textbf{d.} Pump polarization dependence of the THz emission amplitude at 7~K for low and high pump fluence. \textbf{e.} Pump polarization dependence of the THz emission measured at 7~K (ferromagnetic phase), 80~K (just above $T_c$), and 200~K (paramagnetic phase), with the fluence fixed at 100~$\mu$J/cm$^2$. \textbf{f.} Temperature dependence of the THz emission amplitude at a fluence of 100~$\mu$J/cm$^2$. The ferromagnetic phase is shaded in blue, and $T_c$ denotes the Curie temperature. Inset: dashed line indicates the pump polarization angle used for the temperature-dependent measurements.}
    \label{fig:norm incidence}
\end{figure*}

We use time-domain THz emission spectroscopy to demonstrate the proposed magnetic control via an optically generated nonlinear Edelstein-Zeeman field in \CGT, a bulk centrosymmetric van der Waals ferromagnetic semiconductor ~\cite{carteauxCrystallographicMagneticElectronic1995, jiFerromagneticInsulatingSubstrate2013, zeisnerMagneticAnisotropySpinpolarized2019}. \CGT has recently gained renewed interest as a model van der Waals ferromagnet in both its bulk and monolayer forms~\cite{gongDiscoveryIntrinsicFerromagnetism2017,fangLargeMagnetoopticalEffects2018,liElectronicStructureFerromagnetic2018,hanTopologicalMagneticSpinTextures2019,yuPressureInducedStructuralPhase2019,renTunableElectronicStructure2021,fengSpinFilteringEffect2022,sutcliffeTransientMagnetoopticalSpectrum2023,khelaLaserinducedTopologicalSpin2023,leeMagneticStatesVan2024,chakkarBrokenWeakStrong2024,trzaskaChargeDopingSpin2025}. We photo-excite \CGT using an ultrafast near-infrared (NIR) pulse with energy 1.2~eV (near-resonant with the direct-gap at the $\Gamma$ point \cite{jiFerromagneticInsulatingSubstrate2013, liElectronicStructureFerromagnetic2018, fangLargeMagnetoopticalEffects2018, sutcliffeTransientMagnetoopticalSpectrum2023}). The resulting transient photocurrent flows through a globally centrosymmetric, but locally non-centrosymmetric crystal field environment given that none of the occupied Wyckoff positions in \CGT has an inversion center \cite{jiFerromagneticInsulatingSubstrate2013}. This opens up the possibility for strong local spin-orbit effects, allowing for the photocurrent to carry spin and orbital angular momentum, thus producing an effective itinerant ``Edelstein-Zeeman field'' that couples to the ferromagnetism within \CGT's sub-valent \CrMoment band (Supplementary Information). We monitor the resulting change in the magnetization by measuring the consequent magnetic dipole radiation from the sample.

Figure~\ref{fig:norm incidence}c shows the time-domain THz emission from Cr$_2$Ge$_2$Te$_6$ at a temperature of 7~K upon photoexcitation with the NIR pump at normal incidence, and fluence 100~$\mu$J/cm$^2$. The emitted signal $S_{THz}$ is detected along the crystal \textit{a}-axis while the pump polarization, defined by angle $\varphi$, is rotated in the \textit{ab}-plane. Two representative time traces are shown for $\varphi=0^\circ$ and $\varphi=60^\circ$ in Fig.~\ref{fig:norm incidence}c. 
Overall, the emission amplitude ($|S_{THz}|_{max}$) exhibits a striking fourfold symmetry as a function of $\varphi$ (Fig.~\ref{fig:norm incidence}d). 
The four-fold symmetry of the THz emission with $\varphi$ is robust against changes in pump fluence as shown in Fig.~\ref{fig:norm incidence}d for two different incident NIR pump fluences. A similar insensitivity is observed in the temperature dependence for the four-fold symmetry. Figure~\ref{fig:norm incidence}e shows $|S_{THz}|_{max}$ with $\varphi$ at 7~K, 80~K, and 200~K, well below, near, and well above the ferromagnetic ordering temperature $T_c \sim 66$~K of \CGT. Although the overall THz amplitude increases with temperature, the emission pattern consistently preserves its four-fold character. To study the temperature dependence further, we fix the pump polarization to $\varphi = 60^\circ$ (around the angle of maximal emission) and plot $|S_{THz}|_{max}$ with temperature in Fig.~\ref{fig:norm incidence}f. We find that the THz field decreases gradually with cooling and then undergoes a more rapid reduction upon crossing $T_c$.

\begin{figure*}
    \centering
\includegraphics[width=17cm]{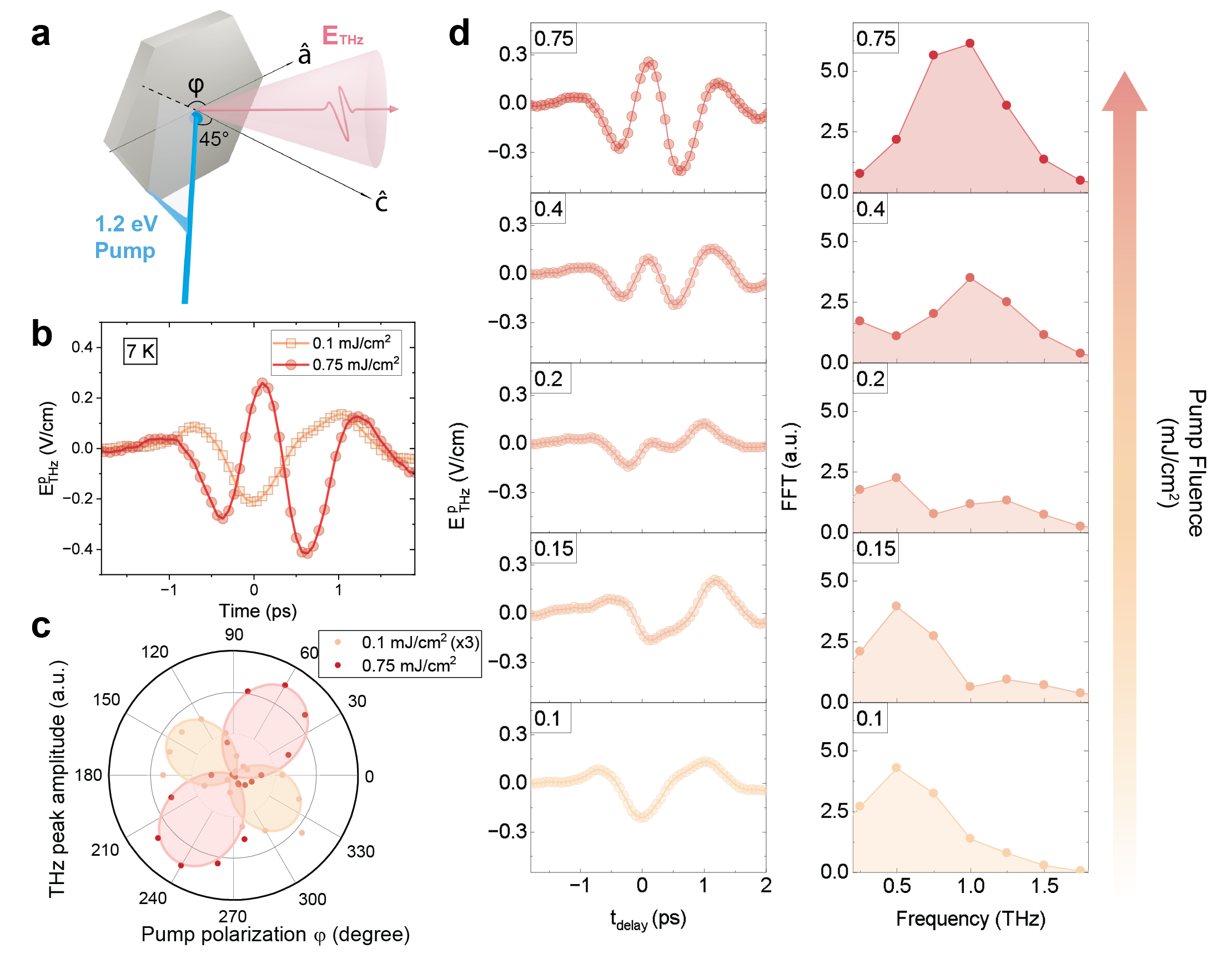}
    \caption{\textbf{THz emission from Cr$_2$Ge$_2$Te$_6$ at 45$^\circ$ incidence.} \textbf{a.} Schematic of the THz emission geometry at 45$^\circ$ pump incidence, with the emitted $p$-polarized THz field detected. \textbf{b.} Time-domain THz waveforms measured at 7~K with the pump polarization aligned parallel to the crystal \textit{a}-axis. Traces at low and high pump fluence are shown in yellow and red, respectively. \textbf{c.} Pump polarization dependence of the THz emission amplitude at low and high pump fluence. \textbf{d.} Time-domain THz emission signal $E_{\mathrm{THz}}$ as a function of pump fluence, with the corresponding frequency spectra shown on the right.}
    \label{fig:45 degree incidence}
\end{figure*}

\begin{figure*}
    \centering
\includegraphics[width=17.5cm]{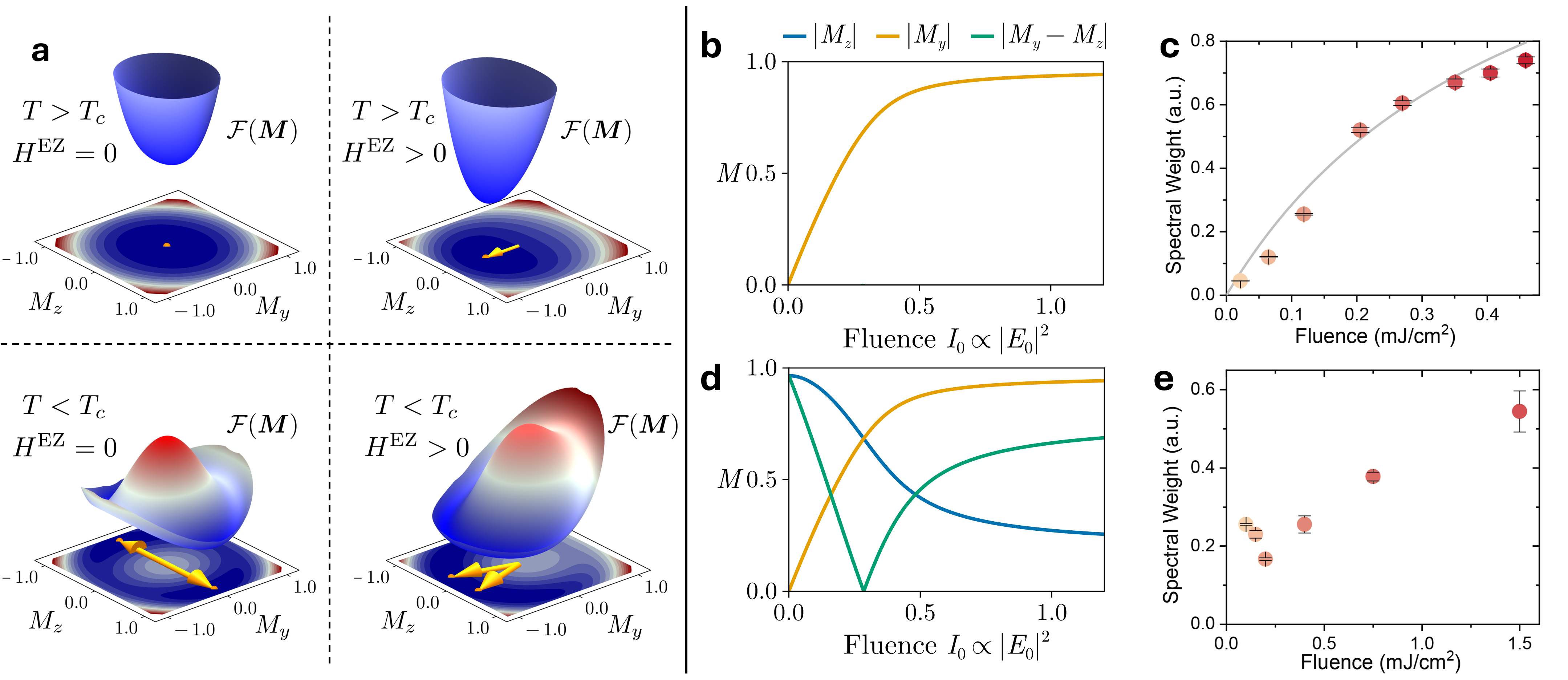}
\caption{\textbf{a.} Deformation of the uniaxial Heisenberg ferromagnetic free energy by the in-plane Edelstein-Zeeman field $H^\EdelZeeman_y$ above the Curie temperature $T_c$ and in the presence of the planar Edelstein-Zeeman field $H^\EdelZeeman_y$. The magnetization vectors in the $(M_y, M_z)$-plane that minimize the free energy is shown below the schematic free energy surfaces. The color scale on the surfaces and contour plots correspond to the value of free energy. \textbf{b.} Simulated light-induced in-plane magnetization as a function of pump-fluence. \textbf{c.} Spectral weight of THz emission as a function of fluence at normal incidence. \textbf{d.} Simulated light-induced magnetization at $45^\circ$ incidence as a function of pump-fluence. \textbf{e.} Spectral weight of THz emission as a function of fluence at $45^\circ$ incidence.}
    \label{fig:data_theory}
\end{figure*}

Such a relatively large-amplitude THz emission (several V/cm) from a centrosymmetric crystal is unexpected since a second-order electric dipole rectification should be symmetry-forbidden. While inversion symmetry is necessarily broken at the surface, any such contribution would (i) not be expected to track the bulk magnetic ordering, and thus would not show the gradual decrease and enhanced suppression below $T_c$ as shown in Fig.~\ref{fig:norm incidence}f, and (ii) scale linearly with pump intensity (since the fluence $\sim$ $E^2$), in contrast with the nonlinear fluence dependence and saturation behavior we observe (shown later in Fig.~\ref{fig:data_theory}c). 

The above considerations rule out a purely surface second-order electric-dipole mechanism as the dominant source of THz emission. Instead, they suggest a bulk rectification process achievable through the nonlinear Edelstein effect:
\begin{align}
    M_\beta(\omega_1 + \omega_2) = \alpha_{\beta,ij}^{(2)}(\omega_1 + \omega_2;\omega_1,\omega_2)E_i\left(\omega_1\right)E_j\left(\omega_2\right).\label{eq:dynamical_NEE}
\end{align}
\noindent with $\omega_1 = -\omega_2$. 
To justify this nonlinear magnetoelectric coupling, we consider the crystal symmetry of \CGT. In the bulk, \CGT crystallizes in the centrosymmetric space group $R\bar{3}$ with the corresponding point group $\bar{3}$ ($\rm{C}_{3i}$ or $\rm{S}_6$) that restricts the allowed components of the nonlinear magnetoelectric tensor $\alpha^{(2)}_{\beta,ij}$ to six independent entries: three associated with purely in-plane electric fields ($E_z = 0$) and the other three involving out-of-place fields ($E_z \neq 0$) (Supplementary Information). In this symmetry, the dyadic product of polar vectors $\boldsymbol{E}\otimes\boldsymbol{E}$ transforms as an axial vector and can therefore couple directly to the magnetization $\boldsymbol{M}$. Within $\bar{3}$, an axial vector is a reducible representation $\Gamma_{\boldsymbol{M}}= A_{g}\oplus E_g $ \cite{hamermeshGroupTheory1989, TinkhamMichael1992GTaQ,dresselhausGroupTheoryApplication2008}. Similarly, the dyad $\boldsymbol{E}\otimes\boldsymbol{E}$ of linearly polarized electric fields decomposes into $\Gamma_{\bm{E} \otimes\bm{E}} = 2A_{g} \oplus 2E_g$. Relating these irreducible representations through magneto-electric coefficients yield the six allowed by symmetry (Supplementary Information). In symmetry-reduced form, the three components of the magnetization follow as:
\begin{equation}
    \begin{aligned}
        M_z &= \alpha^{(2)}_{z,zz}E_z^2 + \alpha^{(2)}_{z,xx}(E_x^2 + E_y^2),
        \\
         M_\pm  &= \alpha^{(2)}_{\pm,\parallel}(E^2_x -E_y^2 \pm 2\mathrm{i}E_x E_y) + \alpha^{(2)}_{\pm,\perp}(E_x \pm \mathrm{i} E_y)E_z.
    \end{aligned}
\end{equation}
Symmetry therefore permits the manipulation of all three magnetic components through a coupling that is tunable via fluence, incidence, and polarization. Given that only the $M_y$ component generates $p$-polarized THz radiation within the experimental geometry at normal incidence, we restrict our attention to that component in particular. Written explicitly, it is given by
\begin{align}
    M_y &= \alpha^{(2)}_{x,xy}(E^2_x -E_y^2) - \alpha^{(2)}_{x,xx}(2E_xE_y) \nonumber \\
    &\phantom{=\;\;} - \alpha^{(2)}_{x,yz}(2E_zE_x) + \alpha^{(2)}_{x,zx}(2E_yE_z).\label{eq:Mx_written_out}
\end{align}
Thus, when the pump is at normal incidence ($E_z = 0$), the magnitude $|M_y|$ is fourfold symmetric with respect to the polarization of the light $\varphi$: $|M_y| \propto |\cos(2\varphi - 2\delta)|$. Here $\delta$ is a material-dependent phase set by the ratio of $\alpha^{(2)}_{x,xx}$ and $\alpha^{(2)}_{x,xy}$, and thus unlike the point group $\bar{3}\rm{m}$ ($\rm{D}_{3d}$) discussed in Refs.~\cite{liElectronicStructureFerromagnetic2018, xiaoIntrinsicNonlinearElectric2022,xiaoTimeReversalEvenNonlinearCurrent2023, xueNonlinearOpticsDrivenSpin2024}, the fourfold symmetric $|M_y|$ is not locked to a particular crystalline axis.


To further probe the observed magnetically mediated rectification via the Edelstein–Zeeman field, we tilt the pump incidence angle to $45^\circ$, such that both the in-plane and out-of-plane components of the induced magnetization contribute to the THz emission. Figure~\ref{fig:45 degree incidence}b shows two representative THz emission waveforms at 7~K for pump fluences of 0.1~mJ/cm$^2$ and 0.75~mJ/cm$^2$, with the pump linearly polarized along the $x$-axis ($\varphi = 0^\circ$). Remarkably, the two waveforms in Fig.~\ref{fig:45 degree incidence}b have comparable amplitudes but opposite polarity, indicating a reversal in the dominant magnetization component contributing to the THz emission as the pump fluence is increased.

This change in the waveform shape with fluence is further clarified by plotting the peak THz amplitude as a function of pump polarization ($\varphi$). At high fluence (0.75~mJ/cm$^2$), the emission exhibits a two-fold pattern with maxima near $\varphi = 60^\circ$ and $\varphi = 240^\circ$ (Fig.~\ref{fig:45 degree incidence}c), whereas at low fluence, the two-fold pattern is rotated by $90^\circ$. The disappearance of the signal at intermediate fluence coincides with destructive interference between contributions of opposite polarity, as confirmed by the minimum in the THz spectral weight for data taken at a fluence of $\sim 0.2$~mJ/cm$^2$ (Fig.~\ref{fig:45 degree incidence}d). 

These observations cannot be accounted for by conventional THz generation mechanisms such as the photo-Dember effect or screening of surface depletion fields \cite{dekorsy1996thz,zhang1992optoelectronic, pettineUltrafastTerahertzEmission2023}, which produce surface dipole currents that are largely isotropic with respect to pump polarization. Instead, the fluence-driven reorientation of the two-fold pattern points to a bulk magnetic mechanism in which in-plane and out-of-plane magnetization components compete and are weighted differently as the Edelstein–Zeeman field strength increases.

To capture this internal competition, we model the low-energy magnetization dynamics of \CGT using a minimal Landau description of a weakly uniaxial ferromagnet subject to an effective, fluence-dependent Edelstein–Zeeman field. 
Within Landau theory, the free energy as a function of the magnetization $\bm{M}$ can be written as:
\begin{align}
        \mathcal{F}(\bm{M}) &= \frac{1}{2}r_z M_z^2 + \frac{1}{2}r_0\left( M_x^2 + M_y^2 \right) + \frac{1}{4}u_z M_z^4 \nonumber
        \\
        &\phantom{=\;}  + \frac{1}{2}g M_z^2\left( M_x^2 + M_y^2 \right) - \bm{H}^\EdelZeeman \cdot \bm{M}. \label{eq:Landau_Edelstein_Zeeman}
\end{align}
\noindent where $r_z \propto T - T_c$ is the distance to transition, indicated as the Curie temperature $T_c$, $r_0 > r_z$ for an easy-axis anisotropy, $u_z > 0$ for thermodynamic stability, and $g > 0$ characterizes the competition between the axial and planar projections of the ferromagnetic moment (Supplementary Information). The Edelstein-Zeeman field $\bm{H}^\EdelZeeman$ is formally defined component-wise as:
\begin{equation}
    H^\EdelZeeman_\beta = J \, \chi^{(2)}_{\beta,ij}(0;\omega, -\omega)E_i(\omega)E_j(-\omega),\label{eq:EdelsteinZeeman_Field}
\end{equation}
where $\chi^{(2)}_{\beta,ij}$ is the nonlinear Edelstein susceptibility and $J$ is the coupling between the local and itinerant moments. 
Whereas $\chi^{(2)}$ represents the susceptibility associated with the photo-excited nonlinear Edelstein effect, the total \textit{dynamical} magneto-electric response tensor for the system, $\alpha^{(2)}$ convolves the near-resonant $\chi^{(2)}$ at the IR scale with the magnetic susceptibility of the \CrMoment moments $\chi^{(\mathrm{M})}$ associated with the slow, soft ferromagnetism at the THz scale (Supplementary Information). This composite magneto-electric response can lead to a large magnetic polarizability through the Edelstein-Zeeman field.

In Fig.~\ref{fig:data_theory}a, we illustrate how both the temperature $T$ and the Edelstein–Zeeman field deform the ferromagnetic free energy, thereby tuning the magnetic state and producing the different THz emission profiles observed experimentally through the same underlying mechanism. For temperatures $T > T_c$, the equilibrium magnetic state of the system, with $H^\EdelZeeman_y = 0$, has no magnetization: $|\boldsymbol{M}| = 0$. However, when driven out of equilibrium with a nonzero Edelstein-Zeeman field, the free energy is minimized towards a nonzero in-plane magnetization $M_y \neq 0$. Below the Curie temperature, meanwhile, the uniaxial anisotropy aligns the equilibrium magnetization out of the $ab$-plane: $M_z \neq 0$. When driven out-of-equilibrium by a planar Edelstein-Zeeman field, $H^\EdelZeeman_y \neq 0$, the external drive must compete with the uniaxial magnetic self-energy, and the magnetization is forced to rotate off-axis as a function of fluence, eventually pulling it completely into the plane.

Thus, minimizing $\mathcal{F}$ with respect to the magnetization provides a framework for understanding the temperature- and fluence-dependent THz emission observed experimentally. Importantly, the composite magneto-electric response that enables ferromagnetic control also captures the observed THz emission patterns in \CGT below $T_c$, where $\alpha^{(2)}_{ijk}\neq 0$ in the absence of the light because time-reversal symmetry is spontaneously broken, as well as \textit{above} $T_c$ where $\alpha^{(2)}_{ijk} = 0$ \textit{in equilibrium}. Because the near-infrared pump photo-excites states from the valence to conduction bands through dipole transitions, the nonequilibrium density matrix develops inter-band coherences causally under the drive. Therefore, the photo-excited nonequilibrium state itself dynamically breaks the time-reversal symmetry of \CGT in the high-temperature phase. Subsequently, when the nonequilibrium state interacts again with the incident light, it can accumulate a spin density due to spin-orbit coupling near locally non-centrosymmetric sites within the unit cell which are capable of generating electric-dipole currents and intra-cell electric quadrupole transitions (Supplementary Information). In turn, these ultrafast optical transitions down-convert into the THz regime at second-order, facilitating optical manipulation of ferromagnetism through the Edelstein-Zeeman field. While our results are reported for a ferromagnetic response, we expect that the mechanism proposed here for non-thermal ultrafast optical control of slower modes to be a generic feature within materials with hidden spin texture and sub-valent ordered states.


When viewed within the Landau phenomenology, we note then that the magnetization responds to the light-driven Edelstein-Zeeman field in the same way that it would to a regular Zeeman field. Importantly, in the presence of external time-reversal symmetry-breaking from the Edelstein-Zeeman field, the magnetization will exhibit crossover behavior as a function of temperature (as seen in the weak temperature dependence in Fig.~\ref{fig:norm incidence}f), rather than sharp criticality, since it cannot break time-reversal symmetry again. Moreover, the fact that the temperature and fluence only control the magnitude of the petals in the fourfold pattern for the normal incidence data in Fig.~\ref{fig:norm incidence}d, but not their orientation relative to the crystal axes, is a direct consequence of the magneto-electric coefficients being driven by inter-band electronic transitions at the infrared scale. 

Similarly, the reorientation of the twofold pattern as a function of fluence seen in the THz emission data at $45^\circ$-degree incidence (Fig.~\ref{fig:45 degree incidence}c) can also be explained from the Landau expansion in Eq.~\eqref{eq:Landau_Edelstein_Zeeman}. Because the ferromagnetism is Heisenberg-like with a weak uniaxial anisotropy, there is an intrinsic competition between the easy axis component, $M_z$, and the in-plane component, $M_y$, apparent at quartic order in Eq.~\eqref{eq:Mx_written_out}. At any oblique incidence, a linear combination of both are measured in the experimental geometry shown in Fig.~\ref{fig:45 degree incidence}a. At low fluence, the THz signal is dominated by one projection of the magnetization, resulting in a two-fold angular pattern along a given direction. As the fluence increases and the Edelstein–Zeeman field strengthens, the balance between $M_y$ and $M_z$ shifts, causing the orthogonal magnetic projection to dominate. This crossover leads to an effective reversal of the measured THz polarity and a rotation of the two-fold emission pattern, with destructive interference between the two contributions producing the observed minimum in spectral weight at intermediate fluence.

To further show the consistency of the fluence dependent data with the Landau expansion in Eq.~\eqref{eq:Landau_Edelstein_Zeeman}, we appeal to self-consistent mean-field theory described in the Supplementary Information and compare the simulated and observed THz emission magnitude as a function of fluence in Fig.~\ref{fig:data_theory}. For normal incidence, where the detected THz signal is primarily sensitive to the in-plane magnetization component $M_y$, the Landau model predicts a rapid initial growth of $M_y$ with increasing Edelstein--Zeeman field, followed by saturation at higher pump fluence. This behavior closely mirrors the response of a ferromagnet subjected to an increasing dc Zeeman field and arises from the competition between the quadratic magnetic susceptibility and the stabilizing quartic terms in the free energy $\mathcal{F}$. The resulting saturation of the in-plane magnetization is directly reflected in the measured fluence dependence of the peak THz amplitude $|S_{\mathrm{THz}}|_{\max}$ at normal incidence, as demonstrated by the agreement between theory and experiment in Figs.~\ref{fig:data_theory}a and \ref{fig:data_theory}b. The situation is more complex at $45^\circ$-incidence, as shown in Fig.~\ref{fig:data_theory}c and Fig.~\ref{fig:data_theory}d. Here, the contribution to the measured THz emission from both $M_y$ and $M_z$ results in a non-monotonic response to the laser fluence, as the growing in-plane magnetization necessarily competes with the frozen easy-axis component, initially pulling the magnetic moment off of the $z$-axis to accommodate the new in-plane projection. This internal magnetic competition and subsequent saturation can explain the observed non-monotonicity in the THz spectral weight as a function of fluence shown in Fig.~\ref{fig:data_theory}e.

Our theoretical and experimental results demonstrate a new mechanism for optical manipulation of magnetism based on resonant nonequilibrium  photo-excitations that dynamically break time-reversal symmetry and generate effective internal magnetic fields. In the centrosymmetric van der Waals ferromagnet \CGT, this mechanism enables non-thermal, ultrafast control of ferromagnetic order via a second-order, photocurrent-assisted Edelstein–Zeeman field, directly observed through strong, polarization-dependent THz emission. Unlike conventional magneto-optical or thermal effects, the Edelstein–Zeeman field provides tensorial control over magnetization, with its magnitude governed by optical fluence and its direction set by light polarization. This purely electronic pathway enables femtosecond-to-picosecond manipulation of magnetic order in materials where such control is forbidden in equilibrium. More broadly, the nonlinear Edelstein mechanism should be widely applicable to spin–orbit–coupled materials with locally non-centrosymmetric environments, opening new opportunities for ultrafast, energy-efficient opto-electronic, and spintronic devices.

\vspace{2mm}

\noindent\textbf{Methods}
\vspace{2mm}

\noindent\textbf{Sample preparation:}
Single crystals of Cr$_2$Ge$_2$Te$_6$ were grown in a self-flux of excess Te and Ge in a similar method to that previously reported. Cr, Ge, and Te were mixed in a molar ratio of 1:3:18 and heated in a sealed quartz tube under vacuum to 1000~$^\circ$C. After being held at 1000~$^\circ$C for 12 hours, the temperature was lowered over a week to 450~$^\circ$C and held for four days before centrifuging the product to separate single crystals from the flux. After isolation, crystals were again sealed in quartz tubes and placed in the hot zone of a tube furnace at 400~$^\circ$C while the cold end was kept outside of the heated zone. The samples were annealed in this condition for two days to remove any residual Te flux.

\vspace{2mm}
\noindent\textbf{THz emission spectroscopy:} THz emission measurements were performed with our custom-built time-domain THz spectroscopy setup based on a Yb:KGW amplifier laser (PHAROS, Light Conversion). The fundamental laser pulse wavelength is 1030~nm with a pulse duration of $\sim160$~fs. The fundamental beam is split into a pump and a probe, with the pump incident onto the sample at either 0$^\circ$ (normal) or a 45$^\circ$ angle-of-incidence while the probe is used for electro-optic sampling (EOS) of the emitted THz field. The sample is mounted at the center of either a magneto-optic cryostat (OptiCool, Quantum Design) or an optical-cryostat with He exchange gas (SHI-950, Janis Research). The THz field radiated by the sample is collected and collimated by an off-axis parabolic mirror. The emitted THz is then focused onto a (110)-cut CdTe crystal. The EOS probe beam is made to spatially and temporally overlap with the emitted THz on the CdTe crystal. The THz field $\boldsymbol{E}(t)$ is thus measured by scanning the time delay of the 1030~nm electro-optic sampling beam relative to the emitted THz field~\cite{planken2001measurement_EOsampling}. 

\vspace{2mm}
\noindent\textbf{Acknowledgments:}
F.M., Y.L., A.M., C.L. and Y.S. acknowledge support from the NSF Career Award No.~DMR-2144256.
F.M.~acknowledges support from the EPiQS program of the Gordon and Betty Moore Foundation, Grant GBMF11069. 
J.C.~acknowledges support from the EPiQS initiative of the Gordon and Betty Moore Foundation, Grant No. GBMF9070, and ARO grant no. W911NF-16-1-0034. 
J.F.~and F.M.~acknowledge support from the Center for Emergent Materials, an NSF MRSEC, under Grant No.~DMR-2011876.
The authors acknowledge the use of facilities and instrumentation supported by NSF through the University of Illinois at Urbana-Champaign Materials Research Science and Engineering Center DMR-2309037.

\vspace{2mm}
\noindent\textbf{Author contributions:} Y.L., A.M., J.F., C.L., and Y.S. performed the THz emission experiments and the corresponding data analysis. W.J.M. developed the theoretical description of the nonlinear Edelstein-Zeeman field and modeled its impact on magnetism. J.W., T.K., and J.C. synthesized and characterized the samples. Y.L., W.J.M., and F.M. wrote the manuscript with input from all of the authors. F.M. conceived and supervised this project.

\vspace{2mm}
\noindent\textbf{Competing interests:} The authors declare no competing interests.

\vspace{2mm}
\noindent\textbf{Data availability:} The data in this manuscript are available from the corresponding authors upon request.

\vspace{2mm}
\noindent\textbf{Corresponding authors:} Correspondence to W. Joe Meese~(meesewj@illinois.edu) and/or Fahad Mahmood~(fahad@illinois.edu).

\bibliography{Mainbib}
\end{document}


\preprint{APS/123-QED}

\title{Supplementary Information\\ Manipulation of ferromagnetism with a light-driven nonlinear Edelstein–Zeeman field}

\maketitle
\setcounter{table}{0}

\renewcommand{\theequation}{S\arabic{equation}}
\setcounter{equation}{0}
\renewcommand{\thefigure}{S\arabic{figure}}
\renewcommand{\thetable}{S\arabic{table}}
\setcounter{table}{0}
\renewcommand{\thepage}{S-\arabic{page}}
\setcounter{page}{1}
\setcounter{section}{0}
\setcounter{subsection}{0}
\numberwithin{equation}{section}
\renewcommand{\bibnumfmt}[1]{[S#1]}
\renewcommand{\citenumfont}[1]{S#1}

\onecolumngrid
\tableofcontents
\newpage
\section{Static Magnetic Characterization of Cr$_2$Ge$_2$Te$_6$}
\begin{figure*}[h!]
    \centering
    \includegraphics[width=8cm]{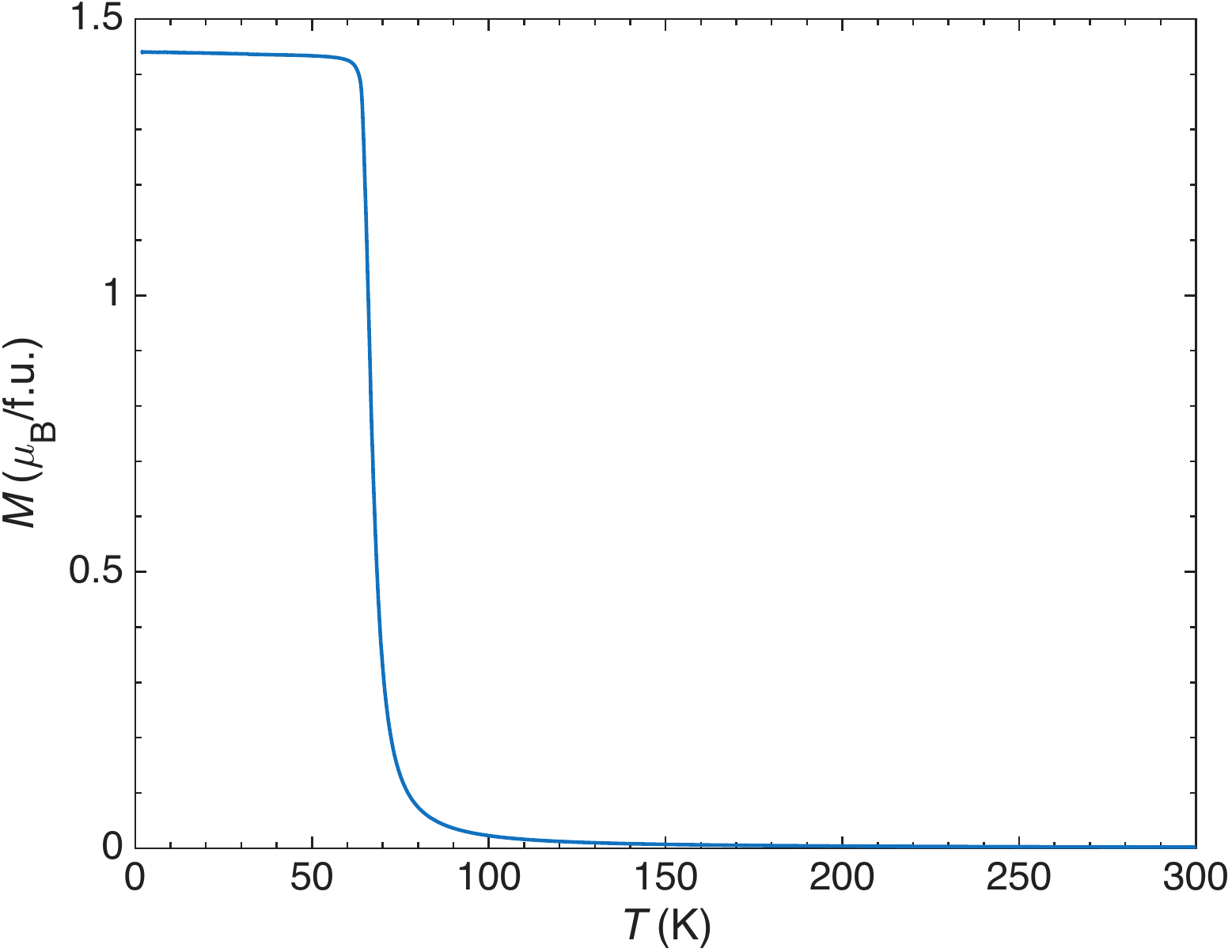}
    \caption{Temperature-dependent magnetization $M(T)$ of Cr$_2$Ge$_2$Te$_6$ (in $\mu_B$ per formula unit). The sharp drop near $T_C \sim 67$ K marks the Curie temperature of the sample.}
\label{fig:M-T plot}
\end{figure*}

\begin{figure*}[h!]
    \centering
    \includegraphics[width=8cm]{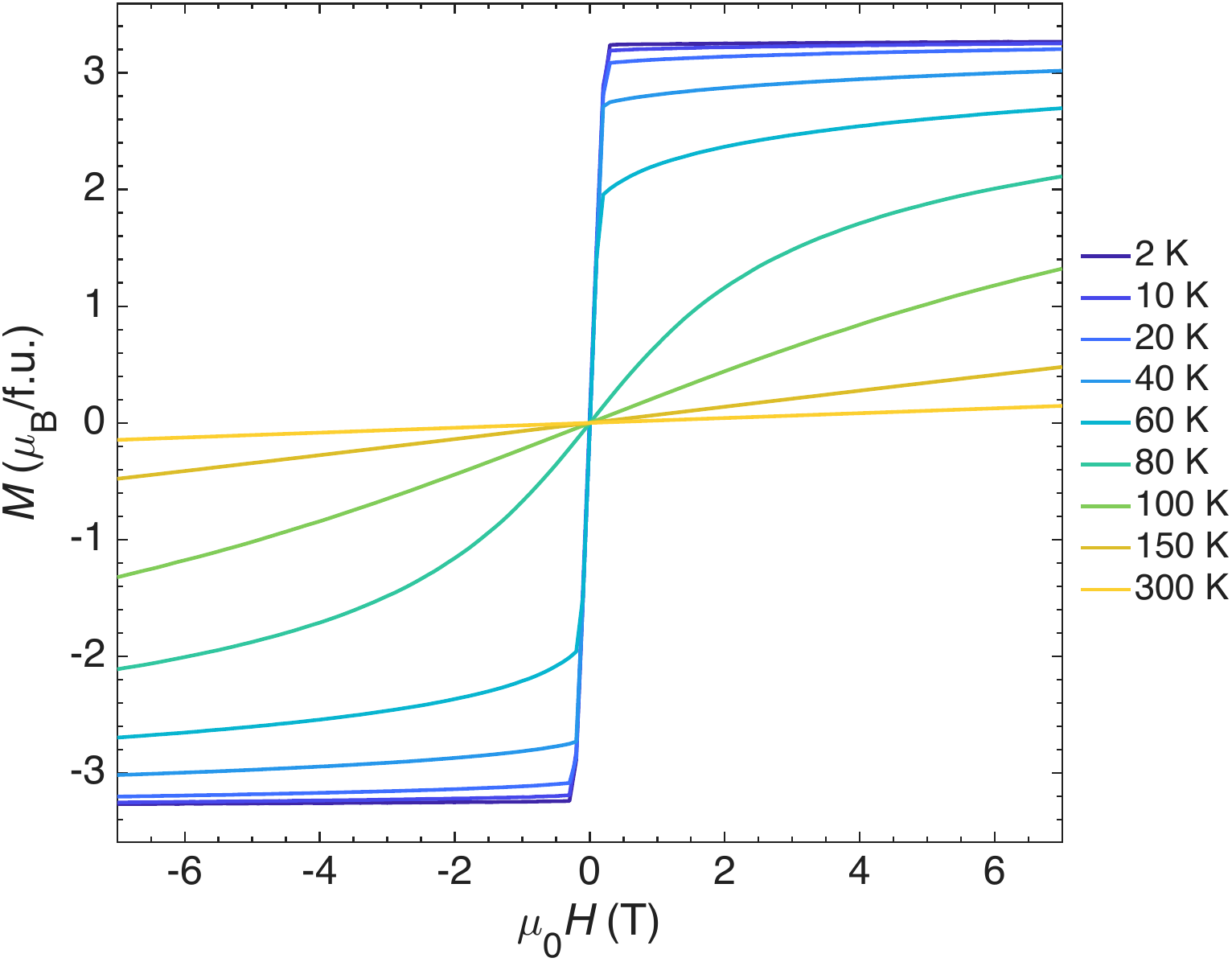}
    \caption{Magnetization $M(H)$ measured as a function of applied magnetic field $H$~(T) at temperatures from 2~K to 300~14K. The absence of a discernible hysteresis loop and a vanishing coercive field across the entire temperature range indicate that the Cr$_2$Ge$_2$Te$_6$ sample is a soft ferromagnet.}
\label{fig:M-H plot}
\end{figure*}

To establish the equilibrium magnetic properties of Cr$_2$Ge$_2$Te$_6$, we first characterize its dc magnetic response using temperature- and field-dependent magnetization measurements. Supplementary~Figure~\ref{fig:M-T plot} shows the magnetization $M(T)$ measured under a constant applied field, revealing a clear ferromagnetic transition at the Curie temperature $T_C \approx 67$~K. The rapid suppression of $M(T)$ above $T_C$ is consistent with a transition to the paramagnetic phase. Field-dependent magnetization measurements $M(H)$ acquired over a wide temperature range (Supplementary~Figure~\ref{fig:M-H plot}) exhibit a nearly linear response with negligible hysteresis and vanishing coercive field, indicating that Cr$_2$Ge$_2$Te$_6$ behaves as a soft ferromagnet with a large dc magnetic susceptibility. These properties provide the magnetic baseline for the optical control experiments discussed in the Main Text.

\newpage
\section{THz Emission Spectroscopy}
\subsection{Experimental Configuration}

\begin{figure*}[h!]
    \centering
    \includegraphics[width=13cm]{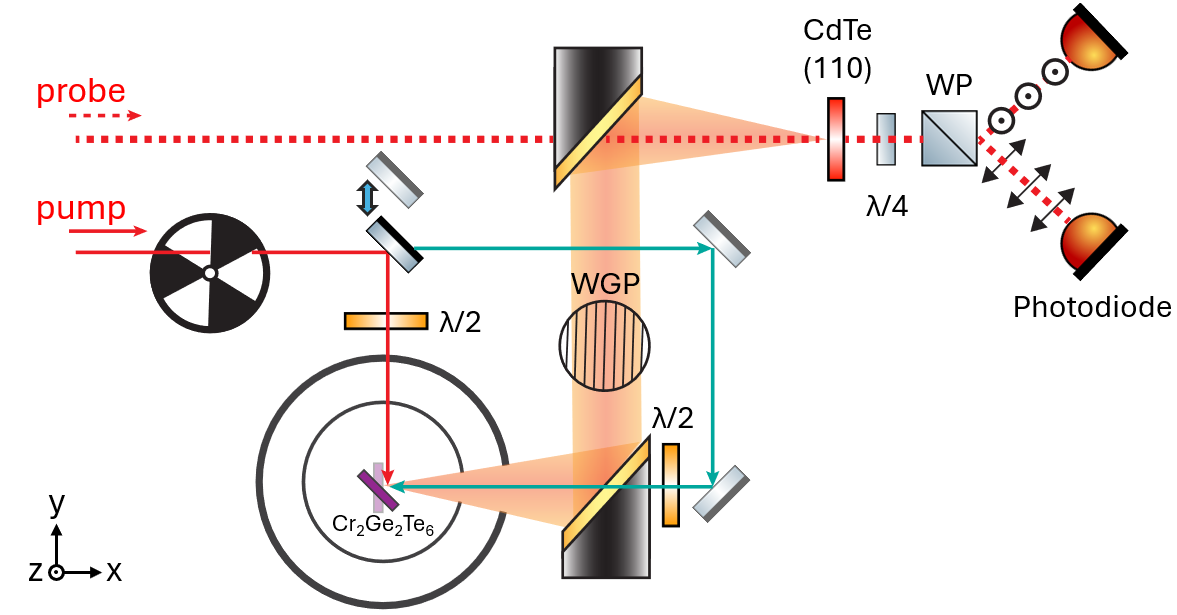}
    \caption{Schematic of the THz emission setup. The Cr$_2$Ge$_2$Te$_6$ sample is placed inside a closed-cycle helium cryostat. $\lambda/2$ and $\lambda/4$ refer to half- and quarter-wave plates, respectively. WPG is a wire-grid polarizer used to analyze the THz polarization. A 1-mm-thick (110)-oriented CdTe crystal is used for electro-optic sampling, and WP is a Wollaston prism used for balanced detection.}
\label{fig:SI_setup_schematic}
\end{figure*}

\noindent THz emission measurements were performed in two complementary experimental geometries:

\noindent (i) Oblique incidence (45$^\circ$): the Cr$_2$Ge$_2$Te$_6$ sample was mounted in a Quantum Design OptiCool closed-cycle helium cryostat and excited using 1.2~eV (1030~nm), 160~fs laser pulses at a repetition rate of 3~kHz.

\noindent (ii) Normal incidence (0$^\circ$): the sample was mounted in a Janis cryostat and pumped with 1.2~eV, 160~fs laser pulses at a repetition rate of 50~kHz.

As shown in Supplementary Fig.~\ref{fig:SI_setup_schematic}, in both configurations the laser output was divided into pump and probe arms with a 97:3 power ratio. The pump beam was collimated and focused onto the sample with a $1/e^2$ intensity diameter of $1.90 \pm 0.20$~mm, and mechanically chopped at half the laser repetition rate. The pump polarization was continuously controlled using a half-wave plate placed immediately before the cryostat window.

The emitted THz radiation was collected and collimated using a pair of 2-inch-diameter, 90$^\circ$ off-axis parabolic mirrors arranged in a 4$f$ geometry. A wire-grid polarizer positioned after the first mirror selected the $x$-component of the THz electric field. The probe beam was delayed using a mechanical delay stage and overlapped collinearly with the THz pulse in a 1-mm-thick (110) CdTe crystal for electro-optic sampling. The temporal waveform of the emitted THz field was recorded via balanced detection (see Sec.~II.C for THz field calibration).

\subsection{Pump Fluence Calibration}

\begin{figure*}[h!]
    \centering
    \includegraphics[width=13cm]{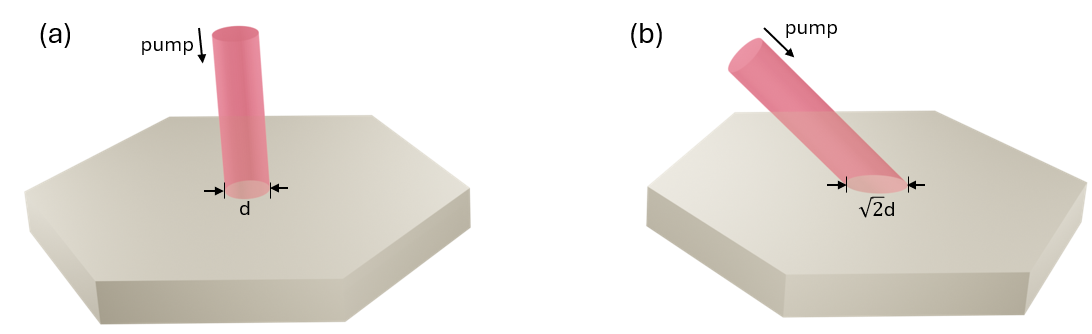}
    \caption{Pump beam profile at the Cr$_2$Ge$_2$Te$_6$ sample surface. (a) Normal-incidence geometry. (b) Oblique-incidence geometry at 45$^\circ$.}
    \label{fig:pump_fluence_schematic}
\end{figure*}

To quantify the pump fluence used in the THz emission measurements, we first characterize the spatial profile of the pump beam at the sample surface. As shown in Supplementary Fig.~\ref{fig:pump_fluence_schematic}a, under normal incidence the $1/e^2$ intensity diameter is determined to be $d = 1.90 \pm 0.20$~mm using the knife-edge method. For a laser repetition rate of 50~kHz (25~kHz after mechanical chopping), the incident fluence is given by:
\begin{equation}
    F=\frac{P}{\pi(d/2)^2f}\approx 0.0014 \cdot P \; [\mathrm{mJ/cm}^2],
    \label{eq:fluence_calculation}
\end{equation}
\noindent where $P$ is the average pump power, $d$ is the beam diameter, and $f$ is the post-chopping repetition rate. In the normal-incidence configuration, pump powers between 10 and 330~mW correspond to fluences ranging from 14.1~$\mu$J/cm$^2$ to 0.46~mJ/cm$^2$.

For the oblique-incidence geometry at 45$^\circ$, the repetition rate is reduced to $f=3$~kHz, and the projected beam area increases by a factor of $\sqrt{2}$. The fluence is therefore:
\begin{equation}
    F=\frac{P}{\pi\sqrt{2}(d/2)^2f}\approx 0.017 \cdot P \; [\mathrm{mJ/cm}^2].
\end{equation}
\noindent For pump powers between 5 and 90~mW, this corresponds to fluences between 83.1~$\mu$J/cm$^2$ and 1.50~mJ/cm$^2$ at the sample surface.

\subsection{Estimation of the Emitted THz Field Strength}

The amplitude and phase of the emitted THz electric field were measured via electro-optic sampling using a 1-mm-thick (110)-oriented CdTe crystal. In this technique, the transient THz field acts as a quasi-static bias that modulates the polarization of the near-infrared probe pulse through the linear Pockels effect. The resulting relative change in probe polarization, $\Delta I/I$, is related to the THz field amplitude $E_T$ by:

\begin{equation}
    \frac{\Delta I}{I}=\sin\left(\frac{\omega n_0^3 r_{41} L E_T}{c}\right),
\end{equation}

\noindent where $\omega$ is the probe angular frequency, $n_0$ is the refractive index of CdTe at the probe wavelength, $r_{41}$ is the electro-optic coefficient, $L$ is the crystal thickness, and $c$ is the speed of light.

In our measurements, a 1030~nm (1.2~eV) probe pulse and a 1-mm-thick CdTe crystal were used. At this wavelength, the refractive index is $n_0 = 2.83$ and the electro-optic coefficient is approximately $r_{41} \approx 7$~pm/V. Using these parameters, we estimate the typical peak emitted THz field from Cr$_2$Ge$_2$Te$_6$ to be on the order of $0.1$~V/cm.

\newpage


\section{Minimal Hamiltonian for the Resonant Interband Nonlinear Edelstein
Effect }\label{supp-sec:Minimal-NEE-model}

Despite the global centrosymmetry of \CGT with crystallographic point group $\overline{3}$, none of its occupied Wyckoff positions have inversion in their local point group \citep{jiFerromagneticInsulatingSubstrate2013,zeisnerMagneticAnisotropySpinpolarized2019}. Because of this, \CGT is a candidate to exhibit a nonlinear Edelstein effect \citep{xuLightinducedStaticMagnetization2021,xiaoIntrinsicNonlinearElectric2022,xiaoTimeReversalEvenNonlinearCurrent2023,oikeImpactElectronCorrelations2024,xueNonlinearOpticsDrivenSpin2024} through a so-called ``hidden spin-texture'' in the band structure
\citep{zhangHiddenSpinPolarization2014,huangHiddenSpinPolarization2020,guanHiddenZeemantypeSpin2023,arnoldiRevealingHiddenSpin2024}. 

Given the richness of nonlinear optical response in a material such as \CGT, it is important to emphasize that the Edelstein-mediated control of ferromagnetism presented in the Main Text is not specific to this compound alone. In this section, we provide a minimal tight-binding model that can give rise to a resonant inter-band nonlinear Edelstein effect in a centrosymmetric material. Generally speaking, the nonlinear Edelstein effect can occur
both through intra- and inter-band processes depending on the microscopic
details of the system in question, as well as the macroscopic details
of the external drive \citep{xuLightinducedStaticMagnetization2021,xiaoIntrinsicNonlinearElectric2022,xiaoTimeReversalEvenNonlinearCurrent2023,oikeImpactElectronCorrelations2024,xueNonlinearOpticsDrivenSpin2024,yoshidaLightinducedMagnetizationQuantum2026}.
For simplicity, we will consider only the spin-response generated by inter-band processes driven by a strictly
homogeneous electric field. Our focus will be on nonlinear magneto-electric
response from the spin sector, though similar arguments can be made
for the orbital sector as well \citep{ryooHiddenOrbitalPolarization2017}.
Applying this model to bulk \CGT will thereby generate a light-induced
coherent magnetic field which then enables ferromagnetic control of
the underlying localized \ce{Cr^{3+}} moments.

\subsection{Model Hamiltonian with Local Non-Centrosymmetry}

\begin{figure}[h!]
    \centering
    \includegraphics[width=0.25\linewidth]{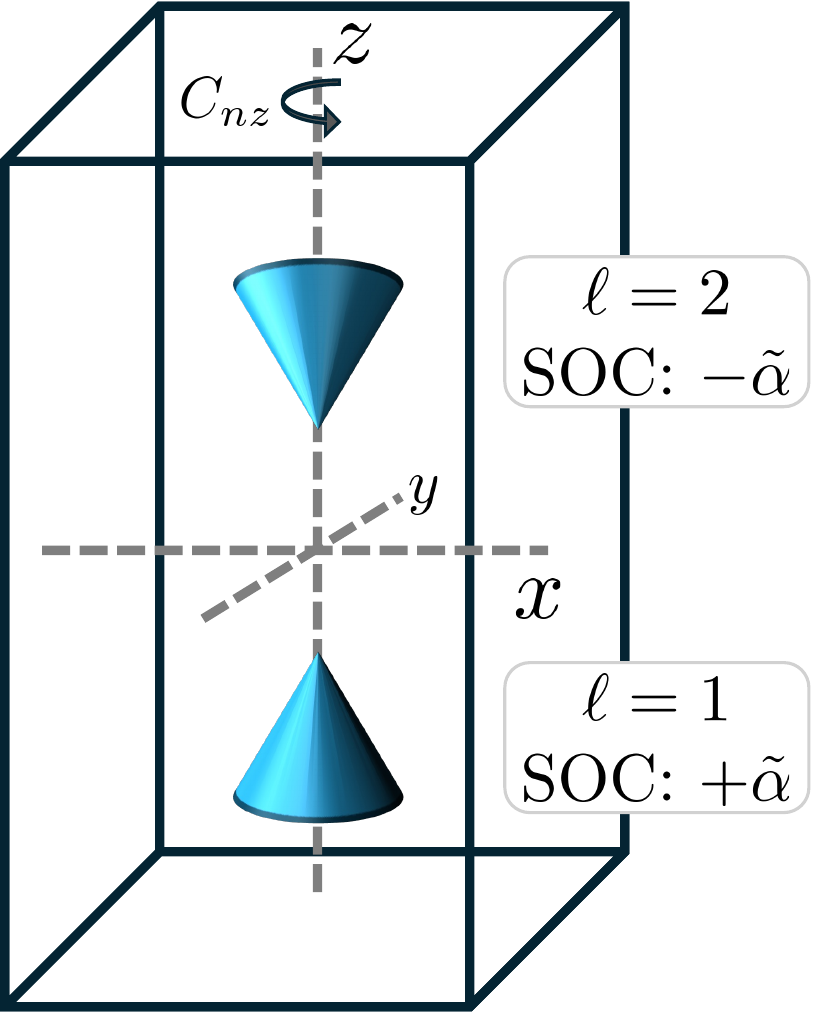}
    \caption{Schematic of a globally centrosymmetric unit cell composed of two locally non-centrosymmetric sites related by global inversion. The $n$-fold rotational symmetry of the unit cell is taken to be about the $z$-axis.  The polarized crystal fields (cones) at each non-centrosymmetric site generate equal and opposite spin–orbit coupling strengths $\pm\tilde{\alpha}$
  on the two sublattices (labeled by $\ell$). Global inversion symmetry guarantees Kramers' degeneracy in the paramagnetic phase, and produces two twofold degenerate bands. Meanwhile, the inversion-compensated spin-orbit coupling (SOC) generates a hidden spin texture between the Kramers pairs in each band. Inter-band electric dipole and quadrupole transitions then convert this hidden spin texture into a net spin density \citep{zhangHiddenSpinPolarization2014, huangHiddenSpinPolarization2020} under a drive.}
    \label{fig:minimal_unit_cell_NEE}
\end{figure}

As shown in Supplementary Fig.~\ref{fig:minimal_unit_cell_NEE}, in a globally centrosymmetric, but locally non-centrosymmetric unit
cell, each occupied non-centrosymmetric Wyckoff position is compensated
by an occupied partner position of opposite parity. Consequently, within a tight-binding
approximation, the atoms within the unit cell are then split into
sublattices that exchange under global inversion. To simplify the
problem, we consider a uniaxial unit cell made of two inversion-partner
atoms aligned along the high-symmetry axis -- taken here to be the
$\hat{z}$-axis. Because of the local non-centrosymmetry, combined
with the rotational symmetry about $\hat{z}$, each site locally gives
rise to Rashba-type spin-orbit coupling of equal and opposite sign.
Thus, the minimum tight-binding Hamiltonian we will consider assumes
the following form

\begin{equation}
\mathcal{H}=\sum_{\boldsymbol{k}}\mathsf{c}_{\boldsymbol{k}}^{\dagger}\cdot\mathsf{h}\left(\boldsymbol{k}\right)\cdot\mathsf{c}_{\boldsymbol{k}}^{\phantom{\dag}},\label{supp-eq:total_tight_binding_hamiltonian}
\end{equation}
in a basis where the electron annihilation operator with momentum
$\boldsymbol{k}$, in sublattice $\ell$, and with spin $\sigma$
is written as $c_{\boldsymbol{k}\ell\sigma}$. Taking the Kronecker
product of the sublattice and spin degrees of freedom, represented
as the $\boldsymbol{\tau}$ and $\boldsymbol{\sigma}$ Pauli operators,
respectively, we combine the $\ell=1,2$ and $\sigma=\uparrow,\downarrow$
states in the following single vector
\begin{equation}
\mathsf{c}_{\boldsymbol{k}}\equiv\begin{pmatrix}c_{\boldsymbol{k}1\uparrow} & c_{\boldsymbol{k}1\downarrow} & c_{\boldsymbol{k}2\uparrow} & c_{\boldsymbol{k}2\downarrow}\end{pmatrix}^{\text{T}}.
\end{equation}
The specific tight-binding Hamiltonian matrix we consider is given
by
\begin{align}
\mathsf{h}\left(\boldsymbol{k}\right) & =\left[t_{0}\left(\boldsymbol{k}\right)-\mu\right]\left(\tau^{0}\otimes\sigma^{0}\right)+\left[t_{1}^{\prime}\left(\boldsymbol{k}\right)\tau^{x}+t_{2}^{\prime}\left(\boldsymbol{k}\right)\tau^{y}\right]\otimes\sigma^{0}+\tilde{\alpha}\left[\tau^{z}\otimes\left(\hat{z}\times\boldsymbol{k}\right)\cdot\boldsymbol{\sigma}\right].\label{supp-eq:tight-binding_hamiltonian_matrix}
\end{align}
 Here, the intra-sublattice tunneling amplitude is $t_{0}=t_{0}\left(\boldsymbol{k}\right)$
and centrosymmetry requires $t_{0}\left(-\boldsymbol{k}\right)=t_{0}\left(+\boldsymbol{k}\right)$.
There are two symmetry-allowed inter-sublattice tunneling amplitudes,
$t_{1}^{\prime}\left(\boldsymbol{k}\right)$ and $t_{2}^{\prime}\left(\boldsymbol{k}\right)$,
which must be even and odd functions of $\boldsymbol{k}$
for Eq.~(\ref{supp-eq:tight-binding_hamiltonian_matrix}) to respect
both inversion and time-reversal symmetries. The chemical potential
is $\mu$, and the spin-orbit coupling strength is given by the parameter
$\tilde{\alpha}$. The $n$-fold uniaxial rotational symmetry, $C_{n}$,
will rotate the momentum as $\boldsymbol{k}\rightarrow C_{n}\boldsymbol{k}$,
and due to spin-orbit coupling, it will also rotate the spins by a
unitary that acts in spin space. Thus, under rotations, the spins
operators transform under the following unitary
\begin{align}
\mathcal{U}\left(C_{n}\right) & \equiv\left[\tau^{0}\otimes\exp\left(-\frac{\text{i}\pi}{n}\sigma^{z}\right)\right]\hat{\mathcal{U}}_{n},\\
\hat{\mathcal{U}}_{n}f\left(\boldsymbol{k}\right) & =f\left(C_{n}\boldsymbol{k}\right).
\end{align}

The inter-sublattice tunnelings $t_{1,2}^{\prime}$ both hybridize
the two sublattices, favoring eigenstates that are equally projected
onto either atom in the unit cell with any spin projection. The tunneling
$t_{1\left(2\right)}^{\prime}$ favors the electronic projection to
be in-phase (phase-shifted) between sublattices. Meanwhile, the spin-orbit
coupling $\tilde{\alpha}$ instead prefers eigenstates with stationary
sublattice index in order to lock the electron spin to its momentum.
The presence of these non-commuting terms in (\ref{supp-eq:tight-binding_hamiltonian_matrix})
eliminates both the sublattice and spin degrees of freedom as ``good
quantum numbers'' to label the eigenstates. 

Global inversion symmetry, $\mathcal{P}$, and time-reversal symmetry,
$\mathcal{T}$, both act in the $\boldsymbol{k}$-local 4-dimensional
sublattice-spin Hilbert space, as well as scalar-valued functions
of momentum, $f\left(\boldsymbol{k}\right)$. They are defined by 

\begin{equation}
\begin{aligned}\mathcal{P} & \equiv\left(\tau^{x}\otimes\sigma^{0}\right)\hat{\mathcal{P}}, & \mathcal{T} & \equiv\left(\tau^{0}\otimes\text{i}\sigma^{y}\right)\hat{\mathcal{T}},\\
\hat{\mathcal{P}}f\left(\boldsymbol{k}\right) & \equiv f\left(-\boldsymbol{k}\right), & \hat{\mathcal{T}}f\left(\boldsymbol{k}\right) & \equiv f^{*}\left(-\boldsymbol{k}\right),
\end{aligned}
\end{equation}
 where the asterisk is used for complex conjugation. Since Kramer's
Theorem guarantees that the energy eigenstates are doubly degenerate,
it is clear that if there is a spin-texture in the bands, it is completely
compensated by its Kramer's conjugate pair. It is most straightforward
to show this explicitly using gauge-invariant band projectors \citep{mitscherlingGaugeinvariantProjectorCalculus2025}.

The three $4\times4$ matrices that appear in Eq.~(\ref{supp-eq:tight-binding_hamiltonian_matrix})
can be used to define a four-dimensional Clifford algebra, with Dirac
matrices given by
\begin{equation}
\begin{aligned}\mathds{1} & \equiv\tau^{0}\otimes\sigma^{0}, & \gamma^{1} & \equiv\tau^{z}\otimes\sigma^{x}, & \gamma^{2} & \equiv\tau^{z}\otimes\sigma^{y}, & \gamma^{3} & \equiv\tau^{x}\otimes\sigma^{0}, & \gamma^{4} & \equiv\tau^{y}\otimes\sigma^{0}.\end{aligned}
\end{equation}
 Being the generators of the Clifford algebra, it is straightforward
to show that these matrices satisfy 
\begin{equation}
\left\{ \gamma^{a},\gamma^{b}\right\} =2\delta^{ab}\mathds{1}.\label{supp-eq:Clifford_algebra_definition}
\end{equation}
Furthermore, they also obey the following trace identities:
\begin{align}
\mathscr{T}_{1}^{a} & \equiv\text{tr}\left(\gamma^{a}\right)=0,\\
\mathscr{T}_{2}^{ab} & \equiv\text{tr}\left(\gamma^{a}\gamma^{b}\right)=4\delta^{ab},\label{supp-eq:Dirac_matrix_trace_prod_2}\\
\mathscr{T}_{3}^{abc} & \equiv\text{tr}\left(\gamma^{a}\gamma^{b}\gamma^{c}\right)=0,\\
\mathscr{T}_{4}^{abcd} & \equiv\text{tr}\left(\gamma^{a}\gamma^{b}\gamma^{c}\gamma^{d}\right)=4\left(\delta^{ab}\delta^{cd}-\delta^{ac}\delta^{bd}+\delta^{ad}\delta^{bc}\right),\label{supp-eq:Dirac_matrix_trace_prod_4}
\end{align}
 which can be used to recursively write the trace over any even number,
$2m$, of Dirac matrices as 
\begin{equation}
\mathscr{T}_{2m}^{a_{1}a_{2}a_{3}\cdots a_{2m}}\equiv\text{tr}\left(\gamma^{a_{1}}\gamma^{a_{2}}\gamma^{a_{3}}\cdots\gamma^{a_{2m}}\right)=\delta^{a_{1}a_{2}}\mathscr{T}_{2\left(m-1\right)}^{a_{3}\cdots a_{2m}}-\delta^{a_{1}a_{3}}\mathscr{T}_{2\left(m-1\right)}^{a_{2}\cdots a_{2m}}+\dots+\delta^{a_{1}a_{2m}}\mathscr{T}_{2\left(m-1\right)}^{a_{2}a_{3}\cdots a_{2m-1}}.\label{supp-eq:Dirac_matrix_trace_prod_2m}
\end{equation}
The trace over any product with an odd number of distinct Dirac matrices,
meanwhile, will vanish.

Using these generators, one can also write the spin operator at each
momentum as
\begin{equation}
\boldsymbol{S}\equiv\frac{1}{2}\boldsymbol{\Sigma}\equiv\frac{1}{2}\left(\tau^{0}\otimes\boldsymbol{\sigma}\right),
\end{equation}
 which translates to the following strings of Dirac matrices: 
\begin{align}
S_{x} & =\frac{1}{2}\Sigma_{x}=\frac{1}{2}\left(\tau^{0}\otimes\sigma^{x}\right)=-\frac{\text{i}}{2}\,\gamma^{1}\gamma^{3}\gamma^{4},\\
S_{y} & =\frac{1}{2}\Sigma_{y}=\frac{1}{2}\left(\tau^{0}\otimes\sigma^{y}\right)=-\frac{\text{i}}{2}\,\gamma^{2}\gamma^{3}\gamma^{4},\\
S_{z} & =\frac{1}{2}\Sigma_{z}=\frac{1}{2}\left(\tau^{0}\otimes\sigma^{z}\right)=-\frac{\text{i}}{2}\,\gamma^{1}\gamma^{2}.
\end{align}
Moreover, Eq.~(\ref{supp-eq:tight-binding_hamiltonian_matrix}) then
assumes a Dirac form given by
\begin{align}
\mathsf{h}\left(\boldsymbol{k}\right) & =d_{0}\left(\boldsymbol{k}\right)\,\mathds{1}+\boldsymbol{d}\left(\boldsymbol{k}\right)\cdot\boldsymbol{\gamma},\label{supp-eq:tight-binding_Hamiltonian_Dirac_form}\\
d_{0}\left(\boldsymbol{k}\right) & \equiv t_{0}\left(\boldsymbol{k}\right)-\mu,\\
\boldsymbol{d}\left(\boldsymbol{k}\right) & \equiv\begin{bmatrix}-\tilde{\alpha}k_{y} & \tilde{\alpha}k_{x} & t_{1}^{\prime}\left(\boldsymbol{k}\right) & t_{2}^{\prime}\left(\boldsymbol{k}\right)\end{bmatrix}^{\text{T}}.\label{supp-eq:dvector}
\end{align}
While the components $d_{1,2,4}\left(\boldsymbol{k}\right)$ are all
odd under both inversion and time-reversal, the single component,
$d_{3}\left(\boldsymbol{k}\right)$, is even with respect to both.
From Eqs.~(\ref{supp-eq:Clifford_algebra_definition}), it is possible
to diagonalize Eq.~(\ref{supp-eq:tight-binding_Hamiltonian_Dirac_form})
to yield the following two twofold energy eigenvalues 
\begin{equation}
\varepsilon_{c/v}\left(\boldsymbol{k}\right)=d_{0}\left(\boldsymbol{k}\right)\pm\left|\boldsymbol{d}\left(\boldsymbol{k}\right)\right|.\label{supp-eq:conduction_valence_band_eigenvalues}
\end{equation}
These correspond to the conduction and valence bands, respectively,
each doubly degenerate at each $\boldsymbol{k}$ from Kramer's theorem.
Moreover, one can substitute Eq.~(\ref{supp-eq:conduction_valence_band_eigenvalues})
into Eq.~(\ref{supp-eq:tight-binding_Hamiltonian_Dirac_form}) to
write 
\begin{equation}
\mathsf{h}\left(\boldsymbol{k}\right)=\sum_{b=v,c}\varepsilon_{b}\left(\boldsymbol{k}\right)\mathsf{P}_{b}\left(\boldsymbol{k}\right),\label{supp-eq:hamiltonian_band_projections}
\end{equation}
 which defines the band projectors as 
\begin{equation}
\mathsf{P}_{b}\left(\boldsymbol{k}\right)=\frac{1}{2}\left[\mathds{1}+\eta_{b}\boldsymbol{\hat{d}}\left(\boldsymbol{k}\right)\cdot\boldsymbol{\gamma}\right],\quad\eta_{b}=\begin{cases}
+1, & b=c\\
-1, & b=v
\end{cases}.\label{supp-eq:conduction_valence_band_projectors}
\end{equation}
 These projectors satisfy the following two relations:
\begin{align}
\mathsf{P}_{b}\left(\boldsymbol{k}\right)\mathsf{P}_{b^{\prime}}\left(\boldsymbol{k}\right) & =\delta_{bb^{\prime}}\mathsf{P}_{b}\left(\boldsymbol{k}\right),\label{supp-eq:projector-orthogonality}\\
\sum_{b=v,c}\mathsf{P}_{b}\left(\boldsymbol{k}\right) & =\mathds{1}.
\end{align}

\section{Dynamic Spin Generation through the Nonlinear Edelstein Effect}

We introduce external homogeneous electromagnetic radiation with vector
potential $\boldsymbol{A}\left(t\right)$ to the problem within the
dipole approximation \citep{Mukamel1995,oikeImpactElectronCorrelations2024}.
Within the Coulomb gauge, the electric field of the radiation is given
by$\boldsymbol{E}\left(t\right)=-\partial_{t}\boldsymbol{A}\left(t\right)$.
This leads to the following time-dependent perturbation to Eq.~(\ref{supp-eq:tight-binding_Hamiltonian_Dirac_form}):
\begin{equation}
\Delta\mathcal{H}\left(t\right)=+e\sum_{\boldsymbol{k}}\boldsymbol{\mathsf{v}}\left(\boldsymbol{k}\right)\cdot\boldsymbol{A}\left(t\right),\label{supp-eq:perturbation_hamiltonian}
\end{equation}
where $-e$ is the electron charge. 

\subsection{The Interband Dipole Operator}

The velocity operator is given by $\boldsymbol{\mathsf{v}}\left(\boldsymbol{k}\right)\equiv\partial_{\boldsymbol{k}}\mathsf{h}\left(\boldsymbol{k}\right)$
whose band structure can be exposed from Eq.~(\ref{supp-eq:hamiltonian_band_projections})
as
\begin{equation}
\mathsf{P}_{b}\left(\boldsymbol{k}\right)\boldsymbol{\mathsf{v}}\left(\boldsymbol{k}\right)\mathsf{P}_{b^{\prime}}\left(\boldsymbol{k}\right)=\delta_{bb^{\prime}}\boldsymbol{v}_{b}\left(\boldsymbol{k}\right)\mathsf{P}_{b}\left(\boldsymbol{k}\right)+\sum_{b^{\prime\prime}=v,c}\varepsilon_{b^{\prime\prime}}\left(\boldsymbol{k}\right)\mathsf{P}_{b}\left(\boldsymbol{k}\right)\left[\partial_{\boldsymbol{k}}\mathsf{P}_{b^{\prime\prime}}\left(\boldsymbol{k}\right)\right]\mathsf{P}_{b^{\prime}}\left(\boldsymbol{k}\right),
\end{equation}
 where the group velocity within band $b$ is defined as 
\begin{equation}
\boldsymbol{v}_{b}\left(\boldsymbol{k}\right)\equiv\partial_{\boldsymbol{k}}\varepsilon_{b}\left(\boldsymbol{k}\right)=\partial_{\boldsymbol{k}}d_{0}\left(\boldsymbol{k}\right)+\eta_{b}\partial_{\boldsymbol{k}}\left|\boldsymbol{d}\left(\boldsymbol{k}\right)\right|.
\end{equation}
 Using Eq.~(\ref{supp-eq:projector-orthogonality}), the second term
in the velocity operator follows as
\begin{equation}
\sum_{b^{\prime\prime}=v,c}\varepsilon_{b^{\prime\prime}}\left(\boldsymbol{k}\right)\mathsf{P}_{b}\left(\boldsymbol{k}\right)\left[\partial_{\boldsymbol{k}}\mathsf{P}_{b^{\prime\prime}}\left(\boldsymbol{k}\right)\right]\mathsf{P}_{b^{\prime}}\left(\boldsymbol{k}\right)=-\Delta\varepsilon_{bb^{\prime}}\left(\boldsymbol{k}\right)\mathsf{P}_{b}\left(\boldsymbol{k}\right)\partial_{\boldsymbol{k}}\mathsf{P}_{b^{\prime}}\left(\boldsymbol{k}\right),
\end{equation}
 where $\Delta\varepsilon_{bb^{\prime}}\left(\boldsymbol{k}\right)\equiv\varepsilon_{b}\left(\boldsymbol{k}\right)-\varepsilon_{b^{\prime}}\left(\boldsymbol{k}\right)$.
Thus, the only terms that will contribute to the above for the band
off-diagonal elements $b\neq b^{\prime}$. This geometric term --
being related to the inter-band electric dipole moment -- is expressed
through the operator-valued Berry connection \citep{xiaoBerryPhaseEffects2010,meraNontrivialQuantumGeometry2022,avdoshkinMultistateGeometryShift2025,mitscherlingGaugeinvariantProjectorCalculus2025,mitscherlingOrbitalMagnetizationParallel2025},
\begin{equation}
\boldsymbol{\mathsf{A}}\left(\boldsymbol{k}\right)\equiv\text{i}\sum_{b\neq b^{\prime}}\mathsf{P}_{b}\left(\boldsymbol{k}\right)\partial_{\boldsymbol{k}}\mathsf{P}_{b^{\prime}}\left(\boldsymbol{k}\right),\label{supp-eq:interband_dipole_operator}
\end{equation}
 which is non-Abelian whenever the momentum-derivatives fail to commute
with the band projectors: 
\begin{equation}
\left[\mathsf{A}^{i}\left(\boldsymbol{k}\right),\mathsf{A}^{j}\left(\boldsymbol{k}\right)\right]\neq0.
\end{equation}
Its band-resolved matrix elements are given by 
\begin{align}
\boldsymbol{\mathsf{A}}_{bb^{\prime}}\left(\boldsymbol{k}\right) & \equiv\mathsf{P}_{b}\left(\boldsymbol{k}\right)\boldsymbol{\mathsf{A}}\left(\boldsymbol{k}\right)\mathsf{P}_{b^{\prime}}\left(\boldsymbol{k}\right)=\text{i}\mathsf{P}_{b}\left(\boldsymbol{k}\right)\partial_{\boldsymbol{k}}\mathsf{P}_{b^{\prime}}\left(\boldsymbol{k}\right),\quad b\neq b^{\prime},
\end{align}
which, when the projectors are expressed in a particular gauge of
Bloch wavefunctions $\left\{ \ket{u_{b,a}\left(\boldsymbol{k}\right)}\right\} $,
assumes the form 
\begin{equation}
\boldsymbol{\mathsf{A}}_{bb^{\prime}}\left(\boldsymbol{k}\right)=\sum_{a,a^{\prime}}\braket{u_{b,a}\left(\boldsymbol{k}\right)|\text{i}\partial_{\boldsymbol{k}}|u_{b^{\prime},a^{\prime}}\left(\boldsymbol{k}\right)}\,\bigg(\ket{u_{b^{\prime},a^{\prime}}\left(\boldsymbol{k}\right)}\bra{u_{b,a}\left(\boldsymbol{k}\right)}\bigg),
\end{equation}
from which one observes the role of the dipole operator more explicitly
by translating the momentum-space derivative into the position space
operator: $\text{i}\partial_{\boldsymbol{k}}\leftrightarrow\boldsymbol{r}$
\citep{xiaoBerryPhaseEffects2010,meraNontrivialQuantumGeometry2022}.
For the tight-binding model given by Eq.~(\ref{supp-eq:tight-binding_Hamiltonian_Dirac_form}),
the band-resolved elements of the Berry connection components are
\begin{equation}
\boldsymbol{\mathsf{A}}_{bb^{\prime}}\left(\boldsymbol{k}\right)=\frac{1}{4}\text{i}\eta_{b^{\prime}}\left\{ \gamma^{j}+\frac{1}{2}\eta_{b}\left[\gamma^{i},\gamma^{j}\right]\hat{d}_{i}\left(\boldsymbol{k}\right)\right\} \partial_{\boldsymbol{k}}\hat{d}_{j}\left(\boldsymbol{k}\right),\quad b\neq b^{\prime},\label{supp-eq:non-abelian_Berry_connection}
\end{equation}
since $\gamma^{i}\gamma^{\ell}=\delta^{i\ell}\mathds{1}+\frac{1}{2}\left[\gamma^{i},\gamma^{\ell}\right]$
and the unit vectors $\boldsymbol{\hat{d}}\left(\boldsymbol{k}\right)$
satisfy $\boldsymbol{\hat{d}}\left(\boldsymbol{k}\right)\cdot\partial_{k_{j}}\boldsymbol{\hat{d}}\left(\boldsymbol{k}\right)=0$
for all components $k_{j}$. Written explicitly, the inter-band dipole
operator from Eq.~(\ref{supp-eq:interband_dipole_operator}) is
\begin{align}
\boldsymbol{\mathsf{A}}\left(\boldsymbol{k}\right) & =-\frac{1}{4}\text{i}\left[\gamma^{i},\gamma^{j}\right]\hat{d}_{i}\left(\boldsymbol{k}\right)\partial_{\boldsymbol{k}}\hat{d}_{j}\left(\boldsymbol{k}\right),
\end{align}
and the band-resolved velocity operator follows as 
\begin{equation}
\mathsf{P}_{b}\left(\boldsymbol{k}\right)\boldsymbol{\mathsf{v}}\left(\boldsymbol{k}\right)\mathsf{P}_{b^{\prime}}\left(\boldsymbol{k}\right)=\delta_{bb^{\prime}}\,\boldsymbol{v}_{b}\left(\boldsymbol{k}\right)\mathsf{P}_{b}\left(\boldsymbol{k}\right)+\text{i}\Delta\varepsilon_{bb^{\prime}}\left(\boldsymbol{k}\right)\boldsymbol{\mathsf{A}}_{bb^{\prime}}\left(\boldsymbol{k}\right).\label{supp-eq:band-resolved_velocity_operator}
\end{equation}
The geometry of the wavefunctions is related to the distribution
of charge within the unit cell, and therefore transitions induced
by geometry will lead to dynamic \emph{intra-cell} electric multipoles
that can, of course, radiate or couple to other degrees of freedom
within the material \citep{LandauElectrodynamicsContinuousMedia,Mukamel1995}.

\subsection{Perturbative Spin Dynamics through Interband Dipole Transitions}

In this section, we show that electric dipole transitions driven by linearly polarized light will generically generate a spin expectation value at second order. This is possible even in non-magnetic systems with centrosymmetry, where Kramer's Theorem guarantees the twofold degeneracy of each band, eliminating any spin expectation value in equilibrium. We show that while the spin density at each momentum $\boldsymbol{k}$ vanishes in each band in equilibrium, the local spin-orbit coupling still generates a hidden spin texture in the bands. Under illumination by light, the inter-band electric dipole transition -- characterized by the perturbation $\boldsymbol{\mathsf{A}} \cdot \boldsymbol{E}(t)$ -- induces a momentum-resolved spin density along the $z$-axis. Because it is an odd function of the momentum, however, the total spin ultimately vanishes upon Brillouin zone integration. At second order, however, we show that there are two different pathways that generate spin density in all three components. These pathways are the result of either (i) an electric dipole current characterized by the dyad $\boldsymbol{v}_b \otimes \boldsymbol{\mathsf{A}}$ or (ii) an intra-cell electric quadrupole characterized by the dyad $\boldsymbol{\mathsf{A}} \otimes \boldsymbol{\mathsf{A}}$. In either case, these dyads are even functions of the momentum, and will survive Brillouin zone integration. This leads to a transient spin density at second order.

\subsubsection{Band-Resolved Perturbation Theory}

Using standard density matrix perturbation theory in the Schr\"odinger
picture \citep{Mukamel1995}, we find can write the time-dependent
expectation value of any operator, $\mathsf{Q}$, at angular frequency
$\omega$ as a Dyson Series
\begin{equation}
\left\langle \mathsf{Q}^{\left(n\right)}\left(\omega\right)\right\rangle =\sum_{\boldsymbol{k}}\sum_{b,b^{\prime}}\text{tr}\left[\mathsf{Q}\mathsf{P}_{b}\left(\boldsymbol{k}\right)\rho_{\boldsymbol{k}}^{\left(n\right)}\left(\omega\right)\mathsf{P}_{b^{\prime}}\left(\boldsymbol{k}\right)\right],\label{supp-eq:operator_dyson_series}
\end{equation}
where $\rho_{\boldsymbol{k}}^{\left(n\right)}\left(\omega\right)$
is the $n^{\text{th}}$-order perturbative correction to the single-particle
density matrix at momentum $\boldsymbol{k}$. The recursive Dyson
Series solution permits the following band-resolved density matrix:
\begin{equation}
\mathsf{P}_{b}\left(\boldsymbol{k}\right)\rho^{\left(n\right)}\left(\omega\right)\mathsf{P}_{b^{\prime}}\left(\boldsymbol{k}\right)=\frac{1}{\omega-\Delta\varepsilon_{bb^{\prime}}\left(\boldsymbol{k}\right)+\text{i}\delta^{+}}\int\frac{\text{d}\omega^{\prime}}{2\pi}\,\mathsf{P}_{b}\left(\boldsymbol{k}\right)\left[\Delta\mathcal{H}\left(\omega^{\prime}\right),\rho^{\left(n-1\right)}\left(\omega-\omega^{\prime}\right)\right]\mathsf{P}_{b^{\prime}}\left(\boldsymbol{k}\right),
\end{equation}
where $\delta^{+}>0$ is an infinitesimal regularizer that guarantees
causality within the non-interacting disorder-free model. It broadens to a finite lifetime when one considers intra-band relaxation or inter-band decoherence processes
whose inclusions are beyond the scope of the current work. Substituting in Eq.~(\ref{supp-eq:perturbation_hamiltonian})
yields 
\begin{equation}
\mathsf{P}_{b}\left(\boldsymbol{k}\right)\rho^{\left(n\right)}\left(\omega\right)\mathsf{P}_{b^{\prime}}\left(\boldsymbol{k}\right)=\frac{\text{i}e}{\omega-\Delta\varepsilon_{bb^{\prime}}\left(\boldsymbol{k}\right)+\text{i}\delta^{+}}\int\frac{\text{d}\omega^{\prime}}{2\pi}\,\frac{1}{\omega^{\prime}}E_{i}\left(\omega^{\prime}\right)\mathsf{P}_{b}\left(\boldsymbol{k}\right)\left[\mathsf{v}_{i}\left(\boldsymbol{k}\right),\rho^{\left(n-1\right)}\left(\omega-\omega^{\prime}\right)\right]\mathsf{P}_{b^{\prime}}\left(\boldsymbol{k}\right),
\end{equation}
 where we have used the Coulomb gauge to write $\boldsymbol{A}\left(\omega\right)=-\boldsymbol{E}\left(\omega\right)/\text{i}\omega$
and have used Einstein summation on the index $i$. In equilibrium,
the density matrix diagonalizes in the bands as 
\begin{equation}
\mathsf{P}_{b}\left(\boldsymbol{k}\right)\rho^{\left(0\right)}\mathsf{P}_{b^{\prime}}\left(\boldsymbol{k}\right)=\delta_{bb^{\prime}}\,n\left(\varepsilon_{b}\left(\boldsymbol{k}\right)\right)\mathsf{P}_{b}\left(\boldsymbol{k}\right),
\end{equation}
 with $n\left(\varepsilon\right)=\left[1+\text{e}^{\beta\left(\varepsilon-\mu\right)}\right]^{-1}$
is the Fermi function. The first-order correction to the density matrix
follows then as 
\begin{align}
\mathsf{P}_{b}\left(\boldsymbol{k}\right)\rho^{\left(1\right)}\left(\omega\right)\mathsf{P}_{b^{\prime}}\left(\boldsymbol{k}\right) & =\frac{\text{i}e\,\Delta n_{b^{\prime}b}\left(\boldsymbol{k}\right)}{\omega}\,\frac{\mathsf{P}_{b}\left(\boldsymbol{k}\right)\mathsf{v}_{i}\left(\boldsymbol{k}\right)\mathsf{P}_{b^{\prime}}\left(\boldsymbol{k}\right)}{\omega-\Delta\varepsilon_{bb^{\prime}}\left(\boldsymbol{k}\right)+\text{i}\delta^{+}}\,E_{i}\left(\omega\right),\label{supp-eq:1st_order_density_matrix}
\end{align}
with $\Delta n_{b^{\prime}b}\left(\boldsymbol{k}\right)\equiv n\left(\varepsilon_{b^{\prime}}\left(\boldsymbol{k}\right)\right)-n\left(\varepsilon_{b}\left(\boldsymbol{k}\right)\right)$.
Since $\Delta n_{bb}\left(\boldsymbol{k}\right)=0$, the only nonzero
components in the above are inter-band for $b\neq b^{\prime}$ and
resonant for $\omega\rightarrow\Delta\varepsilon_{bb^{\prime}}\left(\boldsymbol{k}\right)$.
From Eq. (\ref{supp-eq:band-resolved_velocity_operator}), it follows
that the inter-band coherence develops at first-order due to the dipole
transition mediated by the non-Abelian Berry connection
\begin{equation}
\mathsf{P}_{b}\left(\boldsymbol{k}\right)\rho^{\left(1\right)}\left(\omega\right)\mathsf{P}_{b^{\prime}}\left(\boldsymbol{k}\right)=-\frac{e\,\Delta n_{b^{\prime}b}\left(\boldsymbol{k}\right)}{\omega}\,\frac{\Delta\varepsilon_{bb^{\prime}}\left(\boldsymbol{k}\right)\mathsf{A}_{bb^{\prime}}^{i}\left(\boldsymbol{k}\right)}{\omega-\Delta\varepsilon_{bb^{\prime}}\left(\boldsymbol{k}\right)+\text{i}\delta^{+}}\,E_{i}\left(\omega\right),\quad b\neq b^{\prime}.\label{supp-eq:1st_order_density_matrix-berry}
\end{equation}

The second-order correction then follows as 

\begin{align}
\mathsf{P}_{b}\left(\boldsymbol{k}\right)\rho^{\left(2\right)}\left(\omega\right)\mathsf{P}_{b^{\prime}}\left(\boldsymbol{k}\right) & =\frac{e^{2}}{\omega-\Delta\varepsilon_{bb^{\prime}}\left(\boldsymbol{k}\right)+\text{i}\delta^{+}}\int\frac{\text{d}\omega^{\prime}}{2\pi}\,\frac{E_{i}\left(\omega^{\prime}\right)E_{j}\left(\omega-\omega^{\prime}\right)}{\omega^{\prime}\left(\omega-\omega^{\prime}\right)}\nonumber \\
 & \phantom{=}\times\sum_{b_{1}\neq b_{2}}\Delta n_{b_{2}b_{1}}\left(\boldsymbol{k}\right)\Delta\varepsilon_{b_{1}b_{2}}\left(\boldsymbol{k}\right)\frac{\mathsf{P}_{b}\left(\boldsymbol{k}\right)\left[\mathsf{v}^{i}\left(\boldsymbol{k}\right),\mathsf{P}_{b_{1}}\left(\boldsymbol{k}\right)\mathsf{A}_{b_{1}b_{2}}^{j}\left(\boldsymbol{k}\right)\mathsf{P}_{b_{2}}\left(\boldsymbol{k}\right)\right]\mathsf{P}_{b^{\prime}}\left(\boldsymbol{k}\right)}{\omega-\omega^{\prime}-\Delta\varepsilon_{b_{1}b_{2}}\left(\boldsymbol{k}\right)+\text{i}\delta^{+}}.
\end{align}
We must, therefore, evaluate the following convolutions of the form:
\begin{equation}
\mathcal{I}_{ij}\left(\omega;\Delta\right)\equiv\int\frac{\text{d}\omega^{\prime}}{2\pi}\,\frac{E_{i}\left(\omega^{\prime}\right)E_{j}\left(\omega-\omega^{\prime}\right)}{\omega^{\prime}\left(\omega-\omega^{\prime}\right)\left(\omega-\omega^{\prime}-\Delta+\text{i}\delta^{+}\right)}.\label{supp-eq:integral_kernel}
\end{equation}
Assuming that the external electric field is an analytic function
of frequency and $\left|E_{i}\left(\omega\rightarrow\infty\right)\right|\rightarrow0$,
the convolution follows from the residue theorem given the simple
poles that exist at $\omega^{\prime}=0$, $\omega^{\prime}=\omega$,
and $\omega^{\prime}=\omega-\Delta\varepsilon_{bb^{\prime}}\left(\boldsymbol{k}\right)+\text{i}\delta^{+}$.
In the absence of dc electric fields, $\boldsymbol{E}_{\text{dc}}\equiv\boldsymbol{E}\left(\omega=0\right)=\boldsymbol{0}$,
the resulting convolution is 
\begin{equation}
\mathcal{I}_{ij}\left(\omega;\Delta\neq0,\boldsymbol{E}_{\text{dc}}=\boldsymbol{0}\right)=\frac{\text{i}E_{i}\left(\omega-\Delta\right)E_{j}\left(\Delta\right)}{\Delta\left(\omega-\Delta+\text{i}\delta^{+}\right)}.
\end{equation}
Being a second-order response, this inter-band resonance is the result
of the first-order inter-band coherence driven by light then interacting
again with the light. The resonance in this integral kernel is exposed
through functional differentiation as
\begin{align}
\frac{\delta^{2}\mathcal{I}_{ij}\left(\omega;\Delta\neq0,\boldsymbol{E}_{\text{dc}}=\boldsymbol{0}\right)}{\delta E_{m}\left(\omega_{m}\right)\delta E_{n}\left(\omega_{n}\right)} & =\text{i}\delta\left(\omega-\left(\omega_{m}+\omega_{n}\right)\right)\delta\left(\Delta-\omega_{n}\right)\,\left(\frac{\delta_{im}\delta_{jn}}{\omega_{m}+\text{i}\delta^{+}}\right)+\left(m\leftrightarrow n\right).\label{supp-eq:second-order_functional_derivative}
\end{align}
Focusing on the first term in the above, resonant second-order response
appears when the system is driven by light at the band gap with $\omega_{n}=\Delta$.
In particular, for monochromatic \emph{linearly polarized} light with
$\boldsymbol{E}\left(t\right)\propto\text{Re}\left[\boldsymbol{E}\left(\omega_{n}\right)\text{e}^{-\text{i}\omega_{n}t}\right]$,
energy conservation will enforce the resonance to occur through pure
second-harmonic generation with $\omega=\omega_{\text{SHG}}=\omega_{m}+\omega_{n}=2\Delta$
or pure rectification $\omega=\omega_{\text{R}}=\omega_{m}+\omega_{n}=0$. 

Substituting (\ref{supp-eq:integral_kernel}) into the second-order
correction to the density matrix finally yields

\begin{align}
\mathsf{P}_{b}\left(\boldsymbol{k}\right)\rho^{\left(2\right)}\left(\omega\right)\mathsf{P}_{b^{\prime}}\left(\boldsymbol{k}\right) & =\frac{\text{i}e^{2}}{\omega-\Delta\varepsilon_{bb^{\prime}}\left(\boldsymbol{k}\right)+\text{i}\delta^{+}}\nonumber \\
 & \phantom{=}\times\sum_{b_{1}\neq b_{2}}\frac{\Delta n_{b_{2}b_{1}}\left(\boldsymbol{k}\right)E_{i}\left(\omega-\Delta\varepsilon_{b_{1}b_{2}}\left(\boldsymbol{k}\right)\right)E_{j}\left(\Delta\varepsilon_{b_{1}b_{2}}\left(\boldsymbol{k}\right)\right)}{\omega-\Delta\varepsilon_{b_{1}b_{2}}\left(\boldsymbol{k}\right)+\text{i}\delta^{+}}\nonumber \\
 & \phantom{=\times\sum}\times\mathsf{P}_{b}\left(\boldsymbol{k}\right)\left[\mathsf{v}^{i}\left(\boldsymbol{k}\right),\mathsf{P}_{b_{1}}\left(\boldsymbol{k}\right)\mathsf{A}_{b_{1}b_{2}}^{j}\left(\boldsymbol{k}\right)\mathsf{P}_{b_{2}}\left(\boldsymbol{k}\right)\right]\mathsf{P}_{b^{\prime}}\left(\boldsymbol{k}\right).\label{supp-eq:2nd_order_density_matrix}
\end{align}
Collecting these non-equilibrium corrections to the single-particle
density matrix, Eqs.~(\ref{supp-eq:1st_order_density_matrix}) and
(\ref{supp-eq:2nd_order_density_matrix}), and substituting them into
Eq.~(\ref{supp-eq:operator_dyson_series}) yields
\begin{equation}
\left\langle \mathsf{Q}^{\left(0\right)}\right\rangle =\sum_{\boldsymbol{k}}\sum_{b=v,c}n\left(\varepsilon_{b}\right)\text{tr}\left(\mathsf{P}_{b}\mathsf{Q}\mathsf{P}_{b}\right),
\end{equation}
\begin{align}
\left\langle \mathsf{Q}^{\left(1\right)}\left(\omega\right)\right\rangle  & =-\sum_{\boldsymbol{k}}\frac{e\,\Delta n_{vc}\left(\boldsymbol{k}\right)\Delta\varepsilon_{cv}\left(\boldsymbol{k}\right)}{\omega}\left\{ \frac{\text{tr}\left[\mathsf{Q}\mathsf{A}_{cv}^{i}\left(\boldsymbol{k}\right)\right]}{\omega-\Delta\varepsilon_{cv}\left(\boldsymbol{k}\right)+\text{i}\delta^{+}}-\frac{\text{tr}\left[\mathsf{Q}\mathsf{A}_{vc}^{i}\left(\boldsymbol{k}\right)\right]}{\omega+\Delta\varepsilon_{cv}\left(\boldsymbol{k}\right)+\text{i}\delta^{+}}\right\} \,E_{i}\left(\omega\right),\label{supp-eq:Q(1)_response}
\end{align}
\begin{align}
\left\langle \mathsf{Q}^{\left(2\right)}\left(\omega\right)\right\rangle  & =\sum_{\boldsymbol{k}}\sum_{bb^{\prime}}\sum_{b_{1}\neq b_{2}}\frac{\text{i}e^{2}}{\omega-\Delta\varepsilon_{bb^{\prime}}\left(\boldsymbol{k}\right)+\text{i}\delta^{+}}\frac{\Delta n_{b_{2}b_{1}}\left(\boldsymbol{k}\right)E_{i}\left(\omega-\Delta\varepsilon_{b_{1}b_{2}}\left(\boldsymbol{k}\right)\right)E_{j}\left(\Delta\varepsilon_{b_{1}b_{2}}\left(\boldsymbol{k}\right)\right)}{\omega-\Delta\varepsilon_{b_{1}b_{2}}\left(\boldsymbol{k}\right)+\text{i}\delta^{+}}\nonumber \\
 & \phantom{=\sum_{\boldsymbol{k}}\sum_{bb^{\prime}}\sum_{b_{1}\neq b_{2}}}\times\text{tr}\left\{ \mathsf{Q}\mathsf{P}_{b}\left[\mathsf{v}^{i}\left(\boldsymbol{k}\right),\mathsf{P}_{b_{1}}\mathsf{A}_{b_{1}b_{2}}^{j}\left(\boldsymbol{k}\right)\mathsf{P}_{b_{2}}\right]\mathsf{P}_{b^{\prime}}\right\} \label{supp-eq:Q(2)_response}
\end{align}
 with the trace operation, $\text{tr}\left(\cdot\right)$, being performed
on the $4\times4$ direct-product space of each Bloch momentum $\boldsymbol{k}$.
In the above, we have suppressed the momentum label on the projectors
for brevity. 

\subsubsection{Hidden Spin Texture in Equilibrium }

We now move to evaluate the dynamical spin response to light by substituting
in $\boldsymbol{\Sigma}$ for $\mathsf{Q}$. Suppressing the $\boldsymbol{k}$-dependence,
we write 
\begin{equation}
\mathsf{P}_{b}\boldsymbol{\Sigma}\mathsf{P}_{b^{\prime}}=\frac{1}{4}\left[\boldsymbol{\Sigma}+\hat{d}_{i}\left(\eta_{b}\gamma^{i}\boldsymbol{\Sigma}+\eta_{b^{\prime}}\boldsymbol{\Sigma}\gamma^{i}\right)+\eta_{b}\eta_{b^{\prime}}\hat{d}_{i}\hat{d}_{j}\gamma^{i}\boldsymbol{\Sigma}\gamma^{j}\right].\label{supp-eq:band_resolved_Sigma_operator}
\end{equation}
 At zeroth order, we only consider $b=b^{\prime}$. Then we have that
\begin{align}
\mathsf{P}_{b}\Sigma_{\alpha}\mathsf{P}_{b} & =-\frac{1}{4}\text{i}\left[\gamma^{\alpha}\gamma^{3}\gamma^{4}+\eta_{b}\hat{d}_{i}\left\{ \gamma^{i},\gamma^{\alpha}\gamma^{3}\gamma^{4}\right\} +\hat{d}_{i}\hat{d}_{j}\gamma^{i}\gamma^{\alpha}\gamma^{3}\gamma^{4}\gamma^{j}\right],\quad\alpha\in\left\{ 1,2\right\} ,\label{supp-eq:PbSigma_alphaPb}\\
\mathsf{P}_{b}\Sigma_{z}\mathsf{P}_{b} & =-\frac{1}{4}\text{i}\left[\gamma^{1}\gamma^{2}+\eta_{b}\hat{d}_{i}\left\{ \gamma^{i},\gamma^{1}\gamma^{2}\right\} +\hat{d}_{i}\hat{d}_{j}\gamma^{i}\gamma^{1}\gamma^{2}\gamma^{j}\right],\label{supp-eq:PbSigma_3Pb}
\end{align}
 from which the traces follow as 
\begin{align}
\text{tr}\left(\mathsf{P}_{b}\Sigma_{\alpha}\mathsf{P}_{b}\right) & =-\frac{1}{4}\text{i}\left[\mathscr{T}_{3}^{\alpha34}+\eta_{b}\hat{d}_{i}\left(\mathscr{T}_{4}^{i\alpha34}+\mathscr{T}_{4}^{\alpha34i}\right)+\hat{d}_{i}\hat{d}_{j}\mathscr{T}_{5}^{i\alpha34j}\right]=0,\quad\alpha\in\left\{ 1,2\right\} \\
\text{tr}\left(\mathsf{P}_{b}\Sigma_{z}\mathsf{P}_{b}\right) & =-\frac{1}{4}\text{i}\left[\mathscr{T}_{2}^{12}+\eta_{b}\hat{d}_{i}\left(\mathscr{T}_{3}^{i12}+\mathscr{T}_{3}^{12i}\right)+\hat{d}_{i}\hat{d}_{j}\mathscr{T}_{4}^{i12j}\right]=0.
\end{align}
In the expressions above, we have used the fact that the trace over
any odd number of Dirac matrices vanishes. Additionally, for the $\alpha=1,2$
spin components, $\mathscr{T}_{4}^{i\alpha34}=0$ identically since
$\alpha\in\left\{ 1,2\right\} $, whereas for the out-of-plane spin,
it is the symmetric summation over $\hat{d}_{i}\hat{d}_{j}$ that
eliminates the last term since $\mathscr{T}_{4}^{i12j}=-\mathscr{T}_{4}^{j12i}$.
Thus, in equilibrium, spin expectation value vanishes within \emph{each}
band and for \emph{each} momentum in the Brillouin zone, again reflecting
the Kramer's degeneracy. However, it is important to recognize that
there is a momentum-dependent ``hidden spin texture'' \citep{zhangHiddenSpinPolarization2014,huangHiddenSpinPolarization2020,arnoldiRevealingHiddenSpin2024}
within each band revealed in Eqs.~(\ref{supp-eq:PbSigma_alphaPb})
and (\ref{supp-eq:PbSigma_3Pb}). 

Since the spin at each momentum in each band is compensated, then
the total equilibrium spin also vanishes
\begin{equation}
\left\langle \boldsymbol{S}^{\left(0\right)}\right\rangle =\frac{1}{2}\sum_{\boldsymbol{k}}\sum_{b=v,c}n\left(\varepsilon_{b}\right)\text{tr}\left(\mathsf{P}_{b}\boldsymbol{\Sigma}\mathsf{P}_{b}\right)=\boldsymbol{0},
\end{equation}
 again reflecting the Kramer's degeneracy of the model.

\subsubsection{First-Order Response}

The first-order spin response is computed from the trace in Eq.~(\ref{supp-eq:Q(1)_response}).
For the trace to survive, the Dirac matrices in the product must all
come in pairs. The dipole transitions are induced by a non-Abelian
Berry connection of the form $\gamma+\gamma\gamma$ (see Eq.~(\ref{supp-eq:non-abelian_Berry_connection})),
and therefore only the first term may contribute a nonzero trace for
the in-plane spin components. For the in-plane components, we find
that 
\begin{equation}
\text{tr}\left[\Sigma_{\alpha}\mathsf{A}_{b^{\prime}b}^{i}\left(\boldsymbol{k}\right)\right]=\frac{1}{4}\eta_{b}\partial_{k_{i}}\hat{d}_{\ell}\left(\boldsymbol{k}\right)\,\text{tr}\left(\gamma^{\alpha}\gamma^{3}\gamma^{4}\gamma^{\ell}\right)=0,\quad\alpha=1,2.
\end{equation}
Thus, despite the inter-band resonance, there is no first-order spin
generated in the plane at \emph{any} momentum. The out-of-plane spin
component, meanwhile, will only survive from tracing over the second
term. It follows as
\begin{align}
\text{tr}\left[\Sigma_{z}\mathsf{A}_{b^{\prime}b}^{i}\left(\boldsymbol{k}\right)\right] & =\hat{d}_{2}\left(\boldsymbol{k}\right)\partial_{k_{i}}\hat{d}_{1}\left(\boldsymbol{k}\right)-\hat{d}_{1}\left(\boldsymbol{k}\right)\partial_{k_{i}}\hat{d}_{2}\left(\boldsymbol{k}\right),\quad b\neq b^{\prime},
\end{align}
The result is that the momentum-resolved spin-response generated by
the inter-band coherence is
\begin{align}
\left\langle S_{x,y}^{\left(1\right)}\left(\omega;\,\boldsymbol{k}\right)\right\rangle  & =0,\\
\left\langle S_{z}^{\left(1\right)}\left(\omega;\,\boldsymbol{k}\right)\right\rangle  & =-\frac{e}{2\omega}\sum_{b,b^{\prime}\neq b}\frac{\hat{d}_{2}\left(\boldsymbol{k}\right)\partial_{k_{i}}\hat{d}_{1}\left(\boldsymbol{k}\right)-\hat{d}_{1}\left(\boldsymbol{k}\right)\partial_{k_{i}}\hat{d}_{2}\left(\boldsymbol{k}\right)}{\omega-\Delta\varepsilon_{bb^{\prime}}\left(\boldsymbol{k}\right)+\text{i}\delta^{+}}\Delta n_{b^{\prime}b}\left(\boldsymbol{k}\right)\Delta\varepsilon_{b^{\prime}b}\left(\boldsymbol{k}\right)E_{i}\left(\omega\right).
\end{align}
While the in-plane response vanishes at first-order for all momenta,
the spins are polarized out-of-plane at any given momentum. While
it is true that the inter-band resonance that develops at $\omega=\Delta\varepsilon_{bb^{\prime}}\left(\boldsymbol{k}\right)$
will lead to a momentum-resolved magneto-electric effect analogous
to the intra-band \emph{linear} Edelstein effect \citep{edelsteinSpinPolarizationConduction1990},
summing over all momenta will eliminate the response due to global
centrosymmetry. This is because the quantity $\hat{d}_{1}\left(\boldsymbol{k}\right)\partial_{k_{i}}\hat{d}_{2}\left(\boldsymbol{k}\right)-\hat{d}_{2}\left(\boldsymbol{k}\right)\partial_{k_{i}}\hat{d}_{1}\left(\boldsymbol{k}\right)$
is odd under inversion, and will therefore vanish in centrosymmetric
crystals, such as \CGT. Thus, the total dynamic spin response vanishes
at first order:
\begin{equation}
\left\langle \boldsymbol{S}^{\left(1\right)}\left(\omega\right)\right\rangle =\sum_{\boldsymbol{k}}\left\langle \boldsymbol{S}^{\left(1\right)}\left(\omega;\,\boldsymbol{k}\right)\right\rangle =\boldsymbol{0}.
\end{equation}

\subsubsection{Second-Order Nonlinear Edelstein Effect}

The second order effect arises from Eq.~(\ref{supp-eq:Q(2)_response})
with $\mathsf{Q}=\boldsymbol{\Sigma}$. We must evaluate traces of
the following forms:
\begin{equation}
\text{tr}\left[\boldsymbol{\Sigma}\,\mathsf{P}_{b}^{\phantom{j}}\mathsf{v}_{i}^{\phantom{j}}\left(\boldsymbol{k}\right)\mathsf{P}_{b^{\prime\prime}}^{\phantom{j}}\mathsf{A}_{b^{\prime\prime}b^{\prime}}^{j}\left(\boldsymbol{k}\right)\right]\quad\text{and}\quad\text{tr}\left[\boldsymbol{\Sigma}\,\mathsf{A}_{bb^{\prime\prime}}^{j}\left(\boldsymbol{k}\right)\mathsf{P}_{b^{\prime\prime}}^{\phantom{j}}\mathsf{v}_{i}^{\phantom{j}}\left(\boldsymbol{k}\right)\mathsf{P}_{b^{\prime}}^{\phantom{j}}\right].
\end{equation}
Whereas the first-order resonance considered here only arises from
inter-band processes, the second-order response includes contributions
from \emph{both} intra- and inter-band velocities \citep{xuLightinducedStaticMagnetization2021, xiaoIntrinsicNonlinearElectric2022,xiaoTimeReversalEvenNonlinearCurrent2023,xueNonlinearOpticsDrivenSpin2024,yeNonlinearSpinOrbital2024}.
Using the band-resolved velocity operator in Eq.~(\ref{supp-eq:band-resolved_velocity_operator}),
we have then 
\begin{align}
\text{tr}\left[\boldsymbol{\Sigma}\,\mathsf{P}_{b}^{\phantom{j}}\mathsf{v}_{i}^{\phantom{j}}\left(\boldsymbol{k}\right)\mathsf{P}_{b^{\prime\prime}}^{\phantom{j}}\mathsf{A}_{b^{\prime\prime}b^{\prime}}^{j}\left(\boldsymbol{k}\right)\right] & =\delta_{bb^{\prime\prime}}v_{b}^{i}\left(\boldsymbol{k}\right)\,\text{tr}\left[\boldsymbol{\Sigma}\,\mathsf{P}_{b}^{\phantom{j}}\mathsf{A}_{bb^{\prime}}^{j}\left(\boldsymbol{k}\right)\right]+\text{i}\left(1-\delta_{bb^{\prime\prime}}\right)\Delta\varepsilon_{bb^{\prime\prime}}\left(\boldsymbol{k}\right)\,\text{tr}\left[\boldsymbol{\Sigma}\,\mathsf{A}_{bb^{\prime\prime}}^{i}\left(\boldsymbol{k}\right)\mathsf{A}_{b^{\prime\prime}b^{\prime}}^{j}\left(\boldsymbol{k}\right)\right],\\
\nonumber \\\text{tr}\left[\boldsymbol{\Sigma}\,\mathsf{A}_{bb^{\prime\prime}}^{j}\left(\boldsymbol{k}\right)\mathsf{P}_{b^{\prime\prime}}^{\phantom{j}}\mathsf{v}_{i}^{\phantom{j}}\left(\boldsymbol{k}\right)\mathsf{P}_{b^{\prime}}^{\phantom{j}}\right] & =\delta_{b^{\prime\prime}b^{\prime}}v_{b^{\prime}}^{i}\left(\boldsymbol{k}\right)\,\text{tr}\left[\boldsymbol{\Sigma}\,\mathsf{A}_{bb^{\prime\prime}}^{j}\left(\boldsymbol{k}\right)\mathsf{P}_{b^{\prime}}^{\phantom{j}}\right]+\text{i}\left(1-\delta_{b^{\prime\prime}b^{\prime}}\right)\Delta\varepsilon_{b^{\prime\prime}b^{\prime}}\left(\boldsymbol{k}\right)\,\text{tr}\left[\boldsymbol{\Sigma}\,\mathsf{A}_{bb^{\prime\prime}}^{j}\left(\boldsymbol{k}\right)\mathsf{A}_{b^{\prime\prime}b^{\prime}}^{i}\left(\boldsymbol{k}\right)\right].
\end{align}
 To gain physical insight into the meaning of these terms, we take
$b^{\prime}=v$ to be the valence band. The first-order inter-band
process restrict $b^{\prime\prime}=c$ to the conduction band in the
first trace and $b\neq b^{\prime\prime}$ in the second. We find that
these traces are then 
\begin{align}
\text{tr}\left[\boldsymbol{\Sigma}\,\mathsf{P}_{b}^{\phantom{j}}\mathsf{v}_{i}^{\phantom{j}}\left(\boldsymbol{k}\right)\mathsf{P}_{c}^{\phantom{j}}\mathsf{A}_{cv}^{j}\left(\boldsymbol{k}\right)\right] & =\delta_{bc}\delta_{b^{\prime\prime}c}v_{c}^{i}\left(\boldsymbol{k}\right)\,\text{tr}\left[\boldsymbol{\Sigma}\,\mathsf{P}_{c}^{\phantom{j}}\mathsf{A}_{cv}^{j}\left(\boldsymbol{k}\right)\right]+\text{i}\delta_{bv}\delta_{b^{\prime\prime}c}\Delta\varepsilon_{vc}\left(\boldsymbol{k}\right)\,\text{tr}\left[\boldsymbol{\Sigma}\,\mathsf{A}_{vc}^{i}\left(\boldsymbol{k}\right)\mathsf{A}_{cv}^{j}\left(\boldsymbol{k}\right)\right]\\
\nonumber \\\text{tr}\left[\boldsymbol{\Sigma}\,\mathsf{A}_{bb^{\prime\prime}}^{j}\left(\boldsymbol{k}\right)\mathsf{P}_{b^{\prime\prime}}^{\phantom{j}}\mathsf{v}_{i}^{\phantom{j}}\left(\boldsymbol{k}\right)\mathsf{P}_{v}^{\phantom{j}}\right] & =\delta_{bc}\delta_{b^{\prime\prime}v}v_{v}^{i}\left(\boldsymbol{k}\right)\,\text{tr}\left[\boldsymbol{\Sigma}\,\mathsf{A}_{cv}^{j}\left(\boldsymbol{k}\right)\mathsf{P}_{v}^{\phantom{j}}\right]+\text{i}\delta_{bv}\delta_{b^{\prime\prime}c}\Delta\varepsilon_{cv}\left(\boldsymbol{k}\right)\,\text{tr}\left[\boldsymbol{\Sigma}\,\mathsf{A}_{vc}^{j}\left(\boldsymbol{k}\right)\mathsf{A}_{cv}^{i}\left(\boldsymbol{k}\right)\right]
\end{align}
 Starting with $b^{\prime}=v$ results in a set of four processes:
two resulting in dipole currents within the conduction band $b=c$ and two inter-band virtual dipole-dipole processes with $b=v$. These classes of second-order processes are illustrated in Supplementary Fig.~\ref{fig:2ndOrderProcesses}. The two
inter-band processes are associated with the actual motion of carriers
within the conduction and valence bands moving through an effective
gauge field generated by the inter-band non-Abelian Berry connection. 
As a result, an electric dipole current is generated as a \emph{dyadic}
response of the form $\boldsymbol{v}_{b}\otimes\boldsymbol{\mathsf{A}}$
which has the same point-group character as the electric field dyad
$\boldsymbol{E}\otimes\boldsymbol{E}$ (Supplementary Fig.~\ref{fig:2ndOrderProcesses}a). The two dipole-dipole processes,
meanwhile, are purely geometric and represent interference in the
spin-structure for carriers undergoing the virtual excitation process
from the valence band into the conduction, and then have the same
virtual process return the carrier back to the valence band. Given
that these processes are generated by a dynamic electric dipole, the
second-order response can be viewed generically as a dynamic \emph{quadrupole}
moment (Supplementary Fig.~\ref{fig:2ndOrderProcesses}b). As a result, the material response is again \emph{dyadic}
of the form $\boldsymbol{\mathsf{A}}\otimes\boldsymbol{\mathsf{A}}$
which will again have the same point-group character as $\boldsymbol{E}\otimes\boldsymbol{E}$. 

\begin{figure*}
    \centering
    \includegraphics[width=0.55\linewidth]{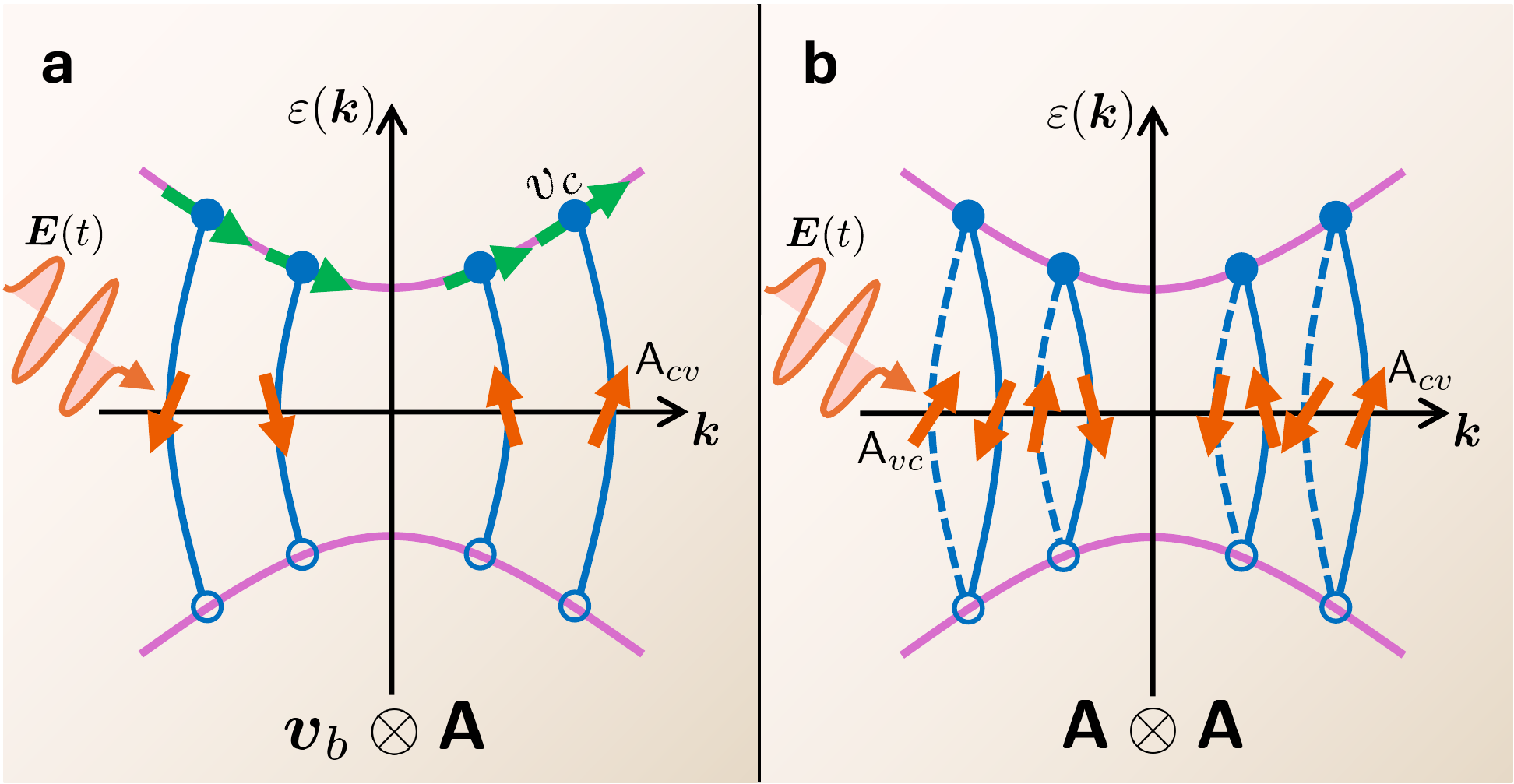}
    \caption{The two second-order processes that give rise to a transient spin density driven by the interband nonlinear Edelstein effect. \textbf{(a)} Spin generation through electric dipole current. The dipole moment is captured by the non-Abelian Berry connection $\boldsymbol{\mathsf{A}}_{cv}(\boldsymbol{k})$ driven by the valence-to-conduction band transition at first order. The dipole then pairs with the photo-excited carriers in the conduction band with velocity $\boldsymbol{v}_c(\boldsymbol{k}) = \partial_{\boldsymbol{k}}\varepsilon_c(\boldsymbol{k})$ to generate the spin density. \textbf{(b)} Spin generation through bound electric quadrupole fluctuations. The valence-to-conduction-to-valence pathway at second order generates a dyad of the non-Abelian Berry connection: $\boldsymbol{\mathsf{A}}_{vc}(\boldsymbol{k}) \otimes \boldsymbol{\mathsf{A}}_{cv}(\boldsymbol{k})$. This dyad, in turn, corresponds to an intra-cell electric quadrupole which subsequently drives a spin density.  }
    \label{fig:2ndOrderProcesses}
\end{figure*}

What remains is to determine which spin components survive the trace.
Suppressing the momentum labels, we have that the following products
decompose into the Dirac matrices as
\begin{align}
\mathsf{P}_{b}^{\phantom{j}}\mathsf{A}_{bb^{\prime}}^{i} & =\frac{\text{i}}{8}\eta_{b^{\prime}}\left\{ \gamma^{m}+\frac{1}{2}\eta_{b}\left[\gamma^{\ell},\gamma^{m}\right]\hat{d}_{\ell}+\eta_{b}\gamma^{n}\gamma^{m}\hat{d}_{n}+\frac{1}{2}\eta_{b}\eta_{b^{\prime}}\gamma^{n}\left[\gamma^{\ell},\gamma^{m}\right]\hat{d}_{n}\hat{d}_{\ell}\right\} \partial_{k_{i}}\hat{d}_{m},\label{supp-eq:PbAibb'}\\
\mathsf{A}_{bb^{\prime\prime}}^{i}\mathsf{A}_{b^{\prime\prime}b^{\prime}}^{j} & =-\frac{\eta_{b^{\prime}}\eta_{b^{\prime\prime}}}{16}\bigg\{\gamma^{m}\gamma^{n}+\frac{1}{2}\eta_{b}\left[\gamma^{\ell},\gamma^{m}\right]\gamma^{n}\hat{d}_{\ell}\nonumber \\
 & \phantom{=-\frac{\eta_{b^{\prime}}\eta_{b^{\prime\prime}}}{16}\bigg\{}+\frac{1}{2}\eta_{b^{\prime\prime}}\gamma^{m}\left[\gamma^{p},\gamma^{n}\right]\hat{d}_{p}+\frac{1}{4}\eta_{b}\eta_{b^{\prime\prime}}\left[\gamma^{\ell},\gamma^{m}\right]\left[\gamma^{p},\gamma^{n}\right]\hat{d}_{\ell}\hat{d}_{p}\bigg\}\left(\partial_{k_{i}}\hat{d}_{m}\right)\left(\partial_{k_{j}}\hat{d}_{n}\right).\label{sup-eq:Aibb''Ajb''b'}
\end{align}
Multiplying the first product by the planar spin components, $\alpha\in\{1,2\}$,
and taking the trace yields 
\begin{align}
\text{tr}\left(\Sigma_{\alpha}^{\phantom{i}}\mathsf{P}_{b}^{\phantom{j}}\mathsf{A}_{bb^{\prime}}^{i}\right) & =\frac{1}{8}\eta_{b^{\prime}}\left\{ \text{tr}\left(\gamma^{\alpha}\gamma^{3}\gamma^{4}\gamma^{m}\right)+\frac{1}{2}\eta_{b}\eta_{b^{\prime}}\text{tr}\left(\gamma^{\alpha}\gamma^{3}\gamma^{4}\gamma^{n}\left[\gamma^{\ell},\gamma^{m}\right]\right)\hat{d}_{n}\hat{d}_{\ell}\right\} \partial_{k_{i}}\hat{d}_{m}=0.
\end{align}
The purely geometric product becomes
\begin{equation}
\text{tr}\left(\Sigma_{\alpha}^{\phantom{i}}\mathsf{A}_{bb^{\prime\prime}}^{i}\mathsf{A}_{b^{\prime\prime}b^{\prime}}^{j}\right)=\frac{\text{i}}{32}\eta_{b^{\prime}}\eta_{b^{\prime\prime}}\left\{ \eta_{b}\text{tr}\left(\gamma^{\alpha}\gamma^{3}\gamma^{4}\left[\gamma^{\ell},\gamma^{m}\right]\gamma^{n}\right)+\eta_{b^{\prime\prime}}\text{tr}\left(\gamma^{\alpha}\gamma^{3}\gamma^{4}\gamma^{m}\left[\gamma^{\ell},\gamma^{n}\right]\right)\right\} \hat{d}_{\ell}\left(\partial_{k_{i}}\hat{d}_{m}\right)\left(\partial_{k_{j}}\hat{d}_{n}\right),
\end{equation}
 which is nonzero only if the indices $\alpha,3,4$ are matched only
once with $\ell,m,n$. Thus, $\ell\neq m\neq n$ and 
\begin{equation}
\text{tr}\left(\gamma^{\alpha}\gamma^{3}\gamma^{4}\left[\gamma^{\ell},\gamma^{m}\right]\gamma^{n}\right)=2\text{tr}\left(\gamma^{\alpha}\gamma^{3}\gamma^{4}\gamma^{\ell}\gamma^{m}\gamma^{n}\right)=-\text{tr}\left(\gamma^{\alpha}\gamma^{3}\gamma^{4}\gamma^{m}\left[\gamma^{\ell},\gamma^{n}\right]\right),
\end{equation}
 which shows that 
\begin{equation}
\text{tr}\left(\Sigma_{\alpha}^{\phantom{i}}\mathsf{A}_{bb^{\prime\prime}}^{i}\mathsf{A}_{b^{\prime\prime}b^{\prime}}^{j}\right)=\frac{\text{i}}{16}\eta_{b^{\prime}}\eta_{b^{\prime\prime}}\left(\eta_{b}-\eta_{b^{\prime\prime}}\right)\mathscr{T}_{6}^{\alpha34\ell mn}\hat{d}_{\ell}\left(\partial_{k_{i}}\hat{d}_{m}\right)\left(\partial_{k_{j}}\hat{d}_{n}\right),\quad\alpha\in\left\{ 1,2\right\} .
\end{equation}
Unlike the dipole current term, this quadrupolar contribution is nonzero
since there is not an additional contraction over a symmetric tensor
$\hat{d}_{n}\hat{d}_{\ell}$. The in-plane spin generated by the geometric
term is then 
\begin{align}
\text{tr}\left(\Sigma_{\alpha}^{\phantom{i}}\mathsf{A}_{bb^{\prime\prime}}^{i}\mathsf{A}_{b^{\prime\prime}b^{\prime}}^{j}\right) & =\frac{\text{i}}{2}\eta_{b^{\prime}}\eta_{b^{\prime\prime}}\left(\eta_{b}-\eta_{b^{\prime\prime}}\right)\nonumber \\
 & \phantom{=}\times\bigg\{\hat{d}_{\alpha}\left[\left(\partial_{k_{i}}\hat{d}_{4}\right)\left(\partial_{k_{j}}\hat{d}_{3}\right)-\left(\partial_{k_{i}}\hat{d}_{3}\right)\left(\partial_{k_{j}}\hat{d}_{4}\right)\right]+\hat{d}_{3}\left[\left(\partial_{k_{i}}\hat{d}_{\alpha}\right)\left(\partial_{k_{j}}\hat{d}_{4}\right)-\left(\partial_{k_{j}}\hat{d}_{\alpha}\right)\left(\partial_{k_{i}}\hat{d}_{4}\right)\right]\nonumber \\
 & \phantom{=\times\bigg\{}+\hat{d}_{4}\left[\left(\partial_{k_{i}}\hat{d}_{3}\right)\left(\partial_{k_{j}}\hat{d}_{\alpha}\right)-\left(\partial_{k_{i}}\hat{d}_{\alpha}\right)\left(\partial_{k_{j}}\hat{d}_{3}\right)\right]\bigg\}.\label{supp-eq:tr(SigmaAA)}
\end{align}
In centrosymmetric systems, each term is even under inversion, and
it will therefore generically survive the Brillouin zone integration. 

We are left to consider the out-of-plane response. Returning to Eq.~(\ref{supp-eq:PbAibb'}), multiplying by $\Sigma_{z}=-\text{i}\gamma^{1}\gamma^{2}$
and taking the trace yields 
\begin{align}
\text{tr}\left(\Sigma_{z}^{\phantom{i}}\mathsf{P}_{b}^{\phantom{j}}\mathsf{A}_{bb^{\prime}}^{i}\right) & =\hat{d}_{1}\partial_{k_{i}}\hat{d}_{2}-\hat{d}_{2}\partial_{k_{i}}\hat{d}_{1},\quad b\neq b^{\prime}.
\end{align}
Thus, the resonant inter-band out-of-plane spin resonance calculated
at first-order will then flow through the crystal at second-order
in the form of a electric dipole current. The contribution resulting
from the resonant inter-band electric quadrupole from Eq.~(\ref{sup-eq:Aibb''Ajb''b'})
is 
\begin{align}
\text{tr}\left(\Sigma_{z}^{\phantom{i}}\mathsf{A}_{bb^{\prime\prime}}^{i}\mathsf{A}_{b^{\prime\prime}b^{\prime}}^{j}\right) & =\frac{\text{i}\eta_{b^{\prime}}\eta_{b^{\prime\prime}}}{16}\left\{ \text{tr}\left(\gamma^{1}\gamma^{2}\gamma^{m}\gamma^{n}\right)+\frac{1}{4}\eta_{b}\eta_{b^{\prime\prime}}\text{tr}\left(\gamma^{1}\gamma^{2}\left[\gamma^{\ell},\gamma^{m}\right]\left[\gamma^{p},\gamma^{n}\right]\right)\hat{d}_{\ell}\hat{d}_{p}\right\} \left(\partial_{k_{i}}\hat{d}_{m}\right)\left(\partial_{k_{j}}\hat{d}_{n}\right)
\end{align}
This first term is nonzero for $m=1,\,n=2$ and vice versa. We must
evaluate the second trace. Since the commutators restrict $\ell\neq m$
and $p\neq n$ and $\gamma^{1}\gamma^{2}=-\gamma^{2}\gamma^{1}$,
the only surviving traces are over sets of distinct pairs $\left(1\neq2\right)$,
$\left(\ell\neq m\right)$, and $\left(p\neq n\right)$ such as
\begin{align}
\text{tr}\left(\gamma^{1}\gamma^{2}\gamma^{1}\gamma^{m}\gamma^{2}\gamma^{n}\right) & =4\delta^{1\ell}\delta^{2p}\delta^{mn}\left(1-\delta^{1m}\right)\left(1-\delta^{2n}\right),
\end{align}
 which will contribute a term to the response proportional to 
\begin{equation}
\hat{d}_{1}\hat{d}_{2}\left[\left(\partial_{k_{1}}\hat{d}_{3}\right)\left(\partial_{k_{j}}\hat{d}_{3}\right)+\left(\partial_{k_{1}}\hat{d}_{4}\right)\left(\partial_{k_{j}}\hat{d}_{4}\right)\right].
\end{equation}
Another nonzero term assumes the form 
\begin{align}
\text{tr}\left(\gamma^{1}\gamma^{2}\gamma^{\ell}\gamma^{1}\gamma^{p}\gamma^{2}\right) & =4\delta^{1m}\delta^{2n}\delta^{\ell p}\left(1-\delta^{1\ell}\right)\left(1-\delta^{2p}\right),
\end{align}
 which contributes response terms proportional to 
\begin{equation}
\left(\hat{d}_{3}^{2}+\hat{d}_{4}^{2}\right)\left(\partial_{k_{i}}\hat{d}_{1}\right)\left(\partial_{k_{j}}\hat{d}_{2}\right).
\end{equation}
 There are six other terms such as these contained within the full
trace. The essential conclusion is that each of these terms are both
nonvanishing \emph{and} even under inversion. As a result, the second-order
spin expectation value through bound electric quadrupolar fluctuations
is both nonzero at each momentum $\boldsymbol{k}$, \emph{and} survives
the Brillouin zone integration. Thus, 
\begin{equation}
\left\langle \boldsymbol{S}^{\left(2\right)}\left(\omega\right)\right\rangle =\sum_{\boldsymbol{k}}\left\langle \boldsymbol{S}^{\left(2\right)}\left(\omega;\,\boldsymbol{k}\right)\right\rangle \neq\boldsymbol{0},
\end{equation}
 giving rise to the nonlinear Edelstein effect in globally centrosymmetric,
but locally non-centrosymmetric, crystals such as \CGT.

\subsection{Symmetry Analysis of the Second-Order Nonlinear Edelstein Effect}

Given that the second-order nonlinear Edelstein effect is generically
nonzero, it remains now is to characterize the symmetry-allowed components
of $\left\langle \boldsymbol{S}^{\left(2\right)}\left(\omega\right)\right\rangle $
within \CGT. We write the magneto-electric susceptibility as 
\begin{align}
\left\langle S_{\alpha}^{\left(2\right)}\left(\omega\right)\right\rangle  & \equiv\chi_{\alpha,ij}^{\left(2\right)}\left(\omega;\,\omega_{1},\omega_{2}\right)E_{i}\left(\omega_{1}\right)E_{j}\left(\omega_{2}\right),\label{supp-eq:S2_magnetoelectric_tensor}
\end{align}
 where the differential susceptibility itself is an integral over
the Brillouin zone
\begin{align}
\chi_{\alpha,ij}^{\left(2\right)}\left(\omega;\,\omega_{1},\omega_{2}\right) & =\sum_{\boldsymbol{k}}\chi_{\alpha,ij}^{\left(2\right)}\left(\omega,\boldsymbol{k};\,\omega_{1},\omega_{2}\right),\nonumber \\
\chi_{\alpha,ij}^{\left(2\right)}\left(\omega,\boldsymbol{k};\,\omega_{1},\omega_{2}\right) & =\frac{1}{2}\delta\left(\omega-\left(\omega_{1}+\omega_{2}\right)\right)e^{2}\Delta n_{vc}\left(\boldsymbol{k}\right)\nonumber \\
 & \phantom{=}\times\sum_{b,b^{\prime}=v,c}\bigg\{\text{tr}\left[\Sigma_{\alpha}\mathsf{P}_{b}\left(\boldsymbol{k}\right)\left[v^{i}\left(\boldsymbol{k}\right),\mathsf{A}^{j}\left(\boldsymbol{k}\right)\right]\mathsf{P}_{b^{\prime}}\left(\boldsymbol{k}\right)\right]\frac{\delta\left(\omega_{2}-\Delta\varepsilon_{cv}\left(\boldsymbol{k}\right)\right)}{\left(\omega_{1}-\text{i}\delta^{+}\right)\left(\omega_{1}+\omega_{2}-\Delta\varepsilon_{bb^{\prime}}+\text{i}\delta^{+}\right)}\nonumber \\
 & \phantom{\phantom{=}\times\sum_{b,b^{\prime}=v,c}\bigg\{}+\text{tr}\left[\Sigma_{\alpha}\mathsf{P}_{b}\left(\boldsymbol{k}\right)\left[v^{j}\left(\boldsymbol{k}\right),\mathsf{A}^{i}\left(\boldsymbol{k}\right)\right]\mathsf{P}_{b^{\prime}}\left(\boldsymbol{k}\right)\right]\frac{\delta\left(\omega_{1}-\Delta\varepsilon_{cv}\left(\boldsymbol{k}\right)\right)}{\left(\omega_{2}-\text{i}\delta^{+}\right)\left(\omega_{1}+\omega_{2}-\Delta\varepsilon_{bb^{\prime}}+\text{i}\delta^{+}\right)}\bigg\}.\label{supp-eq:dynamical_nonlinear_edelstein_tensor}
\end{align}
Given that the traces survive for all spin components $\alpha\in\left\{ x,y,z\right\} $
at each Bloch momentum $\boldsymbol{k}$, we now determine exactly which components of $\chi_{\alpha,ij}^{\left(2\right)}$
survive the Brillouin zone integration by symmetry. Given that the
dyad is inversion-even, it can be decomposed into the same irreducible
representations as the magnetic degrees of freedom such as the spin
and magnetization. Table~\ref{sup-tab:Irreducible-representations-of-magneto-electric}
shows the specific symmetry reduction for the spin $\boldsymbol{S}$,
magnetization $\boldsymbol{M}$, and electric field dyad $\boldsymbol{E}\otimes\boldsymbol{E}$
within point group $\overline{3}$ of \CGT. Suppressing the frequency-dependence
for now, the possible symmetry-allowed magneto-electric coefficients
therefore assume the following form 
\begin{align}
\left\langle S_{z}\right\rangle  & =\chi_{z,zz}^{\left(2\right)}E_{z}^{2}+\chi_{z,xx}^{\left(2\right)}\left(E_{x}^{2}+E_{y}^{2}\right),\label{supp-eq:Sz_by_symmetry}\\
\left\langle S_{\pm}\right\rangle  & =2\chi_{\pm,\pm z}^{\left(2\right)}\left(E_{x}\pm\text{i}E_{y}\right)E_{z}+\chi_{\pm,xx}^{\left(2\right)}\left[E_{x}^{2}-E_{y}^{2}\pm\text{i}\left(2E_{x}E_{y}\right)\right],
\end{align}
 where $\left\langle S_{\pm}\right\rangle \equiv\left\langle S_{x}\right\rangle \pm\text{i}\left\langle S_{y}\right\rangle $.
In Cartesian coordinates -- where the crystallographic $\hat{a}$-axis
is taken to coincide with the $\hat{x}$-axis -- then the resulting
\begin{align}
\left\langle S_{x}\right\rangle  & =\chi_{x,xx}^{\left(2\right)}\left(E_{x}^{2}-E_{y}^{2}\right)+\chi_{x,xy}^{\left(2\right)}\left(2E_{x}E_{y}\right)+\chi_{x,xz}^{\left(2\right)}\left(2E_{x}E_{z}\right)+\chi_{x,yz}^{\left(2\right)}\left(2E_{y}E_{z}\right),\label{supp-eq:Sx_by_symmetry}\\
\left\langle S_{y}\right\rangle  & =\chi_{x,xy}^{\left(2\right)}\left(E_{x}^{2}-E_{y}^{2}\right)-\chi_{x,xx}^{\left(2\right)}\left(2E_{x}E_{y}\right)-\chi_{x,yz}^{\left(2\right)}\left(2E_{x}E_{z}\right)+\chi_{x,xz}^{\left(2\right)}\left(2E_{y}E_{z}\right),\label{supp-eq:Sy_by_symmetry}
\end{align}
 Taken together, there are six distinct second-order magneto-electric
susceptibilities within $\overline{3}$ which can arise from the nonlinear
Edelstein effect \citep{xuLightinducedStaticMagnetization2021,xiaoIntrinsicNonlinearElectric2022,xiaoTimeReversalEvenNonlinearCurrent2023}.
In summary, the response tensor can be written as a matrix in the
following way 
\begin{equation}
\begin{pmatrix}\left\langle S_{x}\right\rangle \\
\left\langle S_{y}\right\rangle \\
\left\langle S_{z}\right\rangle 
\end{pmatrix}=\begin{pmatrix}\chi_{x,xx}^{\left(2\right)} & -\chi_{x,xx}^{\left(2\right)} & 0 & \chi_{x,yz}^{\left(2\right)} & \chi_{x,xz}^{\left(2\right)} & \chi_{x,xy}^{\left(2\right)}\\
\chi_{x,xy}^{\left(2\right)} & -\chi_{x,xy}^{\left(2\right)} & 0 & \chi_{x,xz}^{\left(2\right)} & -\chi_{x,yz}^{\left(2\right)} & -\chi_{x,xx}^{\left(2\right)}\\
\chi_{z,xx}^{\left(2\right)} & \chi_{z,xx}^{\left(2\right)} & \chi_{z,zz}^{\left(2\right)} & 0 & 0 & 0
\end{pmatrix}\begin{pmatrix}E_{x}^{2}\\
E_{y}^{2}\\
E_{z}^{2}\\
2E_{y}E_{z}\\
2E_{z}E_{x}\\
2E_{x}E_{y}
\end{pmatrix}.
\end{equation}
 Focusing on the important limit of normal incidence with electric
field polarization given by $\boldsymbol{E}=E_{0}\left(\cos\varphi,\sin\varphi,0\right)$,
then we find that 
\begin{align}
\left\langle S_{x}\right\rangle _{\parallel} & \equiv E_{0}^{2}\left[\chi_{x,xx}^{\left(2\right)}\cos\left(2\varphi\right)+\chi_{x,xy}^{\left(2\right)}\sin\left(2\varphi\right)\right]=E_{0}^{2}\sqrt{ \left( \chi_{x,xy}^{\left(2\right)} \right)^2 + \left( \chi_{x,xx}^{\left(2\right)} \right)^2 }\sin\left(2\varphi-2\delta\right),\\
\left\langle S_{y}\right\rangle _{\parallel} & \equiv E_{0}^{2}\left[\chi_{x,xy}^{\left(2\right)}\cos\left(2\varphi\right)-\chi_{x,xx}^{\left(2\right)}\sin\left(2\varphi\right)\right]=E_{0}^{2}\sqrt{ \left( \chi_{x,xy}^{\left(2\right)} \right)^2 + \left( \chi_{x,xx}^{\left(2\right)} \right)^2 }\cos\left(2\varphi-2\delta\right),\\
\left\langle S_{z}\right\rangle _{\parallel} & =\chi_{z,xx}^{\left(2\right)}E_{0}^{2},
\end{align}
 where the angle $\delta$ is material-dependent and defined by 
\begin{equation}
\tan\left(2\delta\right)\equiv-\frac{\chi_{x,xx}^{\left(2\right)}}{\chi_{x,xy}^{\left(2\right)}}.
\end{equation}
 Whereas $\left\langle S_{z}\right\rangle _{\parallel}$ is isotropic
with respect to polarization angle at normal incidence, the $\left\langle S_{x,y}\right\rangle _{\parallel}$
components are anisotropic with a $d$-wave rotational symmetry. Consequently,
the magnetic response changes sign upon rotation of the electric field
polarization by $90^{\text{o}}$: $\varphi\rightarrow\varphi+\pi/2$. Unlike
point groups with mirrors, such as $\overline{3}m$ studied in \citep{xiaoIntrinsicNonlinearElectric2022,xiaoTimeReversalEvenNonlinearCurrent2023,xueNonlinearOpticsDrivenSpin2024,oikeImpactElectronCorrelations2024},
the $d$-wave petal structure is not pinned to the crystallographic
axes in a universal manner. Instead, the petals are aligned along
a non-universal, material-dependent axis determined by the angle $\delta$. 

\begin{table}
\caption{Symmetry reduction of the spin $\boldsymbol{S}$, magnetization $\boldsymbol{M}$,
and electric field dyad $\boldsymbol{E}\otimes\boldsymbol{E}$ representations
$\Gamma$ within the point group $\overline{3}$ for a system with
spin-orbit coupling \citep{dresselhausGroupTheoryApplication2008}.
All three representations are even under inversion and transform within
the $A_{g}$ singlet and $E_{g}$ doublet irreducible representations.
}\label{sup-tab:Irreducible-representations-of-magneto-electric}

\bigskip{}

\centering{}%
\begin{tabular}{ccc}
\toprule 
$\Gamma$$\left(\overline{3}\right)$ & $A_{g}$ & $E_{g}^{\pm}$\tabularnewline
\midrule
\midrule 
$\boldsymbol{S}$ & $S_{z}$ & $S_{x}\pm \text{i}S_{y}$\tabularnewline
\midrule 
$\boldsymbol{M}$ & $M_{z}$ & $M_{x}\pm \text{i}M_{y}$\tabularnewline
\midrule 
$\boldsymbol{E}\otimes\boldsymbol{E}$ & $E_{z}^{2}$, $E_{x}^{2}+E_{y}^{2}$ & $\left(E_{x}\pm\text{i}E_{y}\right)E_{z}$, $E_{x}^{2}-E_{y}^{2}\pm\text{i}\left(2E_{x}E_{y}\right)$\tabularnewline
\bottomrule
\end{tabular}
\end{table}

\section{Interaction of the Edelstein-Zeeman Field with Local Ferromagnetism}

The ferromagnetic behavior of \CGT is well-captured by a classical
Heisenberg spins localized on the \ce{Cr^{3+}} atoms with weak uniaxial
anisotropy \citep{gongDiscoveryIntrinsicFerromagnetism2017,zeisnerMagneticAnisotropySpinpolarized2019,fangLargeMagnetoopticalEffects2018,khelaLaserinducedTopologicalSpin2023}.
The low-energy effective Landau functional respecting the high-temperature
symmetries of its magnetic point group, $\overline{3}'$, is given
by
\begin{align}
\mathcal{F}_{0}\left(\boldsymbol{M}\right) & =\frac{1}{2}r_{0}\left(M_{x}^{2}+M_{y}^{2}\right)+\frac{1}{2}r_{z}M_{z}^{2}+\frac{1}{4}u_{z}M_{z}^{4}+\frac{1}{2}gM_{z}^{2}\left(M_{x}^{2}+M_{y}^{2}\right),\label{supp-eq:landau_free_energy}
\end{align}
 where $r_{0}$, $r_{z}$, $u_{z}$, and $g$ are all Landau phenomenological
constants. Here $M_{z}$ is the out-of-plane magnetic component that
orders when $r_{z}\propto T-T_{c}^{0}=0$. Given the instability is
uniaxial, $r_{0}>r_{z}$, and the free energy is stable with respect
to quadratic order in the planar magnetic components, $M_{x}$ and
$M_{y}$. The coefficient $u_{z}>0$ stabilizes the free energy, and
the coefficient $g>0$ controls the competition between $M_{z}$ and
the planar projection $M_{\parallel}=\sqrt{M_{x}^{2}+M_{y}^{2}}$.

\subsection{Mean-Field Theory of Heisenberg Ferromagnetism with Uniaxial Anisotropy}

The external Zeeman field, $\boldsymbol{H}_{\text{ext}}$, is introduced
thermodynamically as the conjugate field for the magnetization
\begin{equation}
\mathcal{F}\left(\boldsymbol{M},\boldsymbol{H}\right)=\mathcal{F}_{0}\left(\boldsymbol{M}\right)-\boldsymbol{H}_{\text{ext}}\cdot\boldsymbol{M}.
\end{equation}
While Eq.~(\ref{supp-eq:landau_free_energy}) captures the low-field,
low-magnetization physics of the Heisenberg model, it fails to capture
effects associated with magnetic saturation. To this end, we employ
the following classical spin Hamiltonian:
\begin{equation}
\mathcal{H}_{\text{eff}}\equiv-J\sum_{\left\langle ij\right\rangle }\boldsymbol{\sigma}_{i}\cdot\boldsymbol{\sigma}_{j}-\frac{1}{2}K\sum_{i}\left(\sigma_{i}^{z}\right)^{2}-\boldsymbol{H}_{\text{ext}}\cdot\sum_{i}\boldsymbol{\sigma}_{i},
\end{equation}
where $J>0$ is the effective ferromagnetic exchange, notation $\left\langle ij\right\rangle $
represents nearest-neighboring \ce{Cr^{3+}} sites $i$ and $j$,
$\boldsymbol{\sigma}_{i}$ is the local \ce{Cr^{3+}} classical Heisenberg
moment, and $K>0$ is the effective uniaxial anisotropy energy. To
capture the essential physics of optical control over a ferromagnetic
order parameter, we employ a mean-field approach. 

The mean-field Hamiltonian is
\begin{equation}
\mathcal{H}_{\text{MF}}\left(\boldsymbol{M},\boldsymbol{H}_{\text{ext}}\right)=\frac{1}{2}Jz\left(\boldsymbol{M}\cdot\boldsymbol{M}\right)+\frac{1}{2}KM_{z}^{2}-\boldsymbol{H}_{\text{eff}}\left(\boldsymbol{M},\boldsymbol{H}_{\text{ext}}\right)\cdot\boldsymbol{\sigma},
\end{equation}
where the Heisenberg spin has unit norm $\boldsymbol{\sigma}\cdot\boldsymbol{\sigma}\equiv1$
and the effective Zeeman field is given by
\begin{align}
\boldsymbol{H}_{\text{eff}}\left(\boldsymbol{M},\boldsymbol{H}_{\text{ext}}\right) & =Jz\left(\boldsymbol{M}+\kappa M_{z}\hat{z}\right)+\boldsymbol{H}_{\text{ext}},\\
\boldsymbol{M}\left(T,\boldsymbol{H}_{\text{ext}}\right) & \equiv\left\langle \boldsymbol{\sigma}\right\rangle =\frac{\text{e}^{\frac{1}{2}\beta Jz\left[\left(\boldsymbol{M}\cdot\boldsymbol{M}\right)+\kappa M_{z}^{2}\right]}}{\mathcal{Z}_{\text{MF}}\left(\boldsymbol{M},\boldsymbol{H}_{\text{ext}};\beta\right)}\int_{\boldsymbol{\sigma}\cdot\boldsymbol{\sigma}=1}\text{d}\boldsymbol{\sigma}\,\exp\left[\beta\boldsymbol{H}_{\text{eff}}\left(\boldsymbol{M},\boldsymbol{H}_{\text{ext}}\right)\cdot\boldsymbol{\sigma}\right]\,\boldsymbol{\sigma}.
\end{align}
 In the above, the lattice coordination number is $z$, the quantity
$\kappa\equiv K/Jz>0$ is the uniaxial anisotropy and
the absolute temperature is given as $T=1/\beta$. The mean-field
partition function, $\mathcal{Z}_{\text{MF}}\left(\beta\right)$,
for these self consistency equations is given as the average of the
spin \textbf{$\boldsymbol{\sigma}$ }over the unit sphere:
\begin{align}
\mathcal{Z}_{\text{MF}}\left(\boldsymbol{M},\boldsymbol{H}_{\text{ext}};\beta\right) & =\frac{\text{e}^{-\frac{1}{2}\beta Jz\left[\left(\boldsymbol{M}\cdot\boldsymbol{M}\right)+\kappa M_{z}^{2}\right]}}{4\pi}\int_{\boldsymbol{\sigma}\cdot\boldsymbol{\sigma}=1}\text{d}\boldsymbol{\sigma}\,\exp\left[\beta\boldsymbol{H}_{\text{eff}}\left(\boldsymbol{M},\boldsymbol{H}_{\text{ext}}\right)\cdot\boldsymbol{\sigma}\right] \\
 & =\frac{\text{e}^{-\frac{1}{2}\beta Jz\left[\left(\boldsymbol{M}\cdot\boldsymbol{M}\right)+\kappa M_{z}^{2}\right]}\sinh\left[\beta\left|\boldsymbol{H}_{\text{eff}}\left(\boldsymbol{M},\boldsymbol{H}_{\text{ext}}\right)\right|\right]}{2\pi\beta\left|\boldsymbol{H}_{\text{eff}}\left(\boldsymbol{M},\boldsymbol{H}_{\text{ext}}\right)\right|}.
\end{align}
Each component of the mean-field magnetization then satisfies the
self-consistent equation of state
\begin{equation}
\boldsymbol{M}\left(T,\boldsymbol{H}_{\text{ext}}\right)=\frac{\partial\log\mathcal{Z}_{\text{MF}}\left(\boldsymbol{M},\boldsymbol{H}_{\text{ext}};\beta\right)}{\partial\beta\boldsymbol{H}_{\text{eff}}}=\begin{cases}
\boldsymbol{0}, & \left|\boldsymbol{H}_{\text{eff}}\right|=0\\
\boldsymbol{\hat{H}}_{\text{eff}}\left(\boldsymbol{M},\boldsymbol{H}_{\text{ext}}\right)\,\mathcal{L}\left(\beta\left|\boldsymbol{H}_{\text{eff}}\left(\boldsymbol{M},\boldsymbol{H}_{\text{ext}}\right)\right|\right), & \left|\boldsymbol{H}_{\text{eff}}\right|>0
\end{cases}\label{supp-eq:mean-field-magnetization}
\end{equation}
where the unit vector $\boldsymbol{\hat{H}}_{\text{eff}}$ is given
by
\begin{equation}
\boldsymbol{\hat{H}}_{\text{eff}}\left(\boldsymbol{M},\boldsymbol{H}_{\text{ext}}\right)\equiv\frac{\boldsymbol{H}_{\text{eff}}\left(\boldsymbol{M},\boldsymbol{H}_{\text{ext}}\right)}{\left|\boldsymbol{H}_{\text{eff}}\left(\boldsymbol{M},\boldsymbol{H}_{\text{ext}}\right)\right|},
\end{equation}
 and $\mathcal{L}\left(x\right)$ is the Langevin function defined by
\begin{equation}
\mathcal{L}\left(x\right)\equiv\coth\left(x\right)-\frac{1}{x}.
\end{equation}
The Langevin function contains the appropriate symmetry-allowed nonlinear
effects to bound the magnitude of the magnetization on the interval
$\left|\boldsymbol{M}\right|\in\left[0,1\right]$.

In zero field, Eq.~(\ref{supp-eq:mean-field-magnetization}) determines
the transition temperature since the magnetization satisfies 
\begin{align}
\boldsymbol{M}\left(T,\boldsymbol{0}\right) & =\frac{Jz}{3T}\left[\left(1+\kappa\right)M_{z}\hat{z}+M_{x}\hat{x}+M_{y}\hat{y}\right]+\mathcal{O}\left(\beta^{3}\left|\boldsymbol{H}_{\text{eff}}\right|^{3}\right),
\end{align}
since the expansion of the Langevin function is $\mathcal{L}(x)\approx\frac{1}{3}x-\frac{1}{45}x^{3}+\dots$. The system therefore spontaneously magnetizes when cooled to the zero-field
critical temperature of 
\begin{equation}
T_{c}^{0}\equiv\frac{1}{3}Jz\left(1+\kappa\right),
\end{equation}
at which point the magnetic point group symmetry is lowered from $\overline{3}'$
to $\overline{3}$ when $\left\langle M_{z}\left(T<T_{c}^{0}\right)\right\rangle \neq0$.
We use $T_{c}^{0}$ to define the energy scale in what follows.

The mean-field free energy is obtained from the partition function
as 
\begin{align}
\mathcal{F}_{\text{MF}}\left(\boldsymbol{M},T\right) & =-\frac{1}{\beta}\log\mathcal{Z}_{\text{MF}}\left(\boldsymbol{M}\right)=\frac{1}{2}Jz\left[\left(\boldsymbol{M}\cdot\boldsymbol{M}\right)+\kappa M_{z}^{2}\right]-\frac{1}{\beta}\log\left\{ \frac{\sinh\left[\beta\left|\boldsymbol{H}_{\text{eff}}\left(\boldsymbol{M},\boldsymbol{H}_{\text{ext}}\right)\right|\right]}{2\pi\beta\left|\boldsymbol{H}_{\text{eff}}\left(\boldsymbol{M},\boldsymbol{H}_{\text{ext}}\right)\right|}\right\} ,
\end{align}
 which expands for $T\approx T_{c}^{0}$ in zero field as 

\begin{align}
\frac{\mathcal{F}_{\text{MF}}\left(\boldsymbol{M},T\right)}{T_{c}^{0}} & =\frac{1}{3}\left(1+\kappa\right)\left(\frac{T}{T_{c}^{0}}\right)\log\left(2\pi\right)+\frac{1}{2}\left[1-\frac{1}{1+\kappa}\left(\frac{T_{c}^{0}}{T}\right)\right]\left(M_{x}^{2}+M_{y}^{2}\right)+\frac{1+\kappa}{2}\left(1-\frac{T_{c}^{0}}{T}\right)M_{z}^{2}\nonumber \\
 & \phantom{\approx\;}+\frac{3\left(1+\kappa\right)}{20}\left(\frac{T_{c}^{0}}{T}\right)^{3}M_{z}^{4}+\frac{3}{10\left(1+\kappa\right)}\left(\frac{T_{c}^{0}}{T}\right)^{3}M_{z}^{2}\left(M_{x}^{2}+M_{y}^{2}\right)+\dots,\label{supp-eq:mean-field_heisenberg_free_energy_Tc0}
\end{align}
 from which Landau phenomenological coefficients in Eq.~(\ref{supp-eq:landau_free_energy})
are obtained as
\begin{equation}
\begin{aligned}r_{z} & =T_{c}^{0}\left(1+\kappa\right)\left(1-\frac{T_{c}^{0}}{T}\right), & r_{0} & =T_{c}^{0}\left[1-\frac{1}{1+\kappa}\left(\frac{T_{c}^{0}}{T}\right)\right],\\
u_{z} & =\frac{3}{5}\left(1+\kappa\right)\left(\frac{T_{c}^{0}}{T}\right)^{3}, & g & =\frac{3}{5\left(1+\kappa\right)}\left(\frac{T_{c}^{0}}{T}\right)^{3}.
\end{aligned}
\end{equation}

Time-reversal symmetry breaking gives rise to a magnetization in this
model, whether it be spontaneously broken below $T_{c}^{0}$, or if time-reversal
symmetry is externally broken by $\boldsymbol{H}_{\text{ext}}\neq\boldsymbol{0}$.
The magnetic response can be written in terms of a susceptibility
$\chi_{ij}^{\left(\text{M}\right)}$ that generically depends on the
magnetic state. From Eq.~(\ref{supp-eq:mean-field-magnetization}),
we have that
\begin{equation}
\chi_{ij}^{\left(\text{M}\right)}\left(\boldsymbol{M},\boldsymbol{H}_{\text{ext}}\right)\equiv\frac{\partial M_{i}}{\partial H_{\text{ext},j}}=\frac{\partial}{\partial H_{\text{ext},j}}\left[\hat{H}_{\text{eff},i}\left(\boldsymbol{M},\boldsymbol{H}_{\text{ext}}\right)\,\mathcal{L}\left(\beta\left|\boldsymbol{H}_{\text{eff}}\left(\boldsymbol{M},\boldsymbol{H}_{\text{ext}}\right)\right|\right)\right],\label{supp-eq:generic_mag_susceptibility}
\end{equation}
 which generates a closed set of linear equations for the susceptibility
tensor upon implicit differentiation:
\begin{equation}
\frac{\partial}{\partial H_{\text{ext},j}}H_{\text{eff},i}\left(\boldsymbol{M},\boldsymbol{H}_{\text{ext}}\right)=\delta_{ij}+Jz\left(1+\kappa\delta_{i3}\right)\chi_{ij}^{\left(\text{M}\right)}\left(\boldsymbol{M},\boldsymbol{H}_{\text{ext}}\right).
\end{equation}
In the above two equations, we have explicitly written out the magnetic
state dependence through the arguments $\left(\boldsymbol{M},\boldsymbol{H}_{\text{ext}}\right)$.

We simplify the analytics by now considering two limiting cases where these expressions are tractable:
(i) the linear regime at temperatures $T\approx T_{c}^{0}$ and (ii)
the high-field limit when $\left|\boldsymbol{H}_{\text{ext}}\right|\gg Jz\left|\boldsymbol{M}\right|$.
Starting with the linear regime with $T>T_{c}^{0}$, we expand Eq.~(\ref{supp-eq:mean-field-magnetization}) in an infinitesimal field
as
\begin{equation}
\boldsymbol{M}=\frac{Jz}{3T}\left[\left(1+\kappa\right)M_{z}\hat{z}+M_{x}\hat{x}+M_{y}\hat{y}\right]+\frac{1}{3T}\boldsymbol{H}_{\text{ext}}.
\end{equation}
In this regime, we obtain the following longitudinal linear susceptibilities
\begin{equation}
\begin{aligned}\chi_{zz}^{\left(\text{M}\right)} & =\frac{1}{3\left(T-T_{c}^{0}\right)}, & \chi_{\parallel}^{\left(\text{M}\right)} & =\frac{1}{3\left(T-\frac{T_{c}^{0}}{1+\kappa}\right)},\end{aligned}
\end{equation}
where $\chi_{\parallel}^{\left(\text{M}\right)}\equiv\chi_{xx}^{\left(\text{M}\right)}=\chi_{yy}^{\left(\text{M}\right)}$.
Thus, the in-plane susceptibility increases to its maximum \emph{finite}
value when cooled to the phase transition. Below the Curie temperature,
the in-plane longitudinal susceptibility is temperature-independent.
To obtain this result, it is easiest to include an infinitesimal external
field in Eq.~(\ref{supp-eq:mean-field_heisenberg_free_energy_Tc0}),
and obtain following three equations of state:
\begin{equation}
\begin{aligned}H_{\text{ext},x} & =\left(r_{0}+gM_{z}^{2}\right)M_{x},\\
H_{\text{ext},y} & =\left(r_{0}+gM_{z}^{2}\right)M_{y},\\
H_{\text{ext},z} & =\left[r_{z}+g\left(M_{x}^{2}+M_{y}^{2}\right)\right]M_{z}+u_{z}M_{z}^{3}.
\end{aligned}
\end{equation}
Since $M_{x}=M_{y}=0$ for $\boldsymbol{H}_{\text{ext}}=\boldsymbol{0}$,
then 
\begin{equation}
M_{z}=\begin{cases}
0, & r_{z}>0\\
\sqrt{\frac{-r_{z}}{u_{z}}}, & r_{z}<0
\end{cases}.
\end{equation}
The longitudinal linear magnetic susceptibilities within the plane
is $\chi_{\parallel}^{\left(\text{M}\right)}=\chi_{xx}^{\left(\text{M}\right)}=\chi_{yy}^{\left(\text{M}\right)}$
and follows as
\begin{align}
\chi_{\parallel}^{\left(\text{M}\right)}\left(T\right) & =\begin{cases}
\frac{1}{T_{c}^{0}\left[1-\frac{1}{1+\kappa}\left(\frac{T_{c}^{0}}{T}\right)\right]}, & T>T_{c}^{0}\\
\frac{1}{T_{c}^{0}}\left(\frac{1}{\kappa}+1\right), & T<T_{c}^{0}
\end{cases}.
\end{align}
 Thus, the in-plane susceptibility is temperature-independent below
the transition, reflecting the large in-plane magnetic polarizability
below the Curie temperature.

We now consider the second case of large external fields, we appeal
to Eq.~(\ref{supp-eq:mean-field_heisenberg_free_energy_Tc0}) to find
the susceptibility. When $\left|\boldsymbol{H}_{\text{ext}}\right|\gg Jz\left|\boldsymbol{M}\right|$,
then 
\begin{equation}
\boldsymbol{M}\approx\boldsymbol{\hat{H}}_{\text{ext}}\,\mathcal{L}\left(\beta\left|\boldsymbol{H}_{\text{ext}}\right|\right).
\end{equation}
 The magnetic susceptibility tensor near saturation is therefore
\begin{equation}
\chi_{ij}^{\left(\text{M}\right)}\left(T,\boldsymbol{H}_{\text{ext}}\right)=\left(\delta_{ij}-\hat{H}_{\text{ext},i}\hat{H}_{\text{ext},j}\right)\left[ \frac{\mathcal{L}\left(\beta\left|\boldsymbol{H}_{\text{ext}}\right|\right)}{\left|\boldsymbol{H}_{\text{ext}}\right|}\right]+\hat{H}_{\text{ext},i}\hat{H}_{\text{ext},j}\,\beta\mathcal{K}\left(\beta\left|\boldsymbol{H}_{\text{ext}}\right|\right),\label{supp-eq:high-field_susceptibility}
\end{equation}
 where the derivative of the Langevin function is given by
\begin{equation}
\mathcal{K}\left(x\right)\equiv\frac{\text{d}\mathcal{L}\left(x\right)}{\text{d}x}=\frac{1}{x^{2}}\left[1-x^{2}\text{csch}^{2}\left(x\right)\right].
\end{equation}
The first term in Eq.~\eqref{supp-eq:high-field_susceptibility} represents the purely transverse response. The longitudinal
susceptibilities are contained by the second term. Since 
\begin{equation}
\beta\mathcal{K}\left(\beta\left|\boldsymbol{H}_{\text{ext}}\right|\right)=\frac{T}{\left|\boldsymbol{H}_{\text{ext}}\right|^2}\left[1-\left(\beta\left|\boldsymbol{H}_{\text{ext}}\right|\right)^{2}\text{csch}^{2}\left(\beta\left|\boldsymbol{H}_{\text{ext}}\right|\right)\right],
\end{equation}
the longitudinal susceptibilities are seen to be \emph{increasing} functions of temperature, $T$, below a threshold energy scale set
by the condition that $\beta\left|\boldsymbol{H}_{\text{ext}}\right|\sim\mathcal{O}\left(1\right)$.
This temperature-dependence at high fields is a feature of the saturation
behavior of the Heisenberg magnet.

\subsection{The Edelstein-Zeeman Field}

As discussed in the Main Text, the resonant coherent interband Edelstein-Zeeman
field generated by ultrafast itinerant photocarriers will couple locally
to the \ce{Cr^{3+}} spin moments. The minimal coupling Hamiltonian
between the itinerant and localized electronic sectors appears in
the form of an $s$-$d$ model as
\begin{align}
\Delta\mathcal{H} & =-J_{sd}\sum_{i}\sum_{\alpha\beta}\hat{c}_{i,\alpha}^{\dagger}\left(\boldsymbol{S}_{\alpha\beta}\cdot\boldsymbol{\sigma}_{i}\right)\hat{c}_{i,\beta}^{\phantom{\dagger}},
\end{align}
where $J_{sd}$ is local exchange integral, $\hat{c}_{i,\alpha}^{\dagger}$
is projected electronic creation operator at \ce{Cr^{3+}} site $i$
with spin $\alpha$, and $\boldsymbol{S}_{\alpha\beta}$ is the single-particle
spin operator for the itinerant states. The itinerant spin density
in this case is projected only onto the valence and conduction bands,
separated by roughly 2-3~eV from the localized \ce{Cr^{3+}} band
\citep{fangLargeMagnetoopticalEffects2018}. In momentum space on
a periodic lattice of $N$ sites, we find that 
\begin{align}
\Delta\mathcal{H} & =-\frac{J_{sd}}{\sqrt{N}}\sum_{\boldsymbol{k},\boldsymbol{q}}\sum_{\alpha\beta}\hat{c}_{\boldsymbol{k},\alpha}^{\dagger}\left(\boldsymbol{S}_{\alpha\beta}\cdot\boldsymbol{\sigma}_{\boldsymbol{q}}\right)\hat{c}_{\boldsymbol{k}-\boldsymbol{q},\beta}^{\phantom{\dagger}}.
\end{align}
When the local spin is expanded around its spatial average magnetization$\boldsymbol{M}\equiv\frac{1}{N}\sum_{i}\left\langle \boldsymbol{\sigma}_{i}\right\rangle $,
the local spin density is sharply peaked at the zone center:
\begin{equation}
\left\langle \boldsymbol{\sigma}_{\boldsymbol{q}}\right\rangle \approx\boldsymbol{M}\sqrt{N}\,\delta_{\boldsymbol{q},\boldsymbol{0}},
\end{equation}
leading to a coupling of the form
\begin{equation}
\Delta\mathcal{H}\approx-J_{sd}\sum_{\boldsymbol{k}}\sum_{\alpha\beta}\hat{c}_{\boldsymbol{k},\alpha}^{\dagger}\left(\boldsymbol{S}_{\alpha\beta}\cdot\boldsymbol{M}\right)\hat{c}_{\boldsymbol{k},\beta}^{\phantom{\dagger}}=-J_{sd}\sum_{\boldsymbol{k}}\boldsymbol{S}_{\boldsymbol{k}}\cdot\boldsymbol{M}.
\end{equation}
In the second step, we then transitioned from the second-quantized
form back to the single-particle operator. This coupling defines the
mean-field dynamical Edelstein-Zeeman field, $\boldsymbol{H}^{\text{EZ}}$,
as 
\begin{align}
\boldsymbol{H}^{\text{EZ}} & \equiv J_{sd}\sum_{\boldsymbol{k}}\left\langle \boldsymbol{S}_{\boldsymbol{k}}\right\rangle .
\end{align}
If time-reversal symmetry is preserved, then $\boldsymbol{H}^{\text{EZ}}=\boldsymbol{0}$.
However, when the itinerant spins are driven by the radiation, $\boldsymbol{H}^{\text{EZ}}\neq\boldsymbol{0}$
through the nonlinear Edelstein effect discussed in Section \ref{supp-sec:Minimal-NEE-model}.
After including a nonzero dc field, $\boldsymbol{H}_{\text{dc}}$,
the total external Zeeman field at angular frequency $\omega$ is
\begin{equation}
H_{\alpha}\left(\omega\right)=J_{sd}\chi_{\alpha,ij}^{\left(2\right)}\left(\omega;\,\omega_{1},\omega_{2}\right)E_{i}\left(\omega_{1}\right)E_{j}\left(\omega_{2}\right)+2\pi H_{\text{dc},\alpha}\delta\left(\omega\right).
\end{equation}
Combining results and summarizing, it is clear that magneto-electric
control is achieved through the resonant Edelstein-Zeeman field coupling
to the localized \ce{Cr^{3+}} moments. The Edelstein response is
then convolved with the magnetic response of the \ce{Cr^{3+}} moments
which is enhanced by both the near-resonant photo-excitation and the
sub-valent soft ferromagnetism. The resulting nonlinear magneto-electric
susceptibility is 
\begin{equation}
\alpha_{i,jk}^{\left(2\right)}\left(\boldsymbol{M},\boldsymbol{H}_{\text{dc}};T\right)\equiv\frac{\partial^{2}M_{i}\left(T,\boldsymbol{H}_{\text{ext}}\right)}{\partial E_{j}\partial E_{k}}=\frac{\partial^{3}\log\mathcal{Z}_{\text{MF}}\left(\boldsymbol{M},\boldsymbol{H}_{\text{ext}};\beta\right)}{\partial E_{j}\partial E_{k}\partial\beta H_{\text{eff},i}}.
\end{equation}
The full thermodynamic state of the system depends on the temperature
$T$, the internal self-consistent magnetization $\boldsymbol{M}$,
and the static external field $\boldsymbol{H}_{\text{dc}}$. Because
of this, and the intrinsic nonlinearity in the problem generated by
the competition between the three components of the Heisenberg spins,
the magneto-electric tensor $\alpha_{i,jk}^{\left(2\right)}$ is a
complicated function of the magnetic state. 

\subsection{Ferromagnetic Control through the Edelstein-Zeeman Field}

Here we consider the effect of an Edelstein-Zeeman field $\boldsymbol{H}^{\text{EZ}}$
driven by an incident, linearly polarized, electric field. The infrared
incidence can be down-converted to THz emission through magnetic dipole
radiation from the sample. Focusing on THz emission parallel to the
crystallographic $\hat{a}=\hat{x}$ direction as reported in the main
text, we focus on the competing roles of the $M_{y}$ and $M_{z}$
components of the ferromagnetic moment in the sample. Here we will
self-consistently compute the ferromagnetic response to the Edelstein-Zeeman
field as a function of temperature, fluence, and polarization.

Parameterizing the polarization of the incident electric field to
be 
\begin{equation}
\hat{E}=\left(\cos\varphi\sin\psi\right)\hat{x}+\left(\sin\varphi\right)\hat{y}+\left(-\cos\varphi\cos\psi\right)\hat{z},
\end{equation}
 where $\psi$ is the angle of incidence relative to the $ab$-plane,
and $\varphi$ is the transverse polarization angle. Here $\hat{x}$
and $\hat{z}$ are chosen to coincide with the crystallographic $\hat{a}$
and $\hat{c}$ axes. We then write the $H_{y}$ and $H_{z}$ components,
from Eqs.~(\ref{supp-eq:Sy_by_symmetry}) and (\ref{supp-eq:Sz_by_symmetry})
as 
\begin{align}
H_{y}\left(\omega\right) & =J_{sd}\left|\boldsymbol{E}\left(\omega_{1}\right)\right|\left|\boldsymbol{E}\left(\omega_{2}\right)\right|\bigg\{\chi_{x,xy}^{\left(2\right)}\left(\omega;\,\omega_{1},\omega_{2}\right)\left(\cos^{2}\varphi\sin^{2}\psi-\sin^{2}\varphi\right)-\chi_{x,xx}^{\left(2\right)}\left(\omega;\,\omega_{1},\omega_{2}\right)\left(\sin2\varphi\sin\psi\right)\nonumber \\
 & \phantom{=J_{sd}\left|\boldsymbol{E}\left(\omega_{1}\right)\right|\left|\boldsymbol{E}\left(\omega_{2}\right)\right|\bigg\{}+\chi_{x,yz}^{\left(2\right)}\left(\omega;\,\omega_{1},\omega_{2}\right)\left(\cos^{2}\varphi\sin2\psi\right)-\chi_{x,xz}^{\left(2\right)}\left(\omega;\,\omega_{1},\omega_{2}\right)\left(\sin2\varphi\cos\psi\right)\bigg\}+2\pi H_{\text{dc},y}\delta\left(\omega\right),\\
\nonumber \\H_{z}\left(\omega\right) & =J_{sd}\left|\boldsymbol{E}\left(\omega_{1}\right)\right|\left|\boldsymbol{E}\left(\omega_{2}\right)\right|\left\{ \chi_{z,zz}^{\left(2\right)}\left(\omega;\,\omega_{1},\omega_{2}\right)\left(\cos^{2}\varphi\cos^{2}\psi\right)+\chi_{z,xx}^{\left(2\right)}\left(\omega;\,\omega_{1},\omega_{2}\right)\left(\cos^{2}\varphi\sin^{2}\psi+\sin^{2}\varphi\right)\right\} \nonumber
\\
&\phantom{=}+2\pi H_{\text{dc},z}\delta\left(\omega\right).
\end{align}
 Approximating the infrared down-conversion to THz emission as a rectification
process, with $\omega\ll\omega_{1,2}$, then we set $\omega\rightarrow0$.
By energy conservation in the magneto-electric response (see Eq.~(\ref{supp-eq:dynamical_nonlinear_edelstein_tensor}))
, then $\omega_{1}=-\omega_{2}\equiv\omega_{\text{IR}}$ and we find
$\left|\boldsymbol{E}\left(\omega_1\right)\right|\left|\boldsymbol{E}\left(\omega_2\right)\right|\rightarrow\left|\boldsymbol{E}\left(\omega_{\text{IR}}\right)\right|\left|\boldsymbol{E}\left(-\omega_{\text{IR}}\right)\right|\equiv I_{0}\left(\omega_{\text{IR}}\right)$,
where $I_{0}\left(\omega\right)$ is proportional to the intensity
of monochromatic light. Neglecting the external dc field, we find
that 
\begin{align}
 H_{y}^{\text{EZ}}\left(0\right)&=J_{sd}I_{0}\left(\omega_{\text{IR}}\right)\bigg\{\chi_{x,xy}^{\left(2\right)}\left(0;\,\omega_{\text{IR}},-\omega_{\text{IR}}\right)\left(\cos^{2}\varphi\sin^{2}\psi-\sin^{2}\varphi\right)-\chi_{x,xx}^{\left(2\right)}\left(0;\,\omega_{\text{IR}},-\omega_{\text{IR}}\right)\left(\sin2\varphi\sin\psi\right)\nonumber \\
 & \phantom{=_{}J_{sd}I_{0}\left(\omega_{\text{IR}}\right)\bigg\{}+\chi_{x,yz}^{\left(2\right)}\left(0;\,\omega_{\text{IR}},-\omega_{\text{IR}}\right)\left(\cos^{2}\varphi\sin2\psi\right)-\chi_{x,xz}^{\left(2\right)}\left(0;\,\omega_{\text{IR}},-\omega_{\text{IR}}\right)\left(\sin2\varphi\cos\psi\right)\bigg\},\\
\nonumber \\H_{z}^{\text{EZ}}\left(0\right)&=J_{sd}I_{0}\left(\omega_{\text{IR}}\right)\left\{ \chi_{z,zz}^{\left(2\right)}\left(0;\,\omega_{\text{IR}},-\omega_{\text{IR}}\right)\left(\cos^{2}\varphi\cos^{2}\psi\right)+\chi_{z,xx}^{\left(2\right)}\left(0;\,\omega_{\text{IR}},-\omega_{\text{IR}}\right)\left(\cos^{2}\varphi\sin^{2}\psi+\sin^{2}\varphi\right)\right\} .
\end{align}
 Substituting these expressions into the self-consistent mean-field
equations of state, Eq.~(\ref{supp-eq:mean-field-magnetization}),
we obtain the temperature and fluence dependence of the $M_{y}$ and
$M_{z}$ components of the magnetization. At normal incidence, $\psi=\pi/2$,
and the Edelstein-Zeeman field expresses a $d$-wave rotational symmetry
in the $ab$-plane as 
\begin{align}
H_{y}^{\text{EZ}}\left(0;\psi=\frac{\pi}{2}\right) & =J_{sd}I_{0}\left(\omega_{\text{IR}}\right)\left\{ \chi_{x,xy}^{\left(2\right)}\left(0;\,\omega_{\text{IR}},-\omega_{\text{IR}}\right)\cos\left(2\varphi\right)-\chi_{x,xx}^{\left(2\right)}\left(0;\,\omega_{\text{IR}},-\omega_{\text{IR}}\right)\sin\left(2\varphi\right)\right\} \nonumber \\
 & \equiv J_{sd}I_{0}\left(\omega_{\text{IR}}\right)\chi_{x,\parallel}\left(0;\,\omega_{\text{IR}},-\omega_{\text{IR}}\right)\cos\left(2\varphi-2\delta_{\parallel}\right),\label{supp-eq:normal_incidence_HyEZ}
\end{align}
where 
\begin{equation}
\chi_{x,\parallel}\left(0;\,\omega_{\text{IR}},-\omega_{\text{IR}}\right)\equiv\sqrt{\left[\chi_{x,xx}^{\left(2\right)}\left(0;\,\omega_{\text{IR}},-\omega_{\text{IR}}\right)\right]^{2}+\left[\chi_{x,xy}^{\left(2\right)}\left(0;\,\omega_{\text{IR}},-\omega_{\text{IR}}\right)\right]^{2}},\quad\tan\left(2\delta_{\parallel}\right)=-\frac{\chi_{x,xx}^{\left(2\right)}\left(0;\,\omega_{\text{IR}},-\omega_{\text{IR}}\right)}{\chi_{x,xy}^{\left(2\right)}\left(0;\,\omega_{\text{IR}},-\omega_{\text{IR}}\right)}.
\end{equation}
The angle $\delta_{\parallel}$ is obtained by finding the polarization
$\varphi$ such that $H_{y}^{\text{EZ}}$ is maximized, which corresponds
to the polarization that maximizes the emission. Within the Main Text,
it is obtained for normal incidence to be $\delta_{\parallel}\approx50^{\text{o}}$.
The out-of-plane component, meanwhile, is isotropic, and given by
\begin{equation}
H_{z}^{\text{EZ}}\left(0;\psi=\frac{\pi}{2}\right)=J_{sd}I_{0}\left(\omega_{\text{IR}}\right)\chi_{z,xx}^{\left(2\right)}\left(0;\,\omega_{\text{IR}},-\omega_{\text{IR}}\right).\label{supp-eq:normal_incidence_HzEZ}
\end{equation}

\begin{figure}
\begin{centering}
\includegraphics[width=0.6\textwidth]{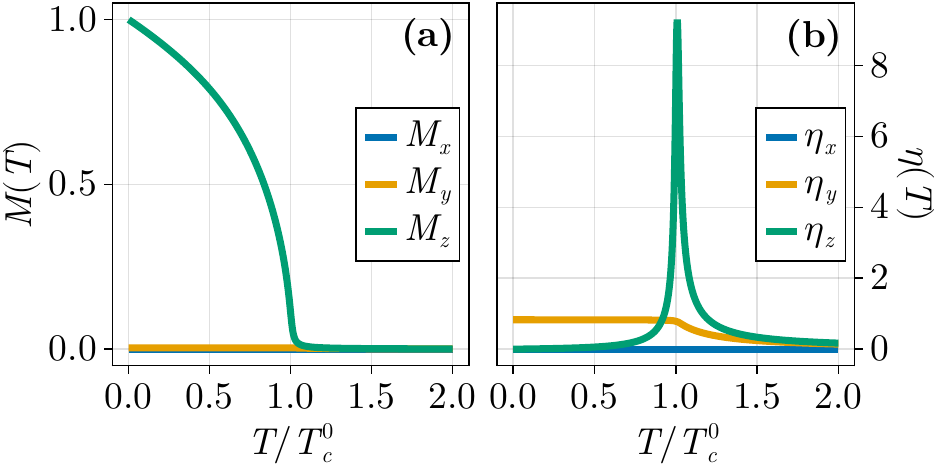}
\par\end{centering}
\caption{Magnetization and fluence susceptibility as a function of temperature
$T$ at normal incidence near equilibrium ($I_{0}=0.005\,T_{c}^{0}$).
The equilibrium Curie temperature is $T_{c}^{0}$. In both plots,
the uniaxial anisotropy is set to be $\kappa=0.2$, and the electric
field polarization is chosen to maximize the $M_{y}$ response. \textbf{(a)}
Magnetization behavior exhibiting a uniaxial magnetic ground state
with nonzero $M_{z}$ at low temperatures. Likewise, \textbf{(b) }the
out-of-plane fluence susceptibility $\eta\equiv\partial M/\partial I_{0}$
exhibits shows a divergence at $T_{c}^{0}$ for the out-of-plane component
$\eta_{z}$. The in-plane susceptibility $\eta_{y}$ remains maximized
throughout the magnetic phase. }\label{supp-fig:Magnetization-and-fluence-susceptibilities-equilibrium}

\end{figure}

Supplementary Figure~\ref{supp-fig:Magnetization-and-fluence-susceptibilities-equilibrium}(a)
shows the mean-field solution for the magnetization $\boldsymbol{M}$
as a function of temperature $T$ obtained by solving Eq.~(\ref{supp-eq:mean-field-magnetization})
self-consistently in the presence of a small Edelstein-Zeeman field
($I_{0}\left(\omega_{\text{IR}}\right)=0.005\,T_{c}^{0}$) and a uniaxial
anisotropy of $\kappa=0.2$. We take the polarization of the light
to be $\varphi=50^{\text{o}}$ relative to the crystallographic $\hat{a}$-axis
to maximize $H_{y}^{\text{EZ}}$. The expected mean-field magnetic
phase is observed at temperatures below $T_{c}^{0}$, which corresponds
to the spontaneous transition in the limit that $\left|\boldsymbol{H}_{\text{ext}}\right|=0$.
Because an increasing fluence, $I_{0}\left(\omega_{\text{IR}}\right)$,
acts as an increasing external Zeeman field, it motivates a ``differential
fluence susceptibility'' defined explicitly by 
\begin{equation}
\eta_{\alpha}\equiv\frac{\partial M_{\alpha}}{\partial I_{0}},
\end{equation}
 which characterizes the polarizability of the ferromagnetic moment
to the Edelstein-Zeeman field. Supplementary Figure~\ref{supp-fig:Magnetization-and-fluence-susceptibilities-equilibrium}(b)
shows the fluence susceptibility obtained for the same system as in
Supplementary Figure~\ref{supp-fig:Magnetization-and-fluence-susceptibilities-equilibrium}(a)
through numerical differentiation using a second-order central difference.
The finite-difference of the fluence $\Delta I_{0}=10^{-6}\,T_{c}^{0}$.
The crossover from the paramagnetic to ferromagnetic phases is indicated
by the peak in $\eta_{z}$, which diverges in the zero-fluence limit
at the spontaneous ferromagnetic phase transition. The in-plane fluence
susceptibility, $\eta_{y}$, is meanwhile nonzero and peaked below
the Curie temperature, indicating the polarizability of the in-plane
degrees of freedom. This constant susceptibility decreases as the
uniaxial anisotropy increases. 

\begin{figure}
\begin{centering}
\includegraphics[width=0.6\textwidth]{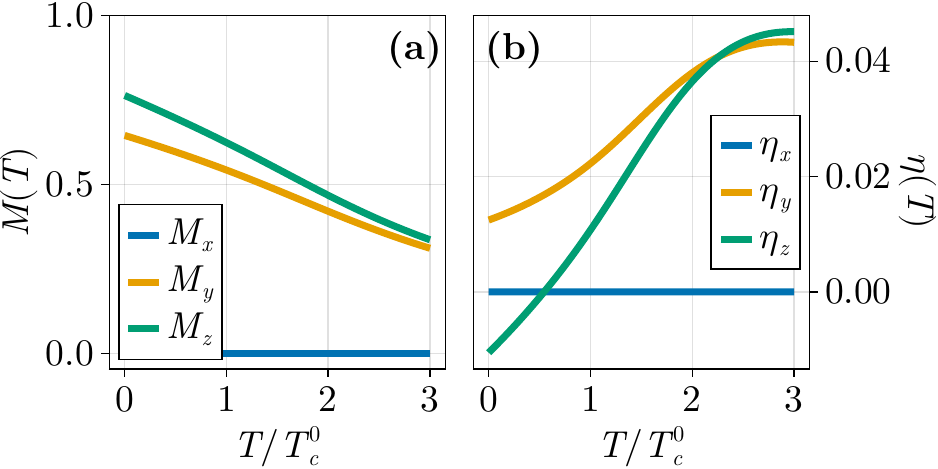}
\par\end{centering}
\caption{Magnetization and fluence susceptibility as a function of temperature
$T$ at normal incidence at high fluence ($I_{0}=5\,T_{c}^{0}$).
The equilibrium Curie temperature is $T_{c}^{0}$. In both plots,
the uniaxial anisotropy is set to be $\kappa=0.2$, and the electric
field polarization is chosen to maximize the $M_{y}$ response. \textbf{(a)}
Nonzero magnetization far above the Curie temperature driven by the
Edelstein-Zeeman field. \textbf{(b) }At high-fluence, the differential
fluence susceptibility $\eta\equiv\partial M/\partial I_{0}$ is an
increasing function of temperature. Whereas $\eta_{y}$ is strictly
positive, $\eta_{z}$ is negative at low temperatures which exemplifies
the intrinsic competition between the $M_{z}$ and $M_{y}$ components
near saturation. }\label{supp-fig:Magnetization-and-fluence-susceptibilities-high-fluence}
\end{figure}

With the low-fluence limit of the phase diagram established, we proceed
with the high-fluence regime. This maps onto a ferromagnetic system
in a strong external field. Supplementary Figure~\ref{supp-fig:Magnetization-and-fluence-susceptibilities-high-fluence}(a)
shows the magnetization as a function of temperature, obtained with
the uniaxial anisotropy ($\kappa=0.2$) for an Edelstein-Zeeman field
with the same incidence and polarization ($\varphi=50^{\text{o}}$).
The high-fluence was chosen to be $I_{0}\left(\omega_{\text{IR}}\right)=5\,T_{c}^{0}$
to probe saturation behavior associated with the Heisenberg ferromagnet
through Edelstein-Zeeman fields with symmetry-allowed nonzero projections
in both the $y$ and $z$ directions (see Eqs.~(\ref{supp-eq:normal_incidence_HyEZ})
and (\ref{supp-eq:normal_incidence_HzEZ})). The second-order Edelstein
susceptibilities were chosen as $\chi_{x,\parallel}=\chi_{z,xx}=0.5\,\left(T_{c}^{0}\right)^{-1}$.
The result in Supplementary Figure~\ref{supp-fig:Magnetization-and-fluence-susceptibilities-high-fluence}(a)
is that the magnetization can be induced well-above the Curie temperature,
as expected from a paramagnet placed in a strong magnetic field. Supplementary Figure~\ref{supp-fig:Magnetization-and-fluence-susceptibilities-high-fluence}(b)
shows the differential fluence susceptibility evaluated under the
same conditions. Both $\eta_{y}$ and $\eta_{z}$ exhibit non-linear
behavior due to the competition between (i) the Heisenberg nature
of the magnetization near saturation, (ii) the high effective Zeeman
field induced by the nonlinear Edelstein effect, and (iii) the temperature.
Whereas $\eta_{y}$ is a strictly nonzero increasing function of temperature,
$\eta_{z}$ is actually negative at low temperatures. This implies
that the $M_{z}$ component tends to \emph{decrease} as the fluence
is \emph{increased }at normal incidence (for sufficiently high fluence).
Ultimately this is due to the magnetic moment $\boldsymbol{M}$ rotating
from the $z$-axis towards the $y$-axis under manipulation through
the Edelstein-Zeeman field.

\begin{figure}
\begin{centering}
\includegraphics[width=0.6\textwidth]{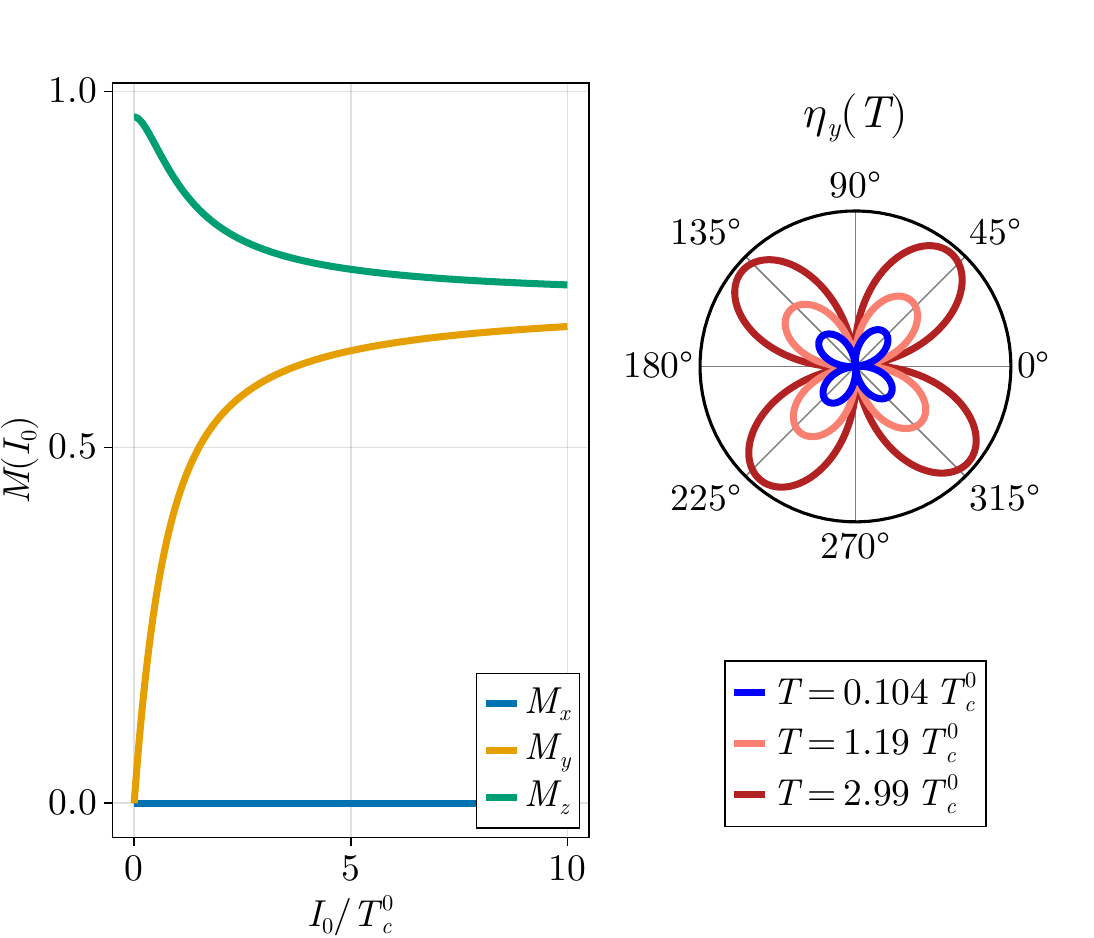}
\par\end{centering}
\caption{Magnetic response to the Edelstein-Zeeman field at normal incidence.
The equilibrium Curie temperature is $T_{c}^{0}$. In both plots,
the uniaxial anisotropy is set to be $\kappa=0.2$. \textbf{\emph{Left:}}
Magnetization as a function of fluence $I_{0}$ at low temperature
($T=0.104\,T_{c}^{0}$) for an electric field polarization that maximizes
the $M_{y}$ response. When the Edelstein-Zeeman field polarizes the
magnetization with nonzero $M_{y}$, the intrinsic competition between
the $M_{y}$ and $M_{z}$ components of the Heisenberg magnet decreases
the out-of-plane component $M_{z}$. This continues as a function
of fluence as the in-plane magnetization continues to increase. \textbf{\emph{Right:}}
Fixed-temperature, high-fluence ($I_{0}=5\,T_{c}^{0}$) differential
fluence susceptibility, $\eta_{y}=\partial M_{y}/\partial I_{0}$,
as a function of in-plane electric field polarization $\varphi$. Here
$\varphi=0$ corresponds to the electric field polarized along the crystallographic
$\hat{a}$-axis. The angle $\delta_{\parallel}=50^{\text{o}}$ was
chosen to match the experimental data in the Main Text. As the temperature
increases, the magnitude of the fluence susceptibility increases,
reflecting the tendency for thermal fluctuations to unpin the ferromagnetic
spin response to strong Edelstein-Zeeman fields.  }\label{supp-fig:Magnetic-response-to-Edelstein-Zeeman-field}

\end{figure}

Supplementary Figure~\ref{supp-fig:Magnetic-response-to-Edelstein-Zeeman-field}
explores the control of ferromagnetism through the Edelstein-Zeeman
field for different parameters within the phase diagram. Holding the
nonlinear Edelstein susceptibilities fixed at $\chi_{x,\parallel}=\chi_{z,xx}=0.5\,\left(T_{c}^{0}\right)^{-1}$
as in Supplementary Figure~\ref{supp-fig:Magnetization-and-fluence-susceptibilities-high-fluence},
the left plot of Supplementary Figure~\ref{supp-fig:Magnetic-response-to-Edelstein-Zeeman-field}
shows the magnetization as a function of fluence $I_{0}$ at the fixed
temperature $T=0.107\,T_{c}^{0}$. Taking $T_{c}^{0}=67\,\text{K}$
for \CGT corresponds to $T=7\,\text{K}$, as studied in the main
text. Just as in Supplementary Figures~\ref{supp-fig:Magnetization-and-fluence-susceptibilities-equilibrium}
and \ref{supp-fig:Magnetization-and-fluence-susceptibilities-high-fluence},
the uniaxial anisotropy was set to be $\kappa=0.2$ while the electric
field polarization was taken to be $\varphi=50^{\text{o}}$ to maximize
the $M_{y}$ response. The fact that $M_{z}$ is a decreasing function
of $I_{0}$, while $M_{y}$ is increasing, demonstrates the negative
fluence susceptibility observed in Supplementary Figure~\ref{supp-fig:Magnetization-and-fluence-susceptibilities-high-fluence}(b),
and again shows that in order polarize a Heisenberg magnet in the
plane, it must \emph{decrease} any initial projection along the $z$-axis.
The $M_{y}$ component, meanwhile, it shown to be an increasing function
of fluence with negative concavity due to the saturating effects of
a strong external Zeeman field. By further decreasing the out-of-plane magnetic response, in-plane magnetization can exceed the out-of-plane component for sufficiently high fluence. This is shown in Fig.~4 of the Main Text.

The right column of Supplementary Figure~\ref{supp-fig:Magnetic-response-to-Edelstein-Zeeman-field}
meanwhile shows the magnitude of the in-plane differential fluence
susceptibility at high-fluence ($I_{0}=5\,T_{c}^{0}$) and fixed temperature
$T$, swept over all polarization angles at normal incidence. The
$d$-wave symmetry is apparent in the response, and for high fluence,
the effect of increasing the temperature results in a larger fluence
susceptibility. The temperatures in the plot correspond to the 7,
80, and 200~K, as shown in Fig.~2 of the Main Text.

\bibliographystyle{apsrev4-2}

\bibliography{SIbib, theory_SI}